\documentclass[12pt]{elsarticle}
\usepackage{nomencl}
\usepackage{ifthen}
\renewcommand{\nomgroup}[1]{%
\ifthenelse{\equal{#1}{S}}{\item[\textbf{Subscripts}]}{%
\ifthenelse{\equal{#1}{V}}{\item[\textbf{Variables}]}{%
\ifthenelse{\equal{#1}{S}}{\item[\textbf{sets}]}{}}}
}
\makenomenclature
\usepackage{amssymb}

\usepackage[a4paper]{geometry}  
\usepackage{times} 
\usepackage{xurl} 
\usepackage[italian,english]{babel}
\usepackage{latexsym} 
\usepackage{xcolor}

\makeatletter
\newcommand{\@BIBLABEL}{\@emptybiblabel}
\newcommand{\@emptybiblabel}[1]{}
\usepackage{graphicx}
\usepackage{latexsym}
\usepackage{amssymb}
\usepackage{bm}
\usepackage{amsmath}
\usepackage{epsfig}
\usepackage{caption}
\usepackage{subcaption}
\usepackage{graphicx}
\usepackage{caption}
\usepackage{subcaption}
\usepackage{mwe} 
\usepackage{morefloats}
\usepackage{devanagari}
\usepackage{xcolor}
\usepackage{epstopdf}
\usepackage[symbol]{footmisc}
\begin{document}\makeatletter
\def\ps@pprintTitle{%
 \let\@oddfoot\@empty
 \let\@evenfoot\@empty
 }
\makeatother
\begin{frontmatter}


\title{New Higher-Order Super-Compact Finite Difference Scheme to Study Three-Dimensional Natural Convection and Entropy Generation in Non-Newtonian Fluids}

\renewcommand{\thefootnote}{*}
\makeatletter
\def\ps@pprintTitle{%
  \let\@oddhead\@empty
  \let\@evenhead\@empty
  \let\@oddfoot\@empty
  \let\@evenfoot\@oddfoot
}
\makeatother
\author{\textbf{Ashwani Punia$^{1}$, Rajendra K. Ray$^{2}$}\footnote[1]{Corresponding author : Rajendra K. Ray, rajendra@iitmandi.ac.in} \\
  1,2. School of Mathematical and Statistical Sciences, Indian Institute of Technology Mandi,\\
  Mandi, Himachal Pradesh, 175005, India \\
 {\tt mr.punia11@gmail.com}, {\tt rajendra@iitmandi.ac.in}}
\begin{abstract}
This work introduces a new higher-order super-compact (HOSC) implicit finite difference scheme for analyzing three-dimensional (3D) natural convection and entropy generation in non-Newtonian fluids. The proposed scheme achieves fourth-order accuracy in space and second-order accuracy in time while utilizing only seven directly adjacent grid points of the compact stencils at the \((n+1)\) time level, making it highly efficient and super compact. To the best of our knowledge, this is the first higher-order accurate finite difference scheme proposed to study 3D natural convection and entropy generation in non-Newtonian fluids. A time-marching technique is applied, where pressure corrections are addressed using a modified artificial compressibility method. The scheme is applied to the power-law model of non-Newtonian fluids, investigating both shear-thinning and shear-thickening effects on natural convection and entropy generation within a 3D cubic cavity. In the numerical simulations, the Prandtl number remains constant at $Pr = 1.0$, while various Rayleigh numbers ($Ra = 10^2$, $10^3$, $10^4$, $10^5$) and power-law indices ($n = 0.75$, $1.0$, $1.25$) are considered. Results are presented in terms of isotherms, streamlines, Nusselt numbers, Bejan numbers, and local entropy generation rates. The validation of the proposed scheme demonstrates excellent agreement with existing benchmark results. The numerical study reveals that the $n$ and $Ra$ have significant impacts on flow dynamics, heat transfer, and entropy generation. As $Ra$ increases, the maximum value of average Nusselt number ($Nu_{\text{avg}}$) also increases, whereas an opposite trend is observed with $n$ values. Shear-thinning fluids demonstrate the highest convection efficiency compared to Newtonian and shear-thickening fluids at any specific $Ra$.
\end{abstract}
\end{frontmatter}
\section{Introduction}
Non-Newtonian fluids, characterized by their variable viscosity that changes with the shear rate, have drawn significant attention in both fundamental research and practical applications. Unlike Newtonian fluids, which maintain a constant viscosity under different shear conditions, non-Newtonian fluids exhibit more complex flow behavior. Fluids inside a cavity are often assumed to be Newtonian, meaning they have constant viscosity. However, in many natural and industrial systems, fluids actually behave as non-Newtonian, making the Newtonian assumption less accurate for real-world cases. Fluids such as molten polymers, food products, paints, organic materials, inks, and adhesives frequently exhibit strong non-Newtonian properties, making them important in various engineering applications. Natural convection, a major mechanism of heat transfer, is widely used in engineering systems. Studying natural convection in non-Newtonian fluids within cavities is essential due to its broad range of applications. Examples include polymer manufacturing, nuclear reactors, food processing, oil drilling, geophysical systems, and electronic cooling systems, where non-Newtonian natural convection plays a key role \cite{Raisi_2016}. Despite its wide range of applications, limited studies have explored their natural convection behavior in 2D and 3D enclosures, mainly because of the complexities associated with shear rate-dependent viscosity fluctuations.\\

To the best of the author's knowledge, Ozoe and Churchill \cite{Ozoe_1972} conducted the first study on natural convection of non-Newtonian fluids within an enclosure. They investigated 2D natural convection in two types of non-Newtonian fluids, specifically Ostwald-de Waele (power law model) and Ellis fluids, within a shallow horizontal cavity heated from below and cooled from above. The results are found to be independent of the initial conditions, and it is also observed that the critical Rayleigh number increases with the flow behavior index. Kim et al. \cite{Kim_2003} used the finite volume approach to investigate transient buoyant convection in a 2D square enclosure filled with an incompressible non-Newtonian power-law fluid. They considered the horizontal wall to be insulated, the left vertical wall to be cold, and the right vertical wall to be maintained at a constant temperature. They found that a drop in the power-law index ($n$) results in more convective activity and improved total heat transfer for high $Ra$ and moderate Prandtl numbers ($Pr$). Lamsaadi et al. \cite{Lamsaadi_2006} used both analytical and numerical simulations to study transient natural convection of non-Newtonian power-law fluids in a shallow 2D cavity. To solve the governing equations, they used a uniform mesh size, second-order central finite difference approach. Their cavity featured long, insulated horizontal walls and short vertical walls, where the vertical walls are subjected to constant heat flux for heating and cooling. Their results show that the non-Newtonian power-law behavior significantly affects the properties of heat transmission and fluid flow. In comparison to Newtonian fluids (where \( n = 1 \)), shear-thinning behavior (with \( 0 < n < 1 \)) enhances fluid circulation and convective heat transfer, whereas shear-thickening behavior (with \( n > 1 \)) results in the opposite effect. Lamsaadi et al. \cite{Lamsaadi_2006_1} also conducted a numerical study on steady natural convection in non-Newtonian power-law fluids within a tilted two-dimensional rectangular slit. The two-dimensional governing equations are solved using the simple central-difference finite difference method. This investigation is carried out in the following ranges: \(-180^\circ \leq \Phi \leq 180^\circ\), \(0.6 \leq n \leq 1.4\) for the power-law index, and \(10 \leq \text{Ra} \leq 10^5\) for the Rayleigh number. They found that the characteristics of fluid flow and heat transfer rate are found to be independent of any increase in the aspect ratio (\( A \)) and the generalized Prandtl number (\( Pr \)) when these parameters are sufficiently large (\( A \geq 12 \) and \( \text{Pr} \geq 100 \)). Additionally, the cavity's rotation has a major impact on the rate of heat transmission for a given $Ra$. The influence of cavity rotation grew as the value of \(n\) dropped, and the maximum rate of heat transfer is achieved when the cavity is heated from the bottom. 2D steady-state simulations of laminar natural convection in square enclosures with differentially heated sides under constant wall temperatures is carried out by Tarun et al. \cite{Turan_2011}. They employed the power-law fluid model and the ANSYS FLUENT commercial package to simulate the behavior of the fluids. They found that for both Newtonian and power-law fluids, the mean Nusselt number \( Nu \) increases as $Ra$ values increase. Notably, the \( Nu \) values for power-law fluids with \( n < 1 \) (indicating shear-thinning behavior) exceed those of Newtonian fluids at the same nominal Rayleigh number \(Ra\), while the \( Nu \) values for power-law fluids with \( n > 1 \) (indicating shear-thickening behavior) are lower. This difference is attributed to the enhanced (for \( n < 1 \)) or diminished (for \( n > 1 \)) convective transport. Ternik and Rudolf \cite{Ternik_2013} explored the natural convection of a non-Newtonian nanofluid in a 2D square cavity with isothermal sidewalls and adiabatic horizontal walls. They solved the governing differential equations using the standard finite volume method, with the hydrodynamic and thermal fields coupled through the Boussinesq approximation. They found that for a given set of solid volume fraction values \( \phi \), an increase in the nanofluid Rayleigh number (\( Ra_{nf} \)) leads to a strengthening of buoyancy forces relative to viscous forces. Later, Ternik and Buchmeister \cite{Ternik_2015} carried out a numerical investigation on laminar natural convection in a 2D square enclosure that was filled with a non-Newtonian fluid. The governing equations are solved using the conventional finite volume approach, and the viscous behavior of the non-Newtonian fluids is described by power-law model. Vinogradov et al. \cite{Vinogradov_2015} conducted a numerical study on 2D steady natural convection of non-Newtonian shear thickening power-law liquids in an inclined 2D cavity. The finite volume approach is used to solve the governing equations. They discovered that in the horizontal position, the heat transfer rate for square-shaped cavities (\( \text{AR} = 1 \)) is much lower than in the vertical 90° position, where the vertical walls are conducting. Additionally, they noted that the thermal behavior of rectangular cavities (\( \text{AR} > 1 \)) becomes more complex due to the formation of Bénard cells. Loenko et al. \cite{Loenko_2019} studied the natural convection of a non-Newtonian power-law fluid in a 2D square cavity containing a heat-generating element. They employed the dimensionless stream function-vorticity formulation of the Navier-Stokes equations, which are solved using the finite difference method. Their findings revealed that as the power-law index \(n\) increases, the flow and heat transfer in the cavity slows down. Consequently, for pseudoplastic fluids, heat is removed more efficiently from the energy source. Additionally, it is noted that for larger values of \(n\), the system reaches a steady state more quickly. Laminar unsteady natural convection in a 2D square cavity with an interior circular cylinder filled with a non-Newtonian fluid is investigated by Pandey et al. \cite{Pandey_2020}. For various Rayleigh number (\( \text{Ra} \)) values, they examined how the inside circular cylinder's vertical movement affected the flow and heat transfer processes. Kim and Reddy \cite{Kim_2021} formulated a mixed least-squares finite element model using spectral/hp approximations to examine 3D natural convection in a non-Newtonian fluid governed by the Carreau–Yasuda model. The study explored how varying the Carreau–Yasuda model parameters affected flow behavior. Their results demonstrated that as the power-law index \(n\) decreases, the heat transfer rate at the surfaces significantly increases. Loenko et al. \cite{Loenko_2021} investigated the unsteady natural convection of a non-Newtonian fluid within a 2D enclosure subjected to sinusoidal wall temperature variations over time. The time derivatives are approximated using a first-order finite difference method, while the spatial partial derivatives are discretized using second-order finite differences. The thermal properties of a power-law fluid in a lid-driven 2D enclosure with discrete heating are examined by Roy et al. \cite{Roy_2023}. They utilized the finite element method to solve the governing Navier-Stokes and energy equations. Their research indicates that the highest heat transfer performance occurs in the natural convection scenario of the pseudo-plastic fluid, while the lowest heat transfer is observed when the fluid is dilatant. Hasan et al. \cite{Hasan_2024} investigated the natural convection of power-law fluids in a differentially heated 3D cubic cavity using a mesoscopic multiple-relaxation-time (MRT) lattice Boltzmann method (LBM) accelerated by a graphics processing unit (GPU). Different $Ra$ and power-law indices ($n$) are incorporated to investigate heat transfer and entropy generation. Their results indicate that the temperature distribution is more prominent near the heated wall, especially for power-law indices \(n = 0.7\) and \(n = 1.0\), while having a less pronounced effect at \(n = 1.4\). Ahmed et al. \cite{Ahmed_2024} investigated natural convection and entropy generation in a non-Newtonian flow inside an irregularly shaped 2D cavity filled with molten polymer, using a method based on the Lattice Boltzmann Method (LBM). They found that as the Rayleigh number (\(Ra\)) increased, heat transfer enhanced. In contrast to dilatant fluids, heat transfer in pseudo-plastic fluids increased as the power-law index (\(n\)) decreased from \(n = 1\) to \(n = 0.5\). Additionally, the Bejan number ($Be$) showed substantial growth across all wall ratios (WRs) with increasing \(n\) index. Fard et al. \cite{Fard_2024} studied free convection heat transfer for non-Newtonian power-law fluids using the meshless local Petrov-Galerkin (MLPG) method. Their enhanced approach applies a unity weighting function and expresses the governing equations in the form of the vorticity-stream function. They displayed the temperature distributions, average Nusselt number, and vertical velocity for various $Pr$ and $Ra$, and compared the findings with those from mesh-based techniques. Based on the comparisons, the graphs are found to be in good agreement with the results of traditional mesh-based approaches. Keddar et al. \cite{Keddar_2024} focused on a numerical investigation of natural convection of a non-Newtonian viscoplastic fluid within a 3D cubic enclosure. The viscoplastic behavior are characterized using the Bingham model, and the 3D steady governing equations are solved using commercial CFD software FLUENT. Their results indicate that as the Bingham number increases, the flow velocity decreases, ultimately causing the flow to stop completely.\\

Over the years, significant attention has been given to the study of natural convection in non-Newtonian fluids using various numerical techniques, including finite element and finite volume methods. However, there remains a scarcity of higher-order finite difference schemes applied to such problems, particularly in three dimensions. Traditional lower-order schemes often lack the accuracy required to resolve complex flow patterns and thermal gradients in non-Newtonian fluids. This limitation leaves a notable research gap that needs to be further investigated. Also, A review of the literature reveals that most of the studies are confined to 2D analysis, mainly due to the high complexity involved in extending such investigations to three dimensions. However, real fluid flow phenomena are inherently three-dimensional in nature, and simplifying the problem to 2D often overlooks critical aspects of the flow behavior. To address these challenges, we propose a new 3D higher-order super-compact (HOSC) implicit finite difference scheme, designed to achieve high spatial and temporal accuracy while maintaining computational efficiency. \\
 Higher-order compact (HOC) finite difference schemes are already well-established as one of the leading methods for capturing complex 2D fluid flow dynamics \cite{Spotz_1995, Kalita_2004, Ray_2010}. For the three-dimensional case, Kalita \cite{Kalita_2014} first introduced the higher-order super-compact scheme to study the Newtonian fluids and found its second-order accuracy in time and fourth-order accuracy in space. Recently, Punia and Ray \cite{Punia_2024} expanded this approach to examine the natural convection of Newtonian fluids within a 3D cubic cavity. They found that this scheme effectively captures 3D heat transfer and fluid dynamics with high accuracy. They also extended this scheme to study power-law fluids in the lid-driven 3D cubic cavity without incorporating the energy/heat equation \cite{Punia_2024_1}. Their results show excellent agreement with existing literature, confirming the remarkable accuracy, consistency, and reliability of the scheme in capturing the complex non-Newtonian fluid behavior. To date, no advanced finite difference technique has been explicitly designed for analyzing the 3D heat transport and natural convection in non-Newtonian power-law fluids.\\

Thus, the novelty of this work lies in the development of new Higher-Order Super-Compact (HOSC) scheme specifically designed to investigate the 3D natural convection and heat transfer of non-Newtonian fluids. This newly developed HOSC scheme is fourth-order accurate in space and second-order accurate in time, utilizing only seven directly adjacent grid points of the compact stencils at the \((n+1)\) time level, making it not only highly accurate but also optimizes computational efficiency. The power-law model of non-Newtonian fluids is employed in this study, which allows for the examination of both shear-thinning and shear-thickening effects. The impact of the power-law index and the Rayleigh number on flow patterns, heat transfer, and entropy generation inside the cubic cavity is thoroughly investigated. Due to the ineffectiveness of conventional iterative solvers such as Successive Over-Relaxation (SOR) and Gauss-Seidel methods for this highly non-linear system of equations, especially when the coefficient matrix is not diagonally dominant, we utilize an advanced iterative solver—the hybrid Bi-Conjugate Gradient Stabilised (BiCGSTAB) method. Pressure is calculated with a pressure-correction approach based on the modified artificial compressibility technique. This approach provides efficiency, simplicity, and ease of implementation. Entropy generation is a critical parameter in thermodynamic optimization and irreversibility analysis, as it provides insight into the efficiency of energy transfer processes within the fluid. By incorporating entropy generation analysis, this study not only advances the understanding of fluid flow and heat transfer in non-Newtonian fluids but also contributes to the field of thermodynamic analysis. The significance of this research is its capacity to accurately capture the natural convection of non-Newtonian shear-thickening and shear-thinning fluids, which are commonly seen in various industrial applications, including polymer processing, food manufacturing, and biomedical engineering.

\section{Problem Statement and Solution Approach}
\label{sec:Problem Description and Discretization of Governing Equations}
\subsection{Problem Statement}
\begin{figure}[htbp]
 \centering
 \vspace*{5pt}%
 \hspace*{\fill}%
\begin{subfigure}{0.50\textwidth}     
    \centering
    \includegraphics[width=\textwidth]{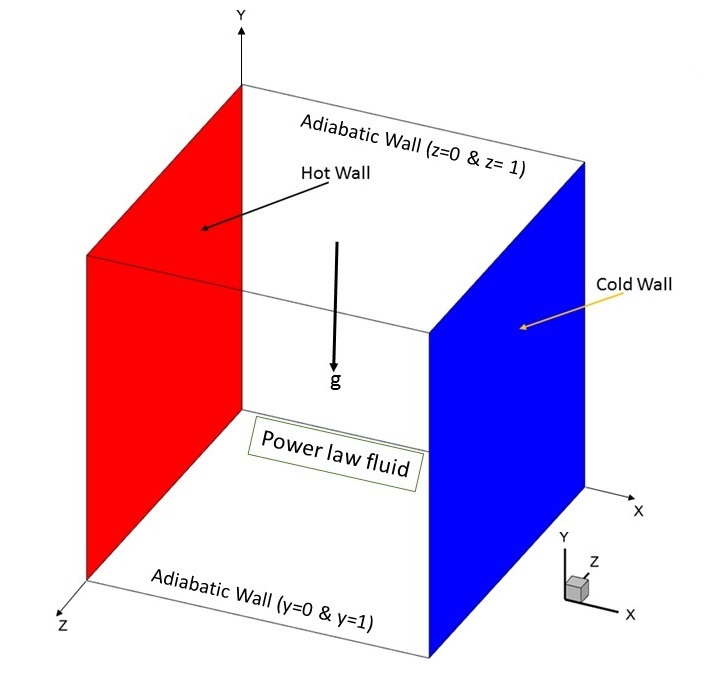}%
    \captionsetup{skip=5pt}%
    \caption{(a)}
    \label{fig:Cavity_3d}
  \end{subfigure}%
 \begin{subfigure}{0.46\textwidth}        
   \centering
    \includegraphics[width=\textwidth]{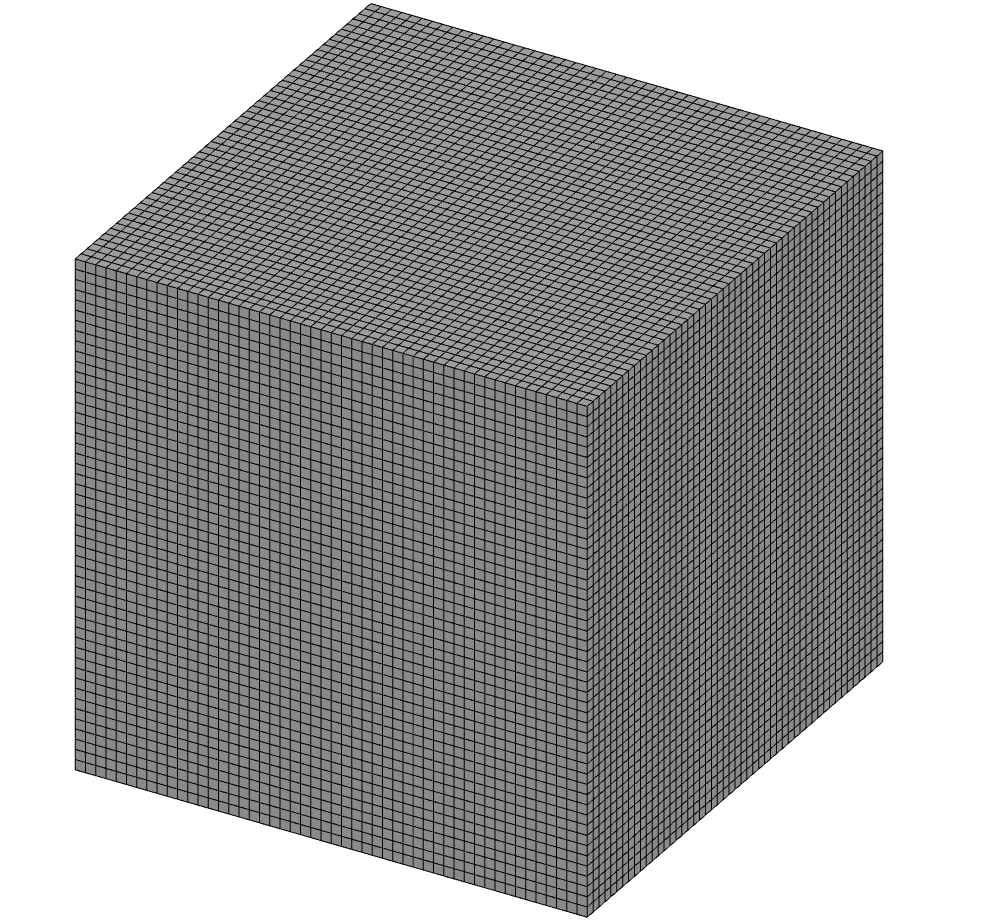}%
    \captionsetup{skip=5pt}%
    \caption{(b)}
    \label{fig:3d_grids}
  \end{subfigure}
  \hspace*{\fill}
  \vspace*{8pt}%
  \hspace*{\fill}%
  \caption{ (a) Schematic diagram of the proposed problem in 3D cavity and (b) Corresponding uniform meshing with a resolution of $51\times51\times51$.}
  \label{fig:Sche_diag}
\end{figure}

This research focuses on the three-dimensional unsteady natural convection of an incompressible power-law fluid confined within a closed cubic cavity. The right wall of the cubic cavity is assumed to be maintained at a cold temperature, while the left wall is held at a constant warm temperature.
 The remaining walls are assumed to be thermally insulated. Figure \ref{fig:Sche_diag} presents a schematic diagram of the problem (Figure \ref{fig:Sche_diag} (a)), as well as the grid layout (Figure \ref{fig:Sche_diag}(b)) in the Cartesian coordinate system. Gravity acts vertically downward, perpendicular to the \( xz \) plane. The cavity contains fluids that follow a power-law behavior. The dimensionless forms of the momentum, energy, and continuity equations for incompressible, three-dimensional, transient flow are presented below \cite{Dhiman_2006, Bilal_2021}.
\begin{equation}\label{Main_governing_eq_1}
\frac{\partial u}{\partial t} + v \frac{\partial u}{\partial y} + w \frac{\partial u}{\partial z} + u \frac{\partial u}{\partial x} = -\frac{\partial p}{\partial x} + Pr \left[ \frac{\partial \tau^{*}_{yx}}{\partial y} + \frac{\partial \tau^{*}_{xx}}{\partial x} + \frac{\partial \tau^{*}_{zx}}{\partial z}\right]
\end{equation}

\begin{equation}\label{Main_governing_eq_2}
\frac{\partial v}{\partial t} + u \frac{\partial v}{\partial x} + w \frac{\partial v}{\partial z} + v \frac{\partial v}{\partial y} = -\frac{\partial p}{\partial y} + Pr \left[ \frac{\partial \tau^{*}_{yy}}{\partial y} + \frac{\partial \tau^{*}_{xy}}{\partial x} + \frac{\partial \tau^{*}_{zy}}{\partial z}\right] + Ra \ Pr \ \theta
\end{equation}

\begin{equation}\label{Main_governing_eq_3}
\frac{\partial w}{\partial t} + u \frac{\partial w}{\partial x} + v \frac{\partial w}{\partial y} + w \frac{\partial w}{\partial z} = -\frac{\partial p}{\partial z} + Pr \left[ \frac{\partial \tau^{*}_{yz}}{\partial y} + \frac{\partial \tau^{*}_{xz}}{\partial x} + \frac{\partial \tau^{*}_{zz}}{\partial z}\right]
\end{equation}

\begin{equation}\label{Main_governing_eq_energy}
\frac{\partial \theta}{\partial t} + v \frac{\partial \theta}{\partial y} + w \frac{\partial \theta}{\partial z} + u \frac{\partial \theta}{\partial x} = \frac{\partial^2 \theta}{\partial y^2} + \frac{\partial^2 \theta}{\partial x^2} + 
 \frac{\partial^2 \theta}{\partial z^2}
\end{equation}

\begin{equation} \label{Main_governing_eq_4} 
\frac{\partial v}{\partial y}+\frac{\partial w}{\partial z}+\frac{\partial u}{\partial x}=0 
\end{equation}
The power-law model, which is expressed by the following equation, describes the fluid behavior:

\begin{equation}\label{Main_governing_eq_6}
\tau^{*}_{ij} = 2 \eta \varepsilon_{i,j}
\end{equation}
$\text{where } \varepsilon_{i j}=\frac{1}{2}\left(\frac{\partial u_j}{\partial x_i}+\frac{\partial u_i}{\partial x_j}\right) \text{is the strain rate tensor.}$
and $\eta$ denotes the apparent non-dimensionalized viscosity, which is defined as 
\begin{equation} \label{Main_governing_eq_7} 
\eta = I_2^{\frac{n-1}{2}} 
\end{equation}
In this context, $I_2$ indicates the second invariant of the strain rate tensor, serving as a measure of the fluid's deformation intensity. 
By using this, the viscosity can be represented as \cite{Zhuang_2018}: 
\begin{equation} \label{Main_governing_eq_8} 
\begin{aligned}
\eta = & \left[2 \left(\left(\frac{\partial w}{\partial z}\right)^2 + \left(\frac{\partial v}{\partial y}\right)^2 + \left(\frac{\partial u}{\partial x}\right)^2 \right) + \left(\frac{\partial u}{\partial z} + \frac{\partial w}{\partial x}\right)^2 \right. \\ 
& \left. \left(\frac{\partial u}{\partial y} + \frac{\partial v}{\partial x}\right)^2 + \left(\frac{\partial v}{\partial z} + \frac{\partial w}{\partial y}\right)^2  \right]^{\frac{n-1}{2}}
\end{aligned}
\end{equation}
The following dimensionless parameters are introduced when the associated governing equations are non-dimensionalized \cite{Bilal_2021}:
$$
Ra = \frac{g \beta \Delta T L^{2n+1} \rho}{K \alpha^n} ,\quad  Pr = \frac{L^{2(1-n)} K}{\rho \alpha^{2-n}}, \quad Nu = \frac{L h}{k}
$$
Here, $K$ represents the thermal conductivity, $L$ indicates the length of the cube, $\alpha$ denotes the thermal diffusivity , and $\rho$ represents the fluid density.
All governing equation variables are non-dimensionalized, and flow in a cube with $L = 1$ is taken into consideration. 
The following momentum equations are obtained by utilizing the continuity equation (\ref{Main_governing_eq_4}) and equations (\ref{Main_governing_eq_6}) - (\ref{Main_governing_eq_8}) into equations (\ref{Main_governing_eq_1}) - (\ref{Main_governing_eq_3}), and then expressing the equations in their conservative form \cite{Dhiman_2006}. \\

\begin{equation}\label{Main_governing_eq_8}
\begin{aligned}
\frac{\partial u}{\partial t} + v \frac{\partial u}{\partial y} + w \frac{\partial u}{\partial z} + u \frac{\partial u}{\partial x}=-\frac{\partial p}{\partial x}+ Pr \cdot \eta \left(\frac{\partial^2 u}{\partial y^2}+ \frac{\partial^2 u}{\partial x^2} +\frac{\partial^2 u}{\partial z^2} \right) \\ + 2 \cdot Pr \left(\varepsilon_{xx} \frac{\partial \eta }{\partial x} +\varepsilon_{yx} \frac{\partial \eta }{\partial y}  + \varepsilon_{zx} \frac{\partial \eta}{\partial z}\right)
\end{aligned}
\end{equation}

\begin{equation}\label{Main_governing_eq_9}
\begin{aligned}
\frac{\partial v}{\partial t} + u \frac{\partial v}{\partial x} + w \frac{\partial v}{\partial z} + v \frac{\partial v}{\partial y}=-\frac{\partial p}{\partial y}+Pr \cdot \eta \left(\frac{\partial^2 v}{\partial y^2}+ \frac{\partial^2 v}{\partial x^2}+ \frac{\partial^2 v}{\partial z^2} \right)  \\ + 2 \cdot Pr \left(\varepsilon_{xy} \frac{\partial \eta }{\partial x} +\varepsilon_{yy} \frac{\partial \eta }{\partial y}  + \varepsilon_{zy} \frac{\partial \eta}{\partial z}\right)
\end{aligned}
\end{equation}

\begin{equation}\label{Main_governing_eq_10}
\begin{aligned}
\frac{\partial w}{\partial t} + u \frac{\partial w}{\partial x} + v \frac{\partial w}{\partial y} + w \frac{\partial w}{\partial z} =-\frac{\partial p}{\partial z}+Pr \cdot \eta \left(\frac{\partial^2 w}{\partial y^2}+ \frac{\partial^2 w}{\partial x^2}+ \frac{\partial^2 w}{\partial z^2}\right)  \\ + 2 \cdot Pr \left(\varepsilon_{xz} \frac{\partial \eta }{\partial x} +\varepsilon_{yz} \frac{\partial \eta }{\partial y}  + \varepsilon_{zz} \frac{\partial \eta}{\partial z}\right)
\end{aligned}
\end{equation}
and, the energy and continuity equations remain as previously defined:
\begin{equation}\label{Main_governing_eq_energy_new}
\frac{\partial \theta}{\partial t} + v \frac{\partial \theta}{\partial y} + w \frac{\partial \theta}{\partial z} + u \frac{\partial \theta}{\partial x} = \frac{\partial^2 \theta}{\partial y^2} + \frac{\partial^2 \theta}{\partial x^2} + 
 \frac{\partial^2 \theta}{\partial z^2}
\end{equation}
\begin{equation} \label{Main_governing_eq_4_new} 
\frac{\partial v}{\partial y}+\frac{\partial w}{\partial z}+\frac{\partial u}{\partial x}=0 
\end{equation}
The following are the velocity, temperature, and pressure dimensionless boundary conditions:\\
1. On the cubic cavity's left heated wall ($x=0$):
\[
w = 0, \quad v = 0, \quad u = 0
\]
\[
\theta = 1, \quad \frac{\partial p}{\partial x}=0
\]
2. On the cubic cavity's right cold wall ($x=1$):
\[
w = 0, \quad v = 0, \quad u = 0
\]
\[
\theta = 0, \quad \frac{\partial p}{\partial x}=0
\]
3. On all other faces of the cavity: \[
V = 0, \quad \frac{\partial \theta}{\partial n}=0, \quad \frac{\partial p}{\partial n}=0
\]
Where, $V=(u,v,w)$ and $n$ is the normal direction to the wall.  \\ 

\subsection{Discretization and Solution Procedure}
This section focuses on the discretization of the nonlinear coupled transport equations (\ref{Main_governing_eq_8})-(\ref{Main_governing_eq_energy_new}) using a uniform mesh. For this, we consider a generalized form of the convection-diffusion-reaction partial differential equation (PDE) with the transport variable $\chi$. By selecting appropriate variable coefficients, this generalized PDE can effectively represent the governing momentum and energy equations, allowing for a flexible and accurate discretization approach. The general PDE can be expressed as follows, capturing the essential features of convection, diffusion, and reaction processes for the transport variable ``$\chi$":
\begin{equation}
\begin{aligned}
& L(x, y, z,t)\frac{\partial \chi}{\partial t}+H^{*}(x, y, z, t) \frac{\partial \chi}{\partial y}+G^{*}(x, y, z, t) \frac{\partial \chi}{\partial x}+I^{*}(x, y, z, t) \frac{\partial \chi}{\partial z} \\
& +K^{*}(x, y, z, t) \chi-\nabla^2 \chi=E^{*}(x, y, z, t),
\end{aligned}\label{eq1}
\end{equation}
Here, all variable coefficients ``$L, H^{*}, G^{*},I^{*}, E^{*}$", and ``$K^{*}$" are function of $ x, y, z,$ and $t$. Since viscosity is constant in Newtonian fluids, it is sufficient to regard ``$L$" as a constant. But $L$ must also be regarded as a variable for non-Newtonian fluids, where the viscosity changes with shear force. This introduces significant complexity in the discretization process, making it much more challenging compared to the Newtonian case. Additionally, the inclusion of the supplementary energy equation and the interaction between temperature and velocity increases the complexity of analyzing heat transfer in non-Newtonian fluids. Consequently, to accurately analyze non-Newtonian fluids, it is essential to treat $L$ as a function of $x$, $y$, $z$, and $t$, reflecting the variable nature of viscosity across space and time. Setting appropriate boundary conditions for the domain is essential to guaranteeing a well-posed and physically relevant problem formulation. A uniform mesh with increments $h$, $k$, and $l$ in the $x$, $y$, and $z$ directions, respectively, is used to discretize the cubical problem domain. During the discretization procedure, we first solve equation (\ref{eq1}) using the Forward-Time Central-Space (FTCS) method, approximating the temporal and spatial derivatives using the Taylor series expansion. By employing the FTCS approximation, Eq. (\ref{eq1}) can be expressed as follows at the generic node $(i, j, k)$:
\begin{equation} \begin{aligned}
\left(L \delta_t^{+}+H^{*}\delta_y+G^{*} \delta_x+I^{*}\delta_z-\delta_z^2-\delta_x^2-\delta_y^2 + K^{*}\right) \chi_{i j k}-\xi_{i j k}=E^{*}_{i j k},
\end{aligned} \label{eq2_} \end{equation} 
For the first and second-order central differences along the spatial variables \(z\), \(y\), and \(x\), respectively, the operators \(\delta_z\), \(\delta_z^2\), \(\delta_y\), \(\delta_y^2\), \(\delta_x\), and \(\delta_x^2\) are represented in the equation above. Meanwhile, \(\delta_t^{+}\) denotes the first-order forward difference in the time variable. At the three-dimensional grid point \((x_i, y_j, z_k)\), the value of the transport variable \(\chi\) is indicated by the term \(\chi_{ijk}\). Using a uniform time step \(\Delta t\), the present numerical approximation introduces a truncation error \(\xi_{ijk}\) that quantifies the error caused by the discretization procedure. This error can be expressed as:
\begin{equation}
\begin{aligned}
\xi_{i j k}= & {\left[ \frac{\Delta t}{2} L \frac{\partial^2 \chi}{\partial t^2}-\frac{h^2}{12}\left(\frac{\partial^4 \chi}{\partial x^4}-2 G^{*} \frac{\partial^3 \chi}{\partial x^3}\right) -\frac{k^2}{12}\left(\frac{\partial^4 \chi}{\partial y^4} - 2 H^{*} \frac{\partial^3 \chi}{\partial y^3}\right)\right.} \\
& \left.-\frac{l^2}{12}\left(\frac{\partial^4 \chi}{\partial z^4}-2 I^{*} \frac{\partial^3 \chi}{\partial z^3}\right)\right]_{i j k}+O\left(\Delta t^2, h^4, k^4, l^4\right).
\end{aligned}\label{eq3}
\end{equation}

For higher spatial and temporal accuracy of Eq. (\ref{eq1}), a compact approximation is employed for the leading term's derivatives, as outlined in Eq. (\ref{eq3}). This strategy significantly reduces the truncation error, leading to a more accurate formulation. This approach uses Eq. (\ref{eq1}) as an auxiliary relationship to calculate higher-order derivatives, specifically the third and fourth spatial derivatives, as well as the second derivative with respect to time. For example, the variables $L, G^{*}, H^{*}, I^{*}, K^{*}$, and $E^{*}$ are subjected to a backward temporal difference technique to calculate the second derivative with respect to time, while the transport variable $\chi$ is subjected to a forward difference approach \cite{Kalita_2014}. This makes it possible to represent the derivatives in the first right-hand side term of Eq. (\ref{eq3}) as follows:
\begin{equation}
\begin{aligned}
\left.L_{i j k} \frac{\partial^2 \chi}{\partial t^2}\right|_{i j k}= & \left(\delta_z^2+\delta_X^2+\delta_Y^2-K^{*}_{i j k}-H^{*}_{i j k} \delta_y-G^{*}_{i j k} \delta_x-I^{*}_{i j k} \delta_z\right) \delta_t^{+} \chi_{i j k} \\
& -\left(\delta_t^{-} H^{*}_{i j k} \delta_y+\delta_t^{-} G^{*}_{i j k} \delta_x+\delta_t^{-} K^{*}_{i j k}+\delta_t^{-} I^{*}_{i j k} \delta_z \right)\chi_{i j k} +\delta_t^{-} E^{*}_{i j k}\\
& \delta_t^{-} L_{i j k} \delta_t^{+} \chi_{i j k} +O\left(\Delta t, h^2, k^2, l^2\right),
\end{aligned} \label{eq4}
\end{equation}
The operators $\delta_t^{+}$ and $\delta_t^{-}$ correspond to the first-order forward and backward time differences, respectively. Similarly, the spatial operators $\delta_x$, $\delta_y$, and $\delta_z$ represent the first-order central differences in the $x$, $y$, and $z$ directions, while $\delta_x^2$, $\delta_y^2$, and $\delta_z^2$ denote the second-order central differences in space for the respective variables. Likewise, the remaining derivatives in equation (\ref{eq3}) can be obtained using equation (\ref{eq1}). The following estimate is obtained by replacing $\xi_{i j k}$ in Eq. (\ref{eq2_}) with the higher derivatives (calculated using Eq.(\ref{eq1})) in equation (\ref{eq3}), obtaining an order of accuracy $O\left(\Delta t^2, h^4, k^4, l^4\right)$ for the primary governing equation (\ref{eq1}).
$$
\begin{aligned}
L_{i j k} \bigg[1 & - \left(\frac{\Delta t}{2 L_{i j k}}-\frac{k^2}{12}\right)\left(\delta_y^2-H^{*}_{i j k} \delta_y\right)
-\left(\frac{\Delta t}{2 L_{i j k}}-\frac{h^2}{12}\right)\left(\delta_x^2-G^{*}_{i j k} \delta_x\right) \\
& -\left(\frac{\Delta t}{2 L_{i j k}}-\frac{l^2}{12}\right)\left(\delta_z^2-I^{*}_{i j k} \delta_z\right) \\
&  + \frac{\Delta t}{2 L_{i j k}} \left( K^{*}_{i j k} + \delta_t^{-} L_{i j k} +  \frac{h^2}{6\Delta t}\delta_x^2 L_{i j k} + \frac{k^2}{6\Delta t}\delta_y^2 L_{i j k}
+ \frac{l^2}{6\Delta t}\delta_z^2 L_{i j k}\right)\\
&  +\frac{\Delta t}{12\Delta t L_{i j k}} \left(- h^2 G^{*}_{i j k}\delta_x L_{i j k}
- l^2 I^{*}_{i j k}\delta_z L_{i j k}
- k^2 H^{*}_{i j k}\delta_y L_{i j k}
\right)\\
& + \left(
+ (\frac{k^2}{6 L_{i j k}}\delta_y L_{i j k})\delta_y
+(\frac{h^2}{6 L_{i j k}}\delta_x L_{i j k})\delta_x
+ (\frac{l^2}{6 L_{i j k}}\delta_z L_{i j k})\delta_z\right) \bigg]\delta_t^{+} \chi_{i j k} \\
& +\left(-\beta_{i j k} \delta_y^2-\alpha_{i j k} \delta_x^2-\gamma_{i j k} \delta_z^2+M2_{i j k} \delta_y+M4_{i j k}+M1_{i j k} \delta_x+M3_{i j k} \delta_z\right) \chi_{i j k} \\
& -\frac{k^2+l^2}{12}\left(\delta_y^2 \delta_z^2-I^{*}_{i j k} \delta_y^2 \delta_z-H^{*}_{i j k} \delta_y \delta_z^2-q1_{i j k} \delta_y \delta_z\right) \chi_{i j k}
\end{aligned}
$$
\begin{equation}
\begin{aligned}
& -\frac{h^2+k^2}{12}\left(\delta_x^2 \delta_y^2-H^{*}_{i j k} \delta_x^2 \delta_y-G^{*}_{i j k} \delta_x \delta_y^2-p1_{i j k} \delta_x \delta_y\right) \chi_{i j k} \\
& -\frac{l^2+h^2}{12}\left(\delta_z^2 \delta_x^2-G^{*}_{i j k} \delta_z^2 \delta_x-I^{*}_{i j k} \delta_z \delta_x^2-r1_{i j k} \delta_z \delta_x\right) \chi_{i j k} = R_{i j k}\\
\end{aligned}\label{eq5}
\end{equation}
The coefficients $ M2_{i j k}, M4_{i j k}, M1_{i j k}, M3_{i j k}, \alpha_{i j k}, \gamma_{i j k}, \beta_{i j k}, R_{i j k}, q1_{i j k},  p1_{i j k}$ and $r1_{i j k}$ are as below:
$$
\begin{aligned}
& \alpha_{i j k}=\left({G^{*}}_{i j k}^2-{K^{*}}_{i j k}-2 \delta_x {G^{*}}_{i j k}\right)\frac{h^2}{12}+1 \text {, } \\
& \gamma_{i j k}=\left({I^{*}}_{i j k}^2-{K^{*}}_{i j k}-2 \delta_z {I^{*}}_{i j k}\right)\frac{l^2}{12}+1\text {, } \\
& \beta_{i j k}=\left({H^{*}}_{i j k}^2-{K^{*}}_{i j k}-2 \delta_y {H^{*}}_{i j k}\right)\frac{k^2}{12}+1 \text {, } \\
& M1_{i j k}=\left[1+\left(\delta_x^2-{G^{*}}_{i j k} \delta_x\right)\frac{h^2}{12}+\left(\delta_y^2-{H^{*}}_{i j k} \delta_y\right)\frac{k^2}{12}+\left(\delta_z^2-{I^{*}}_{i j k} \delta_z\right)\frac{l^2}{12}+\frac{\Delta t}{2} \delta_t^{-}\right] {G^{*}}_{i j k} \\
& -\frac{h^2}{12}\left({G^{*}}_{i j k}-2 \delta_x\right) {K^{*}}_{i j k}, \\
& M2_{i j k}=\left[1+\left(\delta_x^2-{G^{*}}_{i j k} \delta_x\right)\frac{h^2}{12}+\left(\delta_y^2-{H^{*}}_{i j k} \delta_y\right)\frac{k^2}{12}+\left(\delta_z^2-{I^{*}}_{i j k} \delta_z\right)\frac{l^2}{12}+\frac{\Delta t}{2} \delta_t^{-}\right] {H^{*}}_{i j k} \\
& -\frac{k^2}{12}\left({H^{*}}_{i j k}-2 \delta_y\right) {K^{*}}_{i j k} \text {, } \\
& M3_{i j k}=\left[1+\left(\delta_x^2-{G^{*}}_{i j k} \delta_x\right)\frac{h^2}{12}+\left(\delta_y^2-{H^{*}}_{i j k} \delta_y\right)\frac{k^2}{12}+\left(\delta_z^2-{I^{*}}_{i j k} \delta_z\right)\frac{l^2}{12}+\frac{\Delta t}{2} \delta_t^{-}\right] {I^{*}}_{i j k} \\
& -\frac{l^2}{12}\left({I^{*}}_{i j k}-2 \delta_z\right) {K^{*}}_{i j k} \text {, } \\
& M4_{i j k}=\left[1+\left(\delta_x^2-{G^{*}}_{i j k} \delta_x\right)\frac{h^2}{12}+\left(\delta_y^2-{H^{*}}_{i j k} \delta_y\right)\frac{k^2}{12}+\left(\delta_z^2-{I^{*}}_{i j k} \delta_z\right)\frac{l^2}{12}+\frac{\Delta  t}{2} \delta_t^{-}\right] {K^{*}}_{i j k}, \\
& R_{i j k}=\left[1+ \left(\delta_x^2-{G^{*}}_{i j k} \delta_x\right)\frac{h^2}{12}+\left(\delta_y^2-{H^{*}}_{i j k} \delta_y\right)\frac{k^2}{12}+\left(\delta_z^2-{I^{*}}_{i j k} \delta_z\right)\frac{l^2}{12}+\frac{\Delta t}{2} \delta_t^{-}\right] {E^{*}}_{i j k}, \\
& q1_{i j k}=-{H^{*}}_{i j k} {I^{*}}_{i j k}+\frac{2}{k^2+l^2}\left(l^2 \delta_z {H^{*}}_{i j k}+k^2 \delta_y {I^{*}}_{i j k}\right) \text {, } \\
& p1_{i j k}=-{G^{*}}_{i j k} {H^{*}}_{i j k}+\frac{2}{h^2+k^2}\left(k^2 \delta_y {G^{*}}_{i j k}+h^2 \delta_x {H^{*}}_{i j k}\right) \text {, } \\
& r1_{i j k}=-{I^{*}}_{i j k} {G^{*}}_{i j k}+\frac{2}{l^2+h^2}\left(h^2 \delta_x {I^{*}}_{i j k}+l^2 \delta_z {G^{*}}_{i j k}\right) . \\
&
\end{aligned}
$$
In other words, Equation (\ref{eq5}) can be rewritten as follows:
\begin{equation}
\begin{aligned}
L_1\chi_{i+1 j k}^{n+1}+L_2\chi_{i-1 j k}^{n+1}+L_3\chi_{i j+1 k}^{n+1}+L_4\chi_{i j-1 k}^{n+1}+L_5\chi_{i j k+1}^{n+1}+L_6\chi_{i j k-1}^{n+1}+L_7\chi_{i j k}^{n+1}=Q_1\chi_{i+1 j k}^{n}
\\+Q_2\chi_{i j k}^{n}+Q_3\chi_{i-1 j k}^{n}+Q_4\chi_{i j+1 k}^{n}+Q_5\chi_{i j-1 k}^{n}+Q_6\chi_{i j k+1}^{n}+ Q_7\chi_{i j k-1}^{n}+Q_8\chi_{i+1 j+1 k}^{n} \quad \quad  \\
+Q_9\chi_{i-1 j+1 k}^{n}+Q_{10}\chi_{i+1 j-1 k}^{n}+Q_{11}\chi_{i-1 j-1 k}^{n}+Q_{12}\chi_{i j+1 k+1}^{n}+Q_{13}\chi_{i j-1 k+1}^{n}+Q_{14}\chi_{i j+1 k-1}^{n} \\
+Q_{15}\chi_{i j-1 k-1}^{n}+Q_{16}\chi_{i+1 j k+1}^{n} +Q_{17}\chi_{i+1 j k-1}^{n}+Q_{18}\chi_{i-1 j k+1}^{n}+Q_{19}\chi_{i-1 j k-1}^{n}+\Delta t R_{i,j,k} \quad \quad \quad \quad \quad 
 \end{aligned}\label{eq_implicit}
\end{equation}
where,
$
\begin{aligned}
L_1=\left(-\frac{M_1G^{*}_{i j k}}{2h}+\frac{M_1}{h^2}+\frac{M_{15}}{2h}\right)L_{i j k}{, } \quad M_1=\left(-\frac{\Delta t}{2L_{i j k}}+\frac{h^2}{12}\right)\\
\end{aligned}
$

$
\begin{aligned}
L_2=\left(\frac{M_1G^{*}_{i j k}}{2h}+\frac{M_1}{h^2}-\frac{M_{15}}{2h}\right)L_{i j k}\\
\end{aligned}
$

$
\begin{aligned}
L_3=\left(-\frac{M_2H^{*}_{i j k}}{2K}+\frac{M_2}{k^2}+\frac{M_{16}}{2k}\right)L_{i j k}{, } \quad M_2=\left(-\frac{\Delta t}{2L_{i j k}}+\frac{k^2}{12}\right)\\
\end{aligned}
$

$
\begin{aligned}
L_4=\left(\frac{M_2H^{*}_{i j k}}{2K}+\frac{M_2}{k^2}-\frac{M_{16}}{2k}\right)L_{i j k}\\
\end{aligned}
$

$
\begin{aligned}
L_5=\left(-\frac{M_3I^{*}_{i j k}}{2l}+\frac{M_3}{l^2}+\frac{M_{17}}{2l}\right)L_{i j k}{, } \quad M_3=\left(-\frac{\Delta t}{2L_{i j k}}+\frac{l^2}{12}\right)\\
\end{aligned}
$

$
\begin{aligned}
L_6=\left(\frac{M_3I^{*}_{i j k}}{2l}+\frac{M_3}{l^2}-\frac{M_{17}}{2l}\right)L_{i j k}\\
\end{aligned}
$

$
\begin{aligned}
L_7=\left(1-\frac{2M_2}{k^2}-\frac{2M_1}{h^2}+ 2M_4\frac{M18}{\Delta t} + M_4K^{*}_{i j k}-\frac{2M_3}{l^2}\right)L_{i j k} {, } \quad M_4=\left(\frac{\Delta t}{2L_{i j k}}\right)\\
\end{aligned}
$

$
\begin{aligned}
Q_1=\left(L_{i j k}\frac{M_{25}}{2h} \frac{N1_{i j k}}{h^2}-\frac{2\Delta t M_5}{h^2k^2} + \frac{2\Delta t M_5G^{*}_{i j k}}{2hk^2} +\frac{N4_{i j k}}{2h} -\frac{2\Delta t M_7}{l^2k^2}+ \frac{2\Delta t M_7G^{*}_{i j k}}{2hl^2} \right)\\
\end{aligned}
$

$
\begin{aligned}
Q_2=L_{i j k}-2\left(\frac{N2_{i j k}}{k^2}+\frac{N1_{i j k}}{h^2}+\frac{N3_{i j k}}{l^2}\right) + \left(N8_{i j k}M4_{i j k}+N7_{i j k}K^{*}_{i j k}+\right)+ \frac{4\Delta t M_5}{h^2k^2} \\+\frac{4\Delta t M_6}{k^2l^2}+2N7_{i j k}\frac{M18}{\Delta t}+\frac{4\Delta t M_7}{h^2l^2} \quad \quad \quad \quad \quad \quad  \quad \quad \quad \quad \quad \quad \quad \quad \quad \quad \quad \\
\end{aligned}
$

$
\begin{aligned}
Q_3=\left(- L_{i j k}\frac{M_{25}}{2h}\frac{N1_{i j k}}{h^2}-\frac{2\Delta t M_5}{h^2k^2} - \frac{2\Delta t M_5 G^{*}_{i j k}}{2hk^2} +\frac{N4_{i j k}}{2h}-\frac{2\Delta t M_7}{l^2k^2}- \frac{2\Delta t M_7G^{*}_{i j k}}{2hl^2}\right)\\
Q_4=\left( L_{i j k}\frac{M_{26}}{2k}\frac{N2_{i j k}}{k^2}-\frac{2\Delta t M_5}{h^2k^2}+\frac{N5_{i j k}}{2k} + \frac{2\Delta t M_5H^{*}_{i j k}}{2h^2k} -\frac{2\Delta t M_6}{l^2k^2}+ \frac{2\Delta t M_6H^{*}_{i j k}}{2kl^2} \right)\\
Q_5=\left(- L_{i j k}\frac{M_{26}}{2k}\frac{N2_{i j k}}{k^2}-\frac{2\Delta t M_5}{h^2k^2}+\frac{N5_{i j k}}{2k} - \frac{2\Delta t M_5H^{*}_{i j k}}{2h^2k} -\frac{2\Delta t M_6}{l^2k^2}- \frac{2\Delta t M_6H^{*}_{i j k}}{2kl^2}\right)\\
Q_6=\left( L_{i j k}\frac{M_{27}}{2l}\frac{N3_{i j k}}{l^2}-\frac{2\Delta t M_6}{l^2k^2}+\frac{N6_{i j k}}{2l}  -\frac{2\Delta t M_7}{l^2h^2}+ \frac{2\Delta t M_6I^{*}_{i j k}}{2k^2l}+ \frac{2\Delta t M_7I^{*}_{i j k}}{2lh^2} \right)\\
Q_7=\left(- \frac{2\Delta t M_7I^{*}_{i j k}}{2lh^2} - L_{i j k}\frac{M_{27}}{2l}\frac{N3_{i j k}}{l^2}-\frac{2\Delta t M_6}{l^2k^2}-\frac{N6_{i j k}}{2l} - \frac{2\Delta t M_6I^{*}_{i j k}}{2k^2l} -\frac{2\Delta t M_7}{l^2h^2}\right)\\
\end{aligned}
$

$
\begin{aligned}
&&\\
Q_8=\left(-\frac{\Delta t M_5 p1^{*}_{i j k}}{4hk}+\frac{\Delta t M_5}{h^2k^2}- \frac{\Delta t M_5 H^{*}_{i j k}}{2h^2k}- \frac{\Delta t M_5 G^{*}_{i j k}}{2hk^2}\right)\\
Q_9=\left(\frac{\Delta t M_5 p1^{*}_{i j k}}{4hk}+\frac{\Delta t M_5}{h^2k^2}- \frac{\Delta t M_5 H^{*}_{i j k}}{2h^2k}+ \frac{\Delta t M_5 G^{*}_{i j k}}{2hk^2}\right)\\
Q_{10}=\left(\frac{\Delta t M_5 p1^{*}_{i j k}}{4hk}+\frac{\Delta t M_5}{h^2k^2}+ \frac{\Delta t M_5 H^{*}_{i j k}}{2h^2k}- \frac{\Delta t M_5 G^{*}_{i j k}}{2hk^2}\right)\\
Q_{11}=\left(-\frac{\Delta t M_5 p1^{*}_{i j k}}{4hk}+\frac{\Delta t M_5}{h^2k^2}+ \frac{\Delta t M_5 H^{*}_{i j k}}{2h^2k}+ \frac{\Delta t M_5 G^{*}_{i j k}}{2hk^2}\right)\\
Q_{12}=\left(-\frac{\Delta t M_6 q1^{*}_{i j k}}{4kl}+\frac{\Delta t M_6}{l^2k^2}- \frac{\Delta t M_6 I^{*}_{i j k}}{2k^2l}- \frac{\Delta t M_6 H^{*}_{i j k}}{2kl^2}\right)\\
Q_{13}=\left(\frac{\Delta t M_6 q1^{*}_{i j k}}{4kl}+\frac{\Delta t M_6}{l^2k^2}- \frac{\Delta t M_6 I^{*}_{i j k}}{2k^2l}+ \frac{\Delta t M_6 H^{*}_{i j k}}{2kl^2}\right)\\
Q_{14}=\left(\frac{\Delta t M_6 q1^{*}_{i j k}}{4kl}+\frac{\Delta t M_6}{l^2k^2}+ \frac{\Delta t M_6 I^{*}_{i j k}}{2k^2l}- \frac{\Delta t M_6 H^{*}_{i j k}}{2kl^2}\right)\\
Q_{15}=\left(-\frac{\Delta t M_6 q1^{*}_{i j k}}{4kl}+\frac{\Delta t M_6}{l^2k^2}+ \frac{\Delta t M_6 I^{*}_{i j k}}{2k^2l}+ \frac{\Delta t M_6 H^{*}_{i j k}}{2kl^2}\right)\\
Q_{16}=\left(-\frac{\Delta t M_7 r1^{*}_{i j k}}{4lh}+\frac{\Delta t M_7}{l^2h^2}- \frac{\Delta t M_7 G^{*}_{i j k}}{2l^h}- \frac{\Delta t M_7 I^{*}_{i j k}}{2lh^2}\right)\\
Q_{17}=\left(\frac{\Delta t M_7 r1^{*}_{i j k}}{4lh}+\frac{\Delta t M_7}{l^2h^2}- \frac{\Delta t M_7 G^{*}_{i j k}}{2l^h}+ \frac{\Delta t M_7 I^{*}_{i j k}}{2lh^2}\right)\\
Q_{18}=\left(\frac{\Delta t M_7 r1^{*}_{i j k}}{4lh}+\frac{\Delta t M_7}{l^2h^2}+ \frac{\Delta t M_7 G^{*}_{i j k}}{2l^h}- \frac{\Delta t M_7 I^{*}_{i j k}}{2lh^2}\right)\\
Q_{19}=\left(-\frac{\Delta t M_7 r1^{*}_{i j k}}{4lh}+\frac{\Delta t M_7}{l^2h^2}+ \frac{\Delta t M_7 G^{*}_{i j k}}{2l^h}+ \frac{\Delta t M_7 I^{*}_{i j k}}{2lh^2}\right)\\
\end{aligned}
$

$
\begin{aligned}
M_5=\left(\frac{h^2+k^2}{12}\right), \quad M_6=\left(\frac{k^2+l^2}{12}\right), \quad M_7=\left(\frac{l^2+h^2}{12}\right)
\end{aligned}
$

$
\begin{aligned}
N1_{i,j,k}=\left(\Delta t \alpha_{i,j,k}+L_{i j k}M_1 \right), \quad N2_{i,j,k}=\left(\Delta t \beta_{i,j,k} +L_{i j k}M_2\right)
\end{aligned}
$

$
\begin{aligned}
N3_{i,j,k}=\left(\Delta t \gamma_{i,j,k}+L_{i j k}M_3 \right), \quad N4_{i,j,k}=\left(-\Delta t M1_{i j k} -L_{i j k}M_1G^{*}_{i j k}\right)
\end{aligned}
$

$
\begin{aligned}
N5_{i,j,k}=\left(-\Delta t M2_{i j k}-L_{i,j,k}M_2H^{*}_{i j k} \right), \quad N6_{i,j,k}=\left(-\Delta t M3_{i j k} -L_{i,j,k}M_3I^{*}_{i j k}\right)
\end{aligned}
$

$
\begin{aligned}
N7_{i,j,k}=\left(M_4L_{i,j,k}\right), \quad N8_{i,j,k}=\left(-\Delta t \right) 
\end{aligned}
$

$
\begin{aligned}
M15_{i,j,k}=M25_{i,j,k}=\delta_x {L_{i j k}}\left(\frac{h^2}{6L_{i j k}}\right)
\end{aligned}
$

$
\begin{aligned}
M16_{i,j,k}=M26_{i,j,k}=\delta_y {L_{i j k}}\left(\frac{k^2}{6L_{i j k}}\right)
\end{aligned}
$

$
\begin{aligned}
M17_{i,j,k}=M27_{i,j,k}=\delta_z {L_{i j k}}\left(\frac{l^2}{6L_{i j k}}\right)
\end{aligned}
$

$
\begin{aligned}
& M18_{i j k}=\left[\frac{\Delta t \delta_t^{-}}{2} \left(\delta_x^2-{G^{*}}_{i j k} \delta_x\right)\frac{h^2}{12}+\left(\delta_y^2-{H^{*}}_{i j k} \delta_y\right)\frac{k^2}{12}+\left(\delta_z^2-{I^{*}}_{i j k} \delta_z\right)\frac{l^2}{12}\right] {L}_{i j k},\\
&
\end{aligned}
$

\begin{figure}
    \centering
    \includegraphics[width=0.8\textwidth]{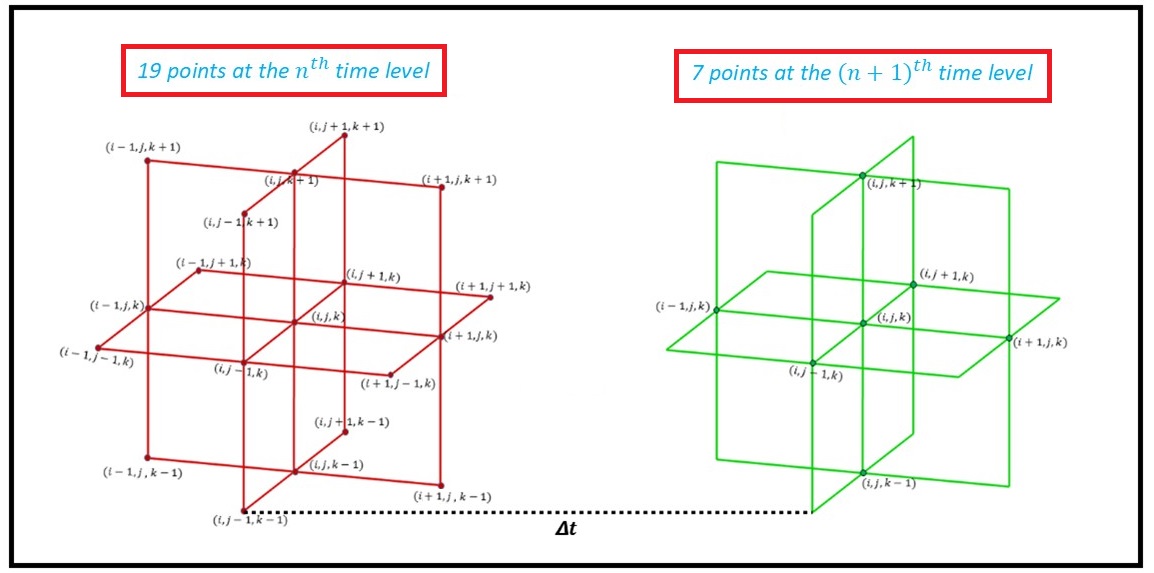}
    \caption{The super-compact unsteady (19, 7) stencil.}
    \label{fig:stencil}
\end{figure}

$\\
\begin{aligned}
 \quad
\end{aligned}
$
This results in the formulation of an implicit higher-order super-compact finite difference method. Utilizing a \((19,7)\) stencil, as shown in Figure \ref{fig:stencil}, the approach reduces computational complexity significantly by producing a compact seven-point stencil at the \((n + 1)^{th}\) time step. Remarkably, many high-order compact techniques designed for two-dimensional convection-diffusion equations, including those described in \cite{Spotz_1995,Kalita_2002,Ray_2017}, often require for a nine-point stencil at the \((n + 1)^{th}\) time level. Moreover, the Newtonian model has been the primary focus of the development of these techniques. On the other hand, even in three-dimensional scenarios, the HOSC method needs only seven-point stencil at the \((n + 1)^{th}\) time level when applied to even non-Newtonian fluids, where fluctuating viscosity contributes extra complexity. There are two major benefits to this scheme. As seen in Figure \ref{fig:stencil}, it first reduces to a seven-point stencil, requiring just the \((i, j, k)\) point and its six neighboring points at the \((n + 1)^{th}\) time level. Also, it removes the need for large number of corner points, which significantly lowers the overall number of points required for the approximation and improves computing efficiency.
In the discretization of the momentum and energy equations (Eqs. (\ref{Main_governing_eq_8}) - (\ref{Main_governing_eq_energy_new})), we define the transport variable ``$\chi$" for each equation as follows: $u$ for the x-momentum, $v$ for the y-momentum, $w$ for the z-momentum, and $\theta$ for the energy equation. The coefficients for momentum equations are defined as:\\
$L$ = $1/ (Pr \cdot \eta$),\\
$G^{*}$ = $u/ (Pr \cdot \eta) $,\\
$H^{*}$ = $v/(Pr \cdot \eta) $, \\
 $I^{*}$ = $w/ (Pr \cdot \eta) $ ,\\
$K^{*}$ = $0$, \\
$E^{*}$ denotes one of the three choices:  -$\frac{1}{ (Pr \cdot \eta)}\frac{\partial p}{\partial x}$ +  $2 \cdot Pr  \left(\varepsilon_{xx} \frac{\partial \eta }{\partial x} +\varepsilon_{yx} \frac{\partial \eta }{\partial y}  + \varepsilon_{zx} \frac{\partial \eta}{\partial z}\right)$ for $x$-momentum equation; 
-$\frac{1}{ (Pr \cdot \eta)}\frac{\partial p}{\partial y}$ +  $2 \cdot Pr \left(\varepsilon_{xy} \frac{\partial \eta }{\partial x} +\varepsilon_{yy} \frac{\partial \eta }{\partial y}  + \varepsilon_{zy} \frac{\partial \eta}{\partial z}\right)+ Ra \cdot Pr \cdot \theta$ for $y$-momentum equation;  
-$\frac{1}{ (Pr \cdot \eta)}\frac{\partial p}{\partial z}$ +  $2 \cdot Pr \left(\varepsilon_{xz} \frac{\partial \eta }{\partial x} +\varepsilon_{yz} \frac{\partial \eta }{\partial y}  + \varepsilon_{zz} \frac{\partial \eta}{\partial z}\right)$ for $z$-momentum equation. For the energy equation, the coefficients are \( L = 1 \), \( G^{*} = u \), \( H^{*} = v \), \( I^{*} = w \), and \( K^{*} = E^{*} = 0 \). In order to approximate pressure gradients, we need numerical techniques since there aren't any clear analytical formulas for pressure. At the interior grid points of the domain, we use central difference techniques, while we apply one-sided approximations at the boundary nodes of the domain. HOSC discretization of the governing equations (Eqs. (\ref{Main_governing_eq_8}) - (\ref{Main_governing_eq_10})) results in an algebraic system of equations with an asymmetric sparse coefficient matrix that lack of diagonal dominance at all grid locations. Consequently, conventional iterative methods such as Gauss-Seidel and Successive Over-Relaxation (SOR) are ineffective in this context. Hence, We use an advance iterative solver, i.e., Hybrid Biconjugate Gradient Stabilized (BiCGSTAB) approach without preconditioning \cite{Spotz_1995, Kalita_2014, Kelley_1995, Saad_2003}. Calculation of the pressure term (\( p \)) for the next time step comes after solving equations (\ref{Main_governing_eq_8}) - (\ref{Main_governing_eq_energy_new}) with constant initial pressure. Since there is no explicit pressure term in the primitive variable form of the Navier-Stokes equations, calculation of the pressure term is a major challenge. To solve the pressure calculation issue, we chose the modified compressibility method given by Cortes and Miller \cite{Cortes_1994}. This approach is chosen due to its effectiveness, simplicity, and convenience of use.
The modified continuity equation using this approach is as follow :
$$
\lambda \nabla \cdot \mathbf{v} + p=0 .
$$
We determine the highest absolute value of the velocity divergence ($\nabla \cdot \mathbf{v}$) at each time step after pressure gradients are calculated and momentum equations are resolved. We consider the pressure to have reached the required level of accuracy if this value is less than a predefined tolerance threshold. Otherwise, if the maximum divergence exceeds the specified threshold, we implement the following pressure correction step to correct the pressure \cite{Cortes_1994}:
$$
p^{n}=p^{o}-\lambda \nabla \cdot \mathbf{v} .
$$
In this context, $p^{n}$ represents the updated pressure, $p^{o}$ indicates the pressure value obtained from the previous pressure iteration, and $\lambda$ denotes the relaxation parameter.
\section{Scheme Validation and Sensitivity Analysis}
\label{sec:SENSITIVITY TESTS AND SCHEME VALIDATION}
\subsection{Grid Independence Test}
A grid independence test is conducted to ensure accuracy and efficiency in the numerical simulations. The computational domain is discretized using uniform grids. For this test, we consider three different grid sizes: \(11 \times 11 \times 11\), \(51 \times 51 \times 51\), and \(91 \times 91 \times 91\). The other parameters are maintained at constant values of \(\Delta t = 0.02\), \(n = 0.75\), and \(Ra = 10^4\). The temperature and velocity readings at two distinct observation positions, \((0.6, 0.6, 0.6)\) and \((0.4, 0.4, 0.4)\), which are close to the core region of cavity, are shown in Table \ref{grid_independent_test}. We selected the shear-thinning fluid case (\( n = 0.75 \)) for the grid independence test because accurately capturing the flow characteristics of shear-thinning fluids is inherently more complex. By ensuring grid adequacy for this more challenging case, we can be confident that the grid resolution will also be sufficient for the Newtonian (\( n = 1 \)) and shear-thickening (\( n = 1.25 \)) cases, which are less sensitive to refinement due to their relatively simpler flow dynamics. According to our assessment, there is only a maximum relative error of \(1.06\%\) between the calculated values on the \(51 \times 51 \times 51\) and \(91 \times 91 \times 91\) grids. This finding implies that the findings have reached grid independence and that further grid size increases have no discernible effect on the calculated results. Therefore, a mesh size of \(51 \times 51 \times 51\) is sufficient to accurately capture the heat transfer and fluid phenomena. Based on the grid independence test, we select a \(51 \times 51 \times 51\) mesh for the subsequent analysis.

{\small\begin{table}[htbp]
\caption{\small Temperature and velocity measurements at two key monitoring positions, ($0.60, 0.60, 0.60$) and ($0.40, 0.40, 0.40$), located near the center of the cavity with constant $Ra = 10^4$, $n=0.75$, $\Delta t = 0.02$, and $t=100$, using three different grid resolutions.}\label{grid_independent_test}
\centering
\resizebox{\textwidth}{!}{%
 \begin{tabular}{ccccccc}  \hline \hline
Location & Grid sizes    &  $\theta$ & $u$   &  $v$  &  $w$ & $\delta_e$(\%)    \\ \hline
\hline
(0.60, 0.60, 0.60) & (11 $\times$ 11 $\times$ 11) &  0.3952&  -11.4398 &   12.7982   &   0.7413  & --- \\
& (51 $\times$ 51 $\times$ 51)                    &  0.3766 &  -9.7684  &  13.9728  &   0.7168 & 14.4 \\
& (91 $\times$ 91 $\times$ 91)                     &  0.3725&  -9.6821 &  14.1210  &   0.7159  & 1.06  \\
\hline
(0.40, 0.40, 0.40) & (11 $\times$ 11 $\times$ 11) &  0.5339&  8.8872 &   -13.5572    &   -0.2822  & --- \\
& (51 $\times$ 51 $\times$ 51)                    &  0.5709 &  9.9121  & -12.1467  &   -0.2979 & 11.5 \\
& (91 $\times$ 91 $\times$ 91)                     &  0.5765 &  9.9362 &  -12.1075 &   -0.2991  & 0.98  \\
\hline
\hline
 \end{tabular}}
\end{table}
}

\subsection{Validation of the Proposed Numerical Scheme}
\begin{figure}
    \centering
    \includegraphics[width=\textwidth]{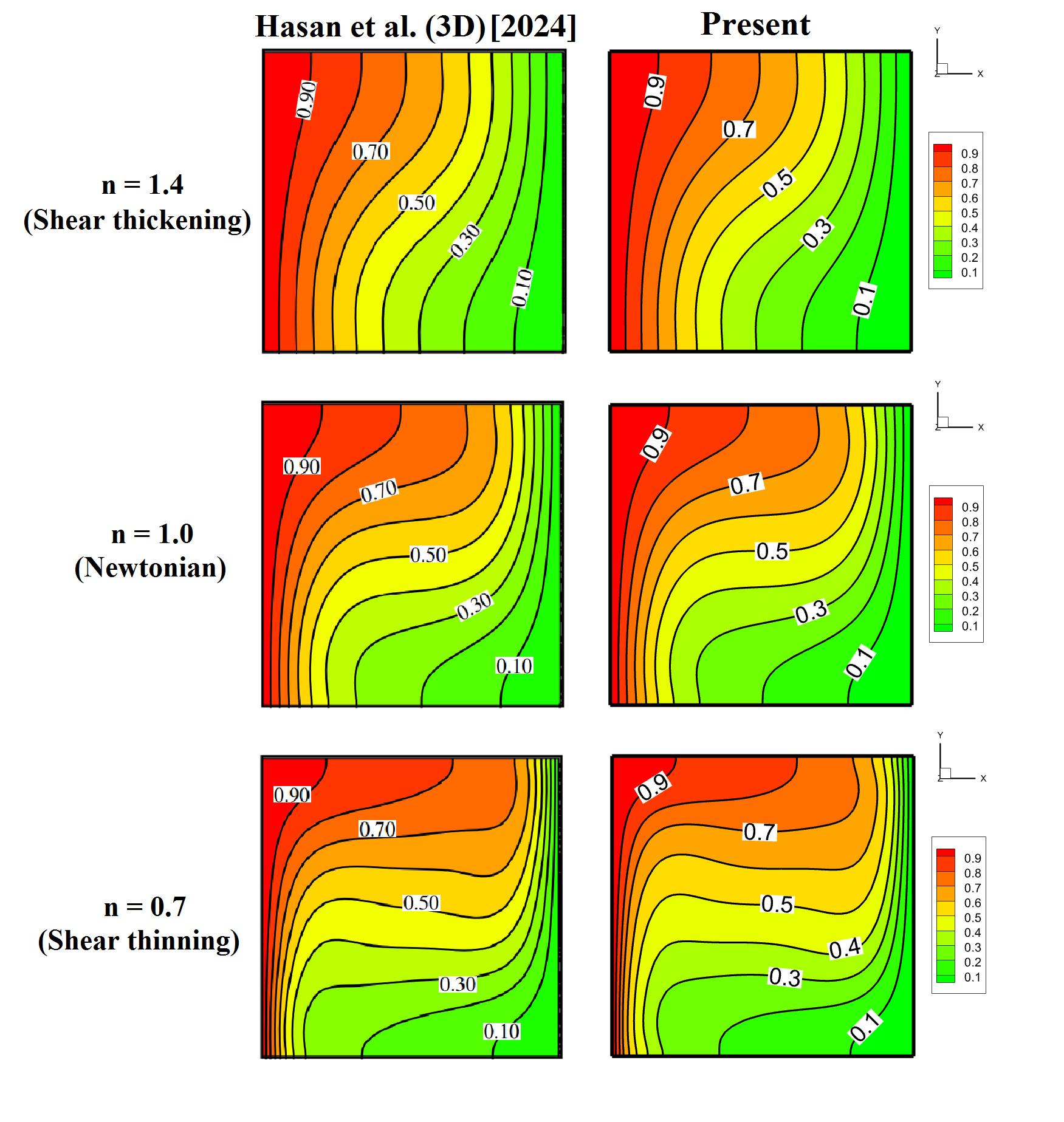}
    \caption{Comparison of the isotherms (on $z=0.5$) between present results and the results of Hasan et al. (3D) \cite{Hasan_2024} for non-Newtonian (shear-thickening $(n=1.4)$, Newtonian $(n=1.0)$, and shear-thinning ($n=0.7)$) fluids at $Ra=10^4$}
    \label{fig:isotherm_comparison}
\end{figure}

\begin{figure}
    \centering
    \includegraphics[width=\textwidth]{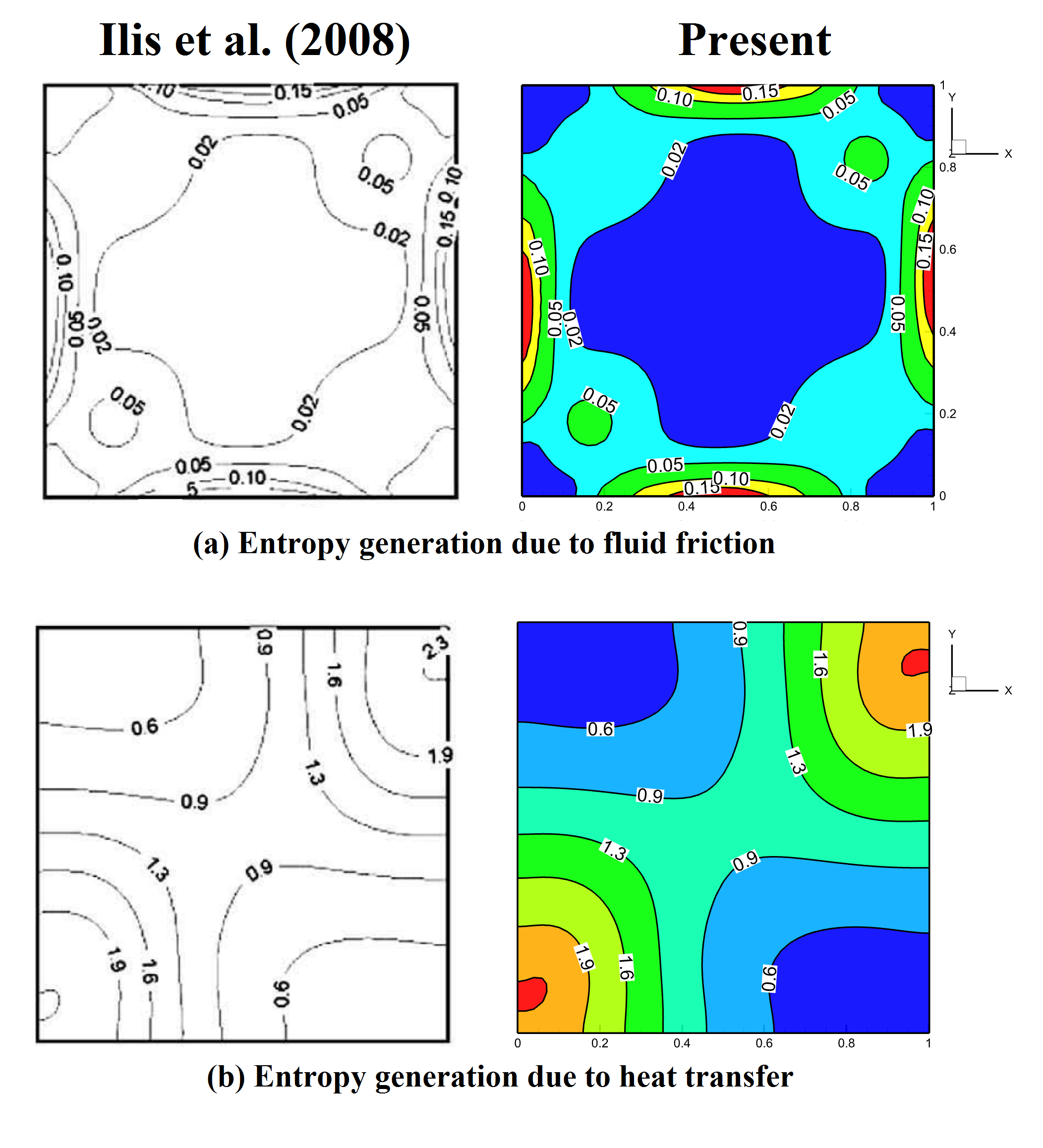}
    \caption{Comparison of (a) local entropy generation from fluid friction and (b) local entropy generation from fluid friction between current findings and the benchmark results of Ilis et al. \cite{Ilis_2008} at $Ra=10^3$ and $Pr=0.71$.}
    \label{fig:Entropy_comparison}
\end{figure} 
To verify the robustness and reliability of the proposed HOSC scheme, present results are compared with well-established 3D benchmark results. Figure \ref{fig:isotherm_comparison} illustrates a comparison of isotherm contours at the symmetric plane \( z = 0.5 \) within the cubic cavity for shear-thickening (\( n = 1.4 \)), Newtonian fluids, and shear-thinning (\( n = 0.7 \)) fluids with the work of Hasan et al. (2024) \cite{Hasan_2024}. From the figure, we observe excellent agreement across all values of the power-law index \( n \), reinforcing the accuracy of our scheme in capturing the behavior of different fluid types. Given that entropy generation is a significant aspect of our study, we have also validated the entropy generation results. Specifically, the contours for entropy generation due to fluid friction and heat transfer have been compared with the benchmark results of Ilis et al. \cite{Ilis_2008}. As evident from the comparison (Figure \ref{fig:Entropy_comparison}), present contour plots are matching well with the existing data, which further demonstrating the reliability of the HOSC scheme for studying entropy generation. For a quantitative assessment, we compare the maximum values of the average Nusselt number obtained from our simulations with the well-established 3D benchmark results of Hinojosa et al. \cite{Hinojosa_2010} and Wang et al. \cite{Wang_2017}. Table \ref{Average_Nusselt_Number_Comparison} presents the comparison, showing that the maximum relative percentage difference is only 0.99\%. This minimal difference indicates a strong agreement with previous results, affirming the accuracy of both our numerical method and code.

{\small\begin{table}[htbp]
\caption{\small Comparison of the max. average Nusselt number at the hot wall for a $51\times51\times51$ grid, along with the relative percentage error ($\delta_e$(\%)), against the results from \cite{Hinojosa_2010} and \cite{Wang_2017} for $Pr=0.71$ at different $Ra$ values. }\label{Average_Nusselt_Number_Comparison}
\centering
 \begin{tabular}{cccccccccc}  \hline \hline
& Rayleigh No.&   Present & Hinojosa et al. \cite{Hinojosa_2010} & $\delta_e$(\%)  &  Wang et al.\cite{Wang_2017}   & $\delta_e$(\%)      \\ \hline 
& $Ra=10^3 $       &  1.098 & 1.091 & 0.63  & 1.088  & 0.91 \\
& $Ra=10^4$          &  2.256 & 2.275 & 0.84 &    2.247  & 0.39  \\
& $Ra=10^5$     &  4.638 & 4.684 & 0.99 & 4.599  &  0.84   \\

\hline\hline
 \end{tabular}
\end{table}
}

\section{Results and Discussion}
\label{sec:Results and Discussion}
In this section, we present our computed results for natural convection of power-law fluids in a 3D cubic cavity, obtained using the proposed HOSC scheme. In Fig. \ref{fig:Sche_diag}, the problem configuration is schematically diagrammed. Given the scarcity of existing results for 3D natural convection of power-law fluids in a cubic cavity, these findings may serve as benchmark solutions for future research. We investigate the effects of varying Rayleigh numbers ($Ra = 10^2, 10^3, 10^4, 10^5$) and different power-law indices, i.e., $n=0.75$ (shear-thinning), $n=1.0$ (Newtonian), and $n=1.25$ (shear-thickening), with a fixed Prandtl number ($Pr=1$). All simulations are performed on a $51 \times 51 \times 51$ grid.
\begin{figure}[htbp]
 \centering
 \vspace*{0pt}%
 \hspace*{\fill}%
\begin{subfigure}{0.33\textwidth}     
    \centering
    \includegraphics[width=\textwidth]{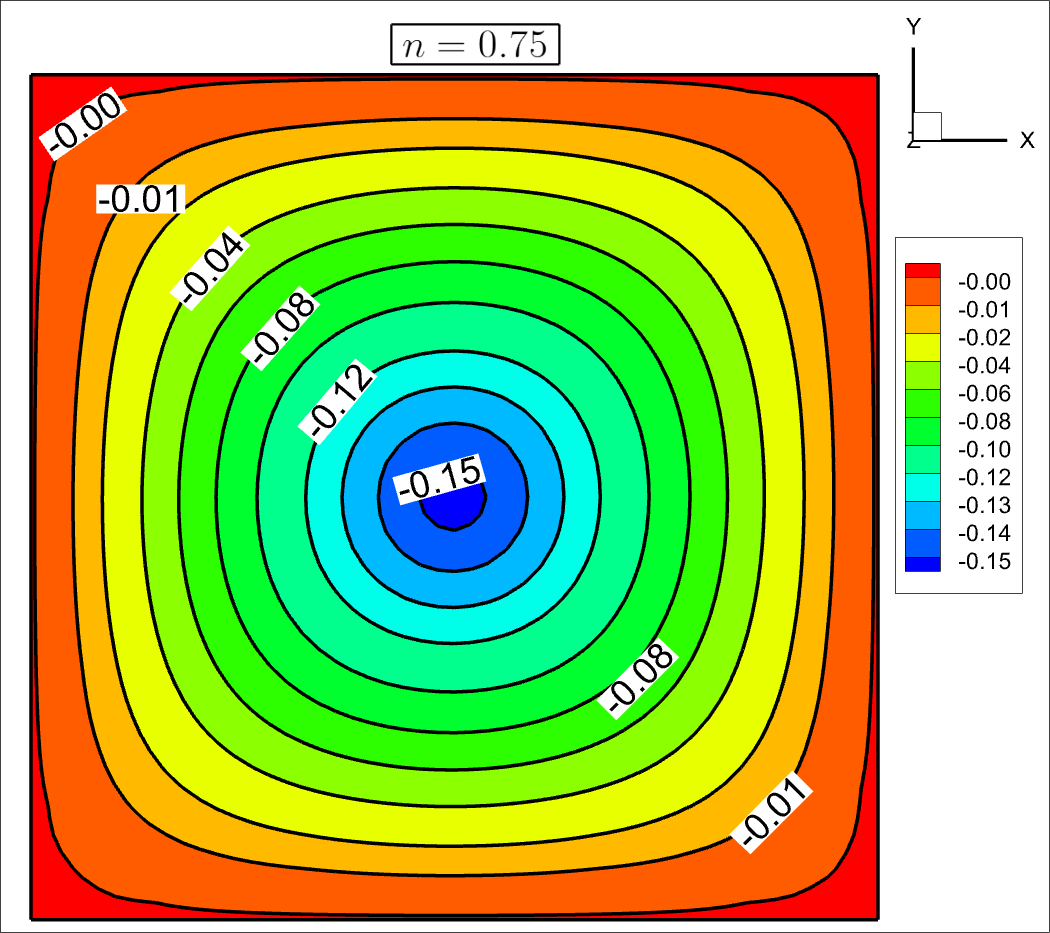}%
    \captionsetup{skip=2pt}%
    \caption{(a) $n=0.75, Ra = 10^2$ }
    \label{fig:2D_streamlines_Ra_10^2_n_0_75.png}
  \end{subfigure}%
 \begin{subfigure}{0.33\textwidth}        
   \centering
    \includegraphics[width=\textwidth]{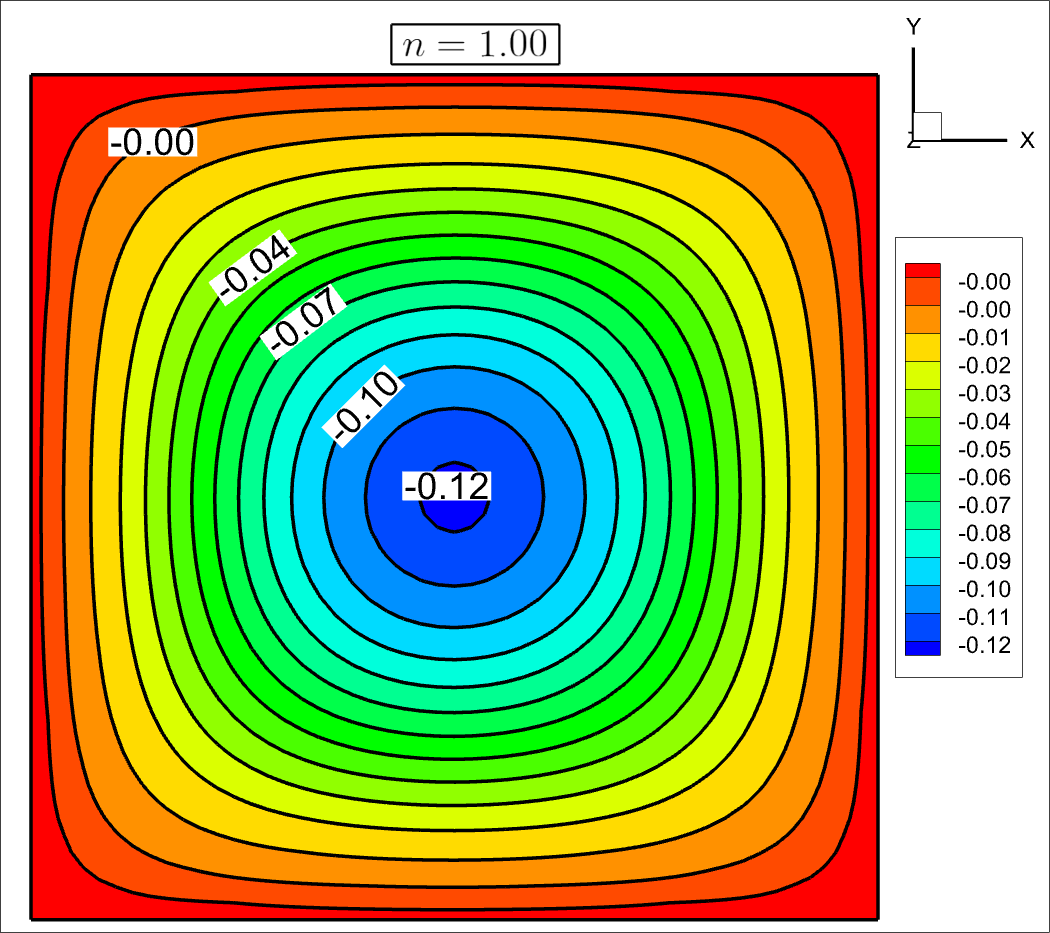}%
    \captionsetup{skip=2pt}%
    \caption{(b) $n=1.00, Ra = 10^2$}
    \label{fig:2D_streamlines_Ra_10^2_n_1_00.png}
  \end{subfigure}
   \begin{subfigure}{0.33\textwidth}        
   \centering
    \includegraphics[width=\textwidth]{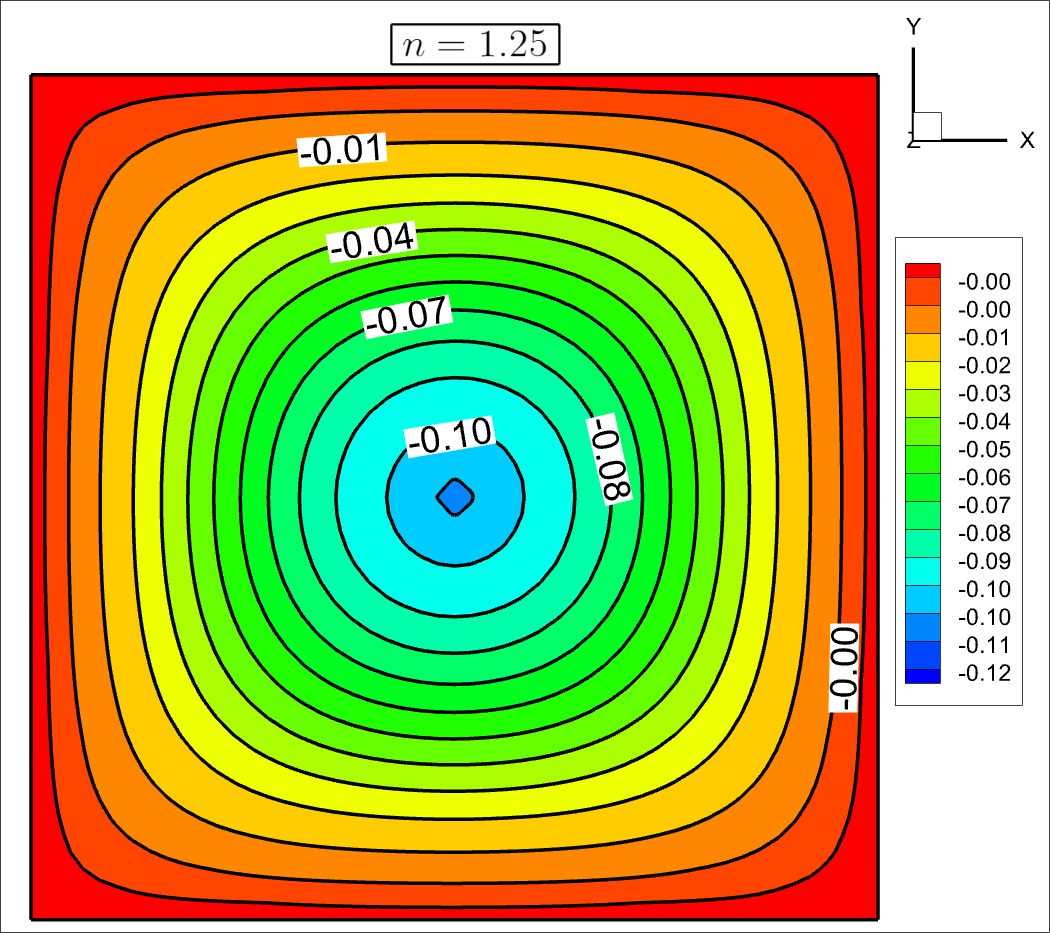}%
    \captionsetup{skip=2pt}%
    \caption{(c) $n=1.25, Ra = 10^2$}
    \label{fig:2D_streamlines_Ra_10^2_n_1_25.png}
  \end{subfigure}%
  \hspace*{\fill}

  \vspace*{8pt}%
  \hspace*{\fill}%
  \begin{subfigure}{0.33\textwidth}     
    \centering
    \includegraphics[width=\textwidth]{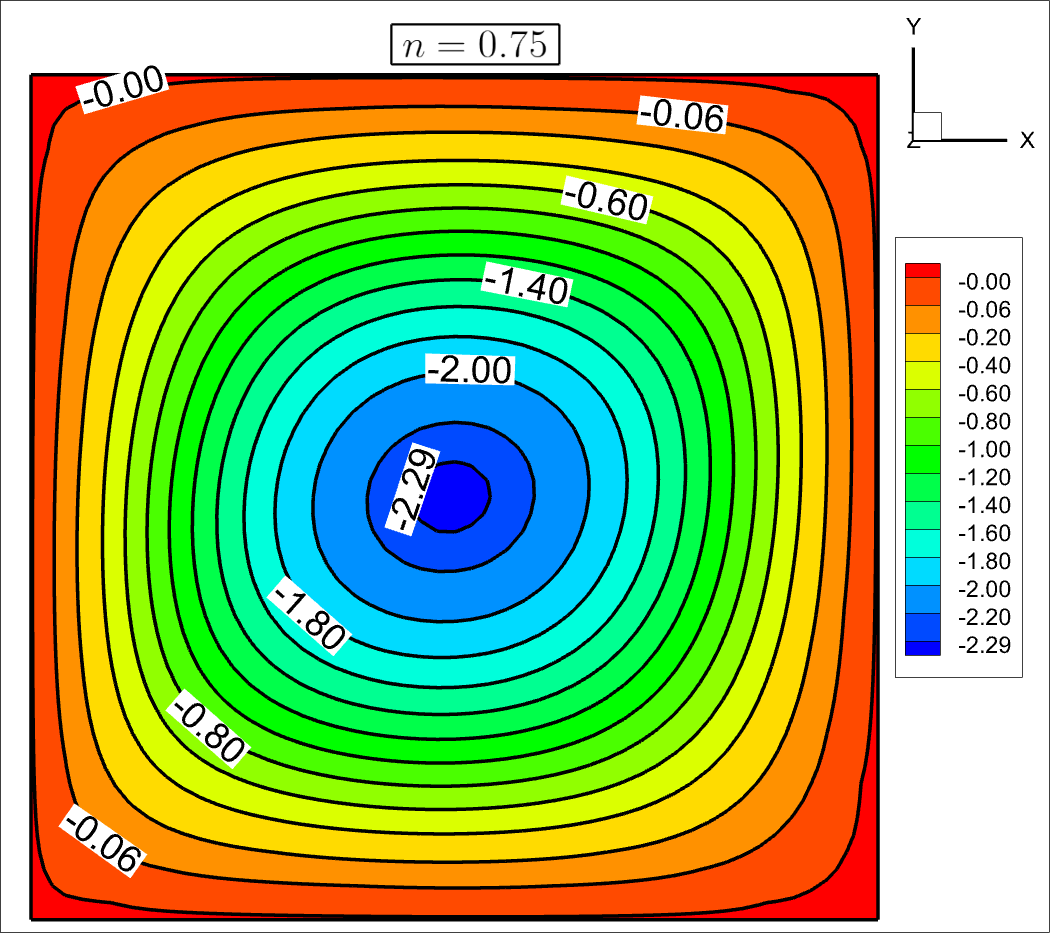}%
    \captionsetup{skip=2pt}%
    \caption{(d) $n=0.75, Ra = 10^3$}
    \label{fig:2D_streamlines_Ra_10^3_n_0_75.png}
  \end{subfigure}%
 \begin{subfigure}{0.33\textwidth}        
   \centering
    \includegraphics[width=\textwidth]{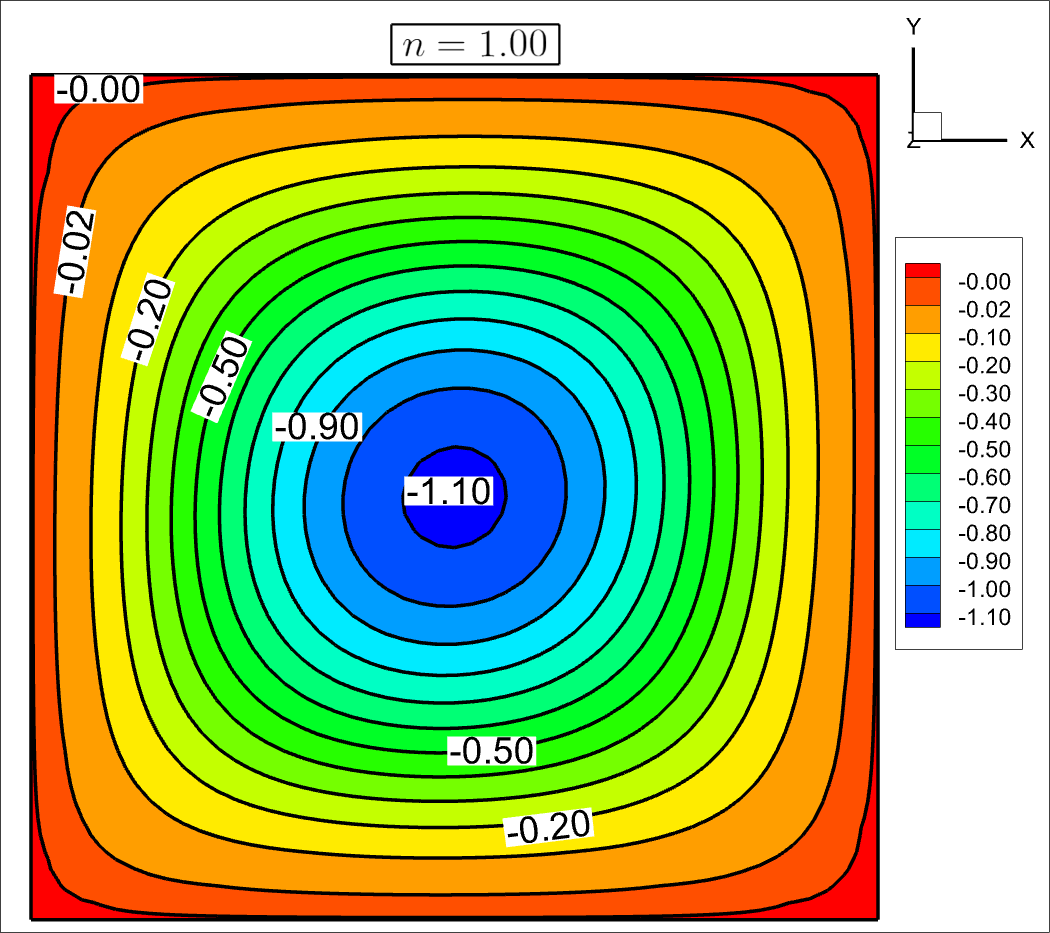}%
    \captionsetup{skip=2pt}%
    \caption{(e) $n=1.00, Ra = 10^3$}
    \label{fig:2D_streamlines_Ra_10^3_n_1_00.png}
  \end{subfigure}
   \begin{subfigure}{0.33\textwidth}        
   \centering
    \includegraphics[width=\textwidth]{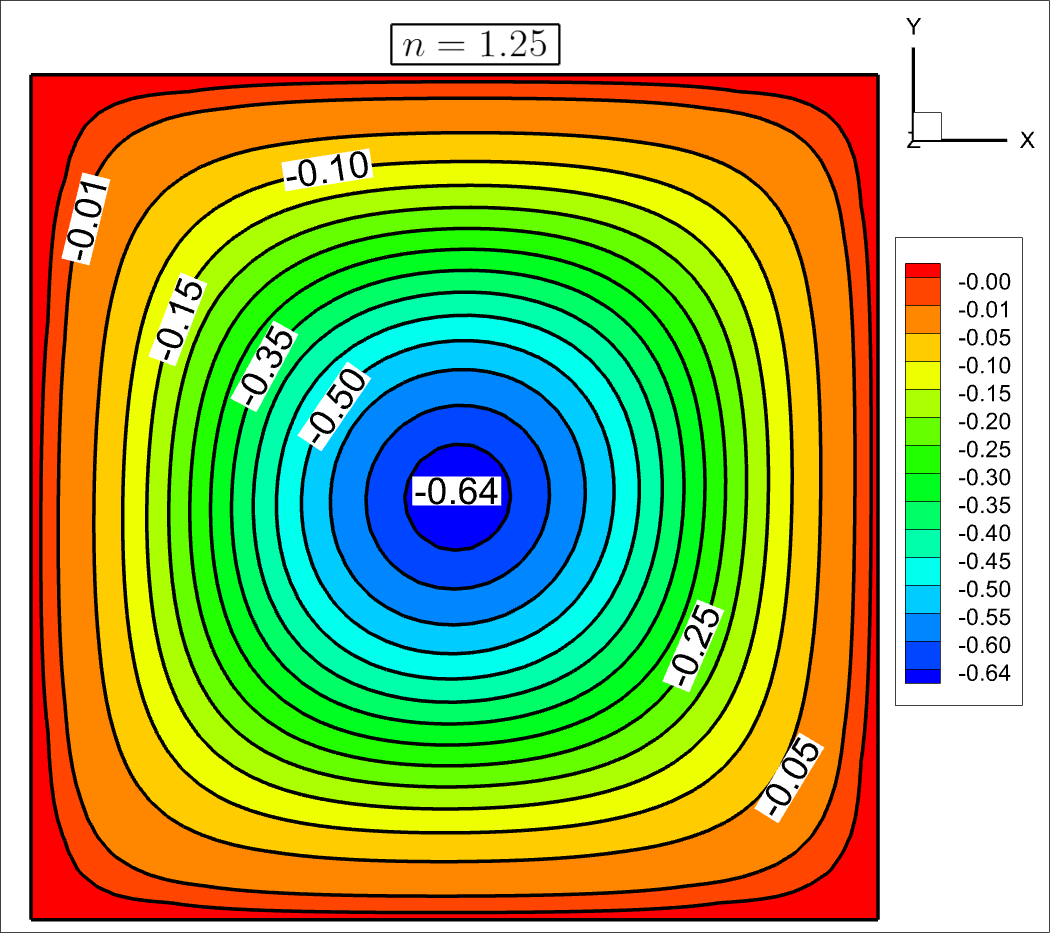}%
    \captionsetup{skip=2pt}%
    \caption{(f) $n=1.25, Ra = 10^3$}
    \label{fig:2D_streamlines_Ra_10^3_n_1_25.png}
  \end{subfigure}%
  \hspace*{\fill}

  \vspace*{8pt}%
  \hspace*{\fill}%
  \begin{subfigure}{0.33\textwidth}     
    \centering
    \includegraphics[width=\textwidth]{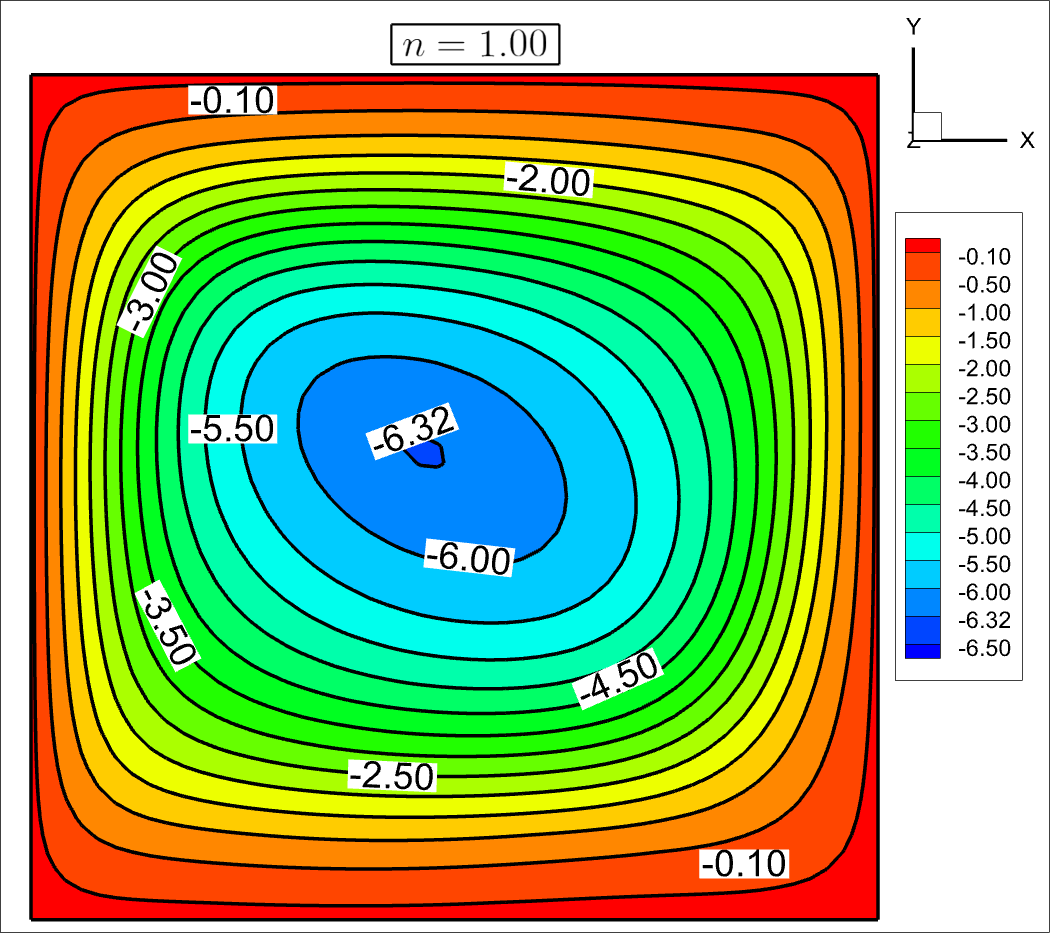}%
    \captionsetup{skip=2pt}%
    \caption{(g) $n=0.75, Ra = 10^4$}
    \label{fig:2D_streamlines_Ra_10^4_n_0_75.png}
  \end{subfigure}%
 \begin{subfigure}{0.33\textwidth}        
   \centering
    \includegraphics[width=\textwidth]{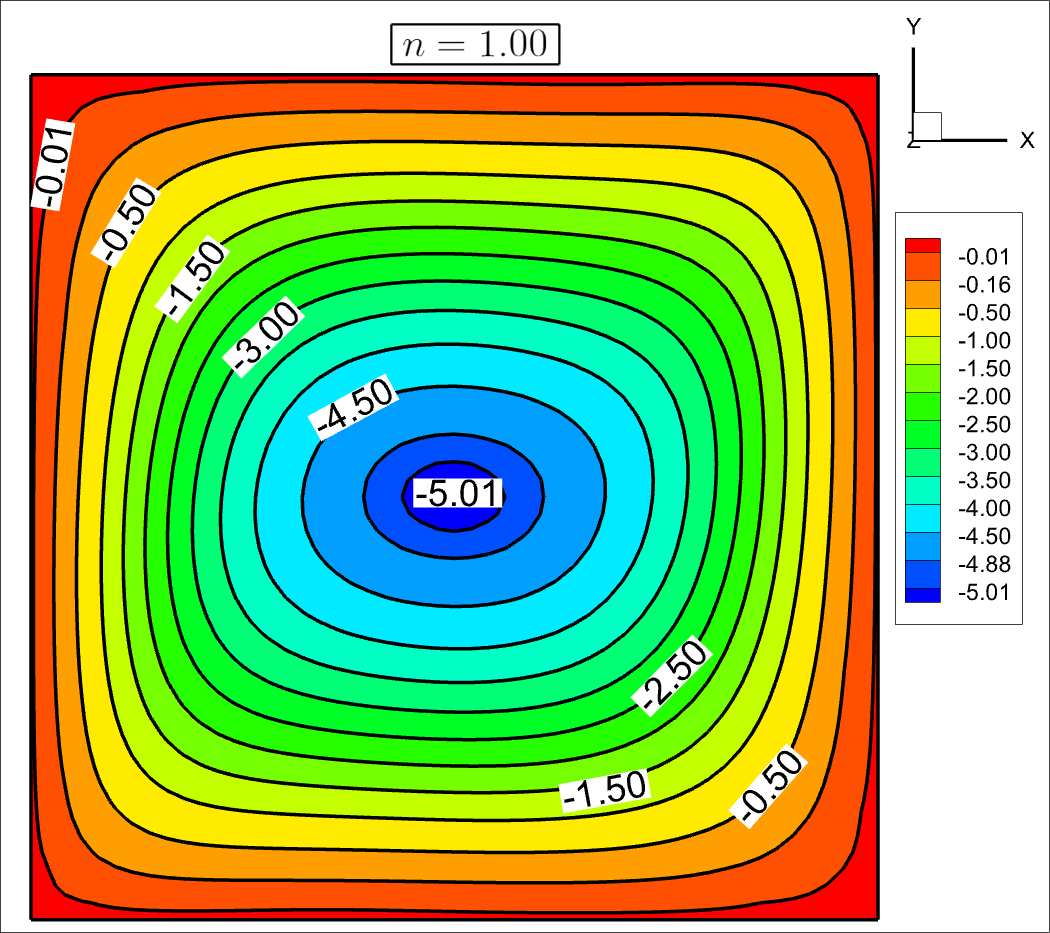}%
    \captionsetup{skip=2pt}%
    \caption{(h) $n=1.00, Ra = 10^4$}
    \label{fig:2D_streamlines_Ra_10^4_n_0_1.png}
  \end{subfigure}
   \begin{subfigure}{0.33\textwidth}        
   \centering
    \includegraphics[width=\textwidth]{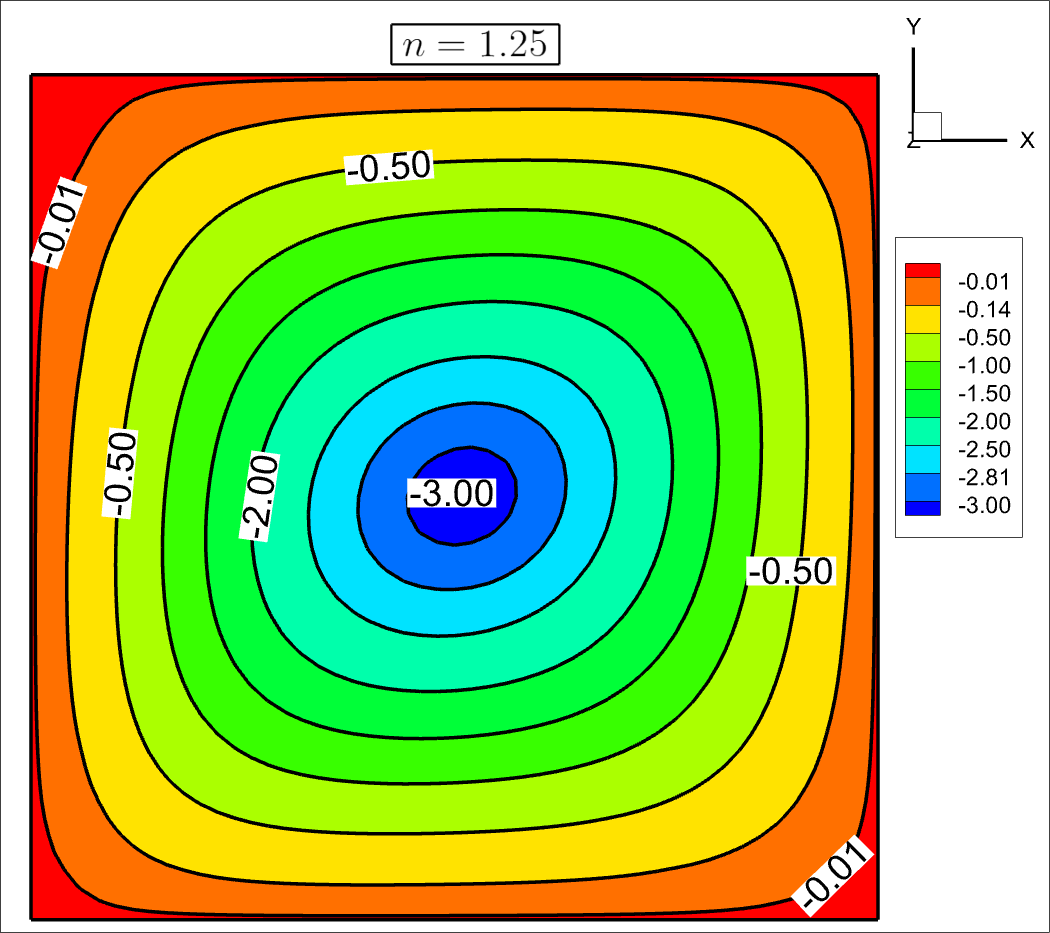}%
    \captionsetup{skip=2pt}%
    \caption{(i) $n=1.25, Ra = 10^4$}
    \label{fig:2D_streamlines_Ra_10^4_n_0_1.25.png}
  \end{subfigure}%
  \hspace*{\fill}

  \vspace*{8pt}%
  \hspace*{\fill}%
  \begin{subfigure}{0.33\textwidth}     
    \centering
    \includegraphics[width=\textwidth]{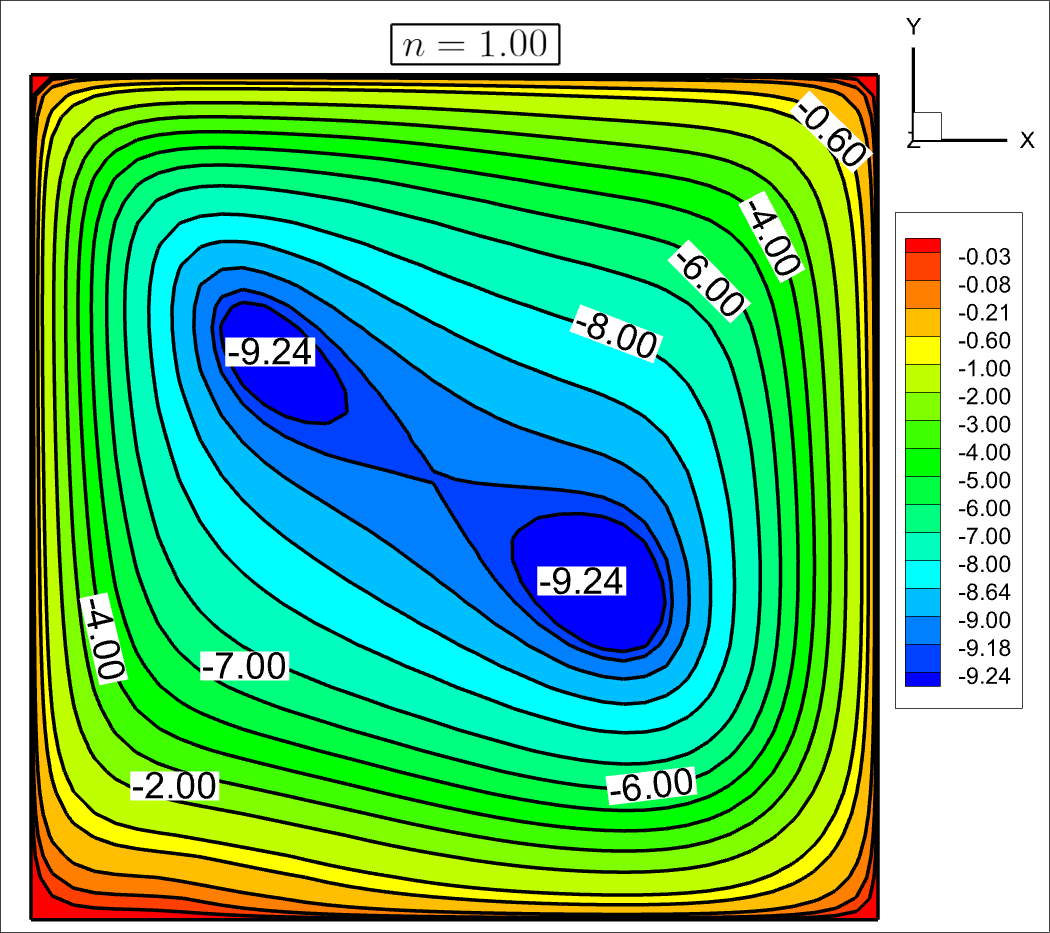}%
    \captionsetup{skip=2pt}%
    \caption{(j) $n=0.75, Ra = 10^5$}
    \label{fig:2D_streamlines_Ra_10^5_n_0_75.png}
  \end{subfigure}%
 \begin{subfigure}{0.33\textwidth}        
   \centering
    \includegraphics[width=\textwidth]{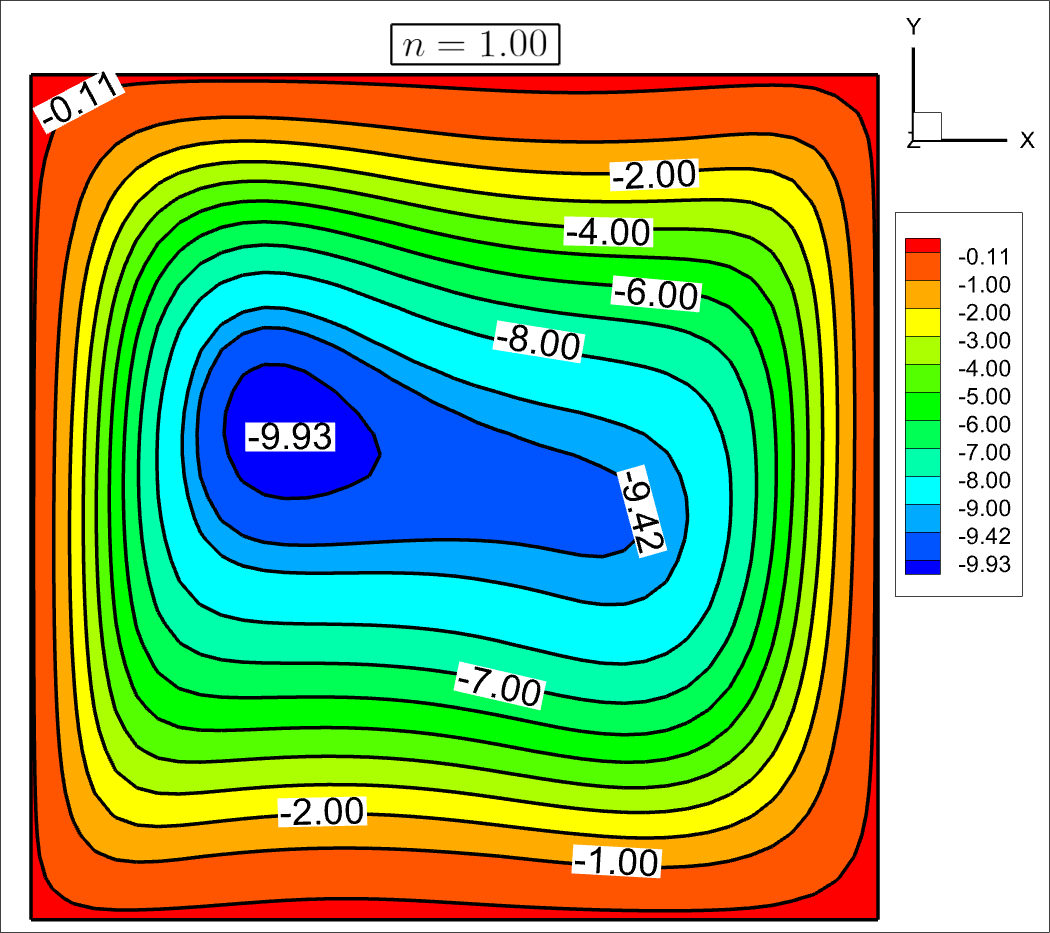}%
    \captionsetup{skip=2pt}%
    \caption{(k) $n=1.00, Ra = 10^5$}
    \label{fig:2D_streamlines_Ra_10^5_n_1_00.png}
  \end{subfigure}
   \begin{subfigure}{0.33\textwidth}        
   \centering
    \includegraphics[width=\textwidth]{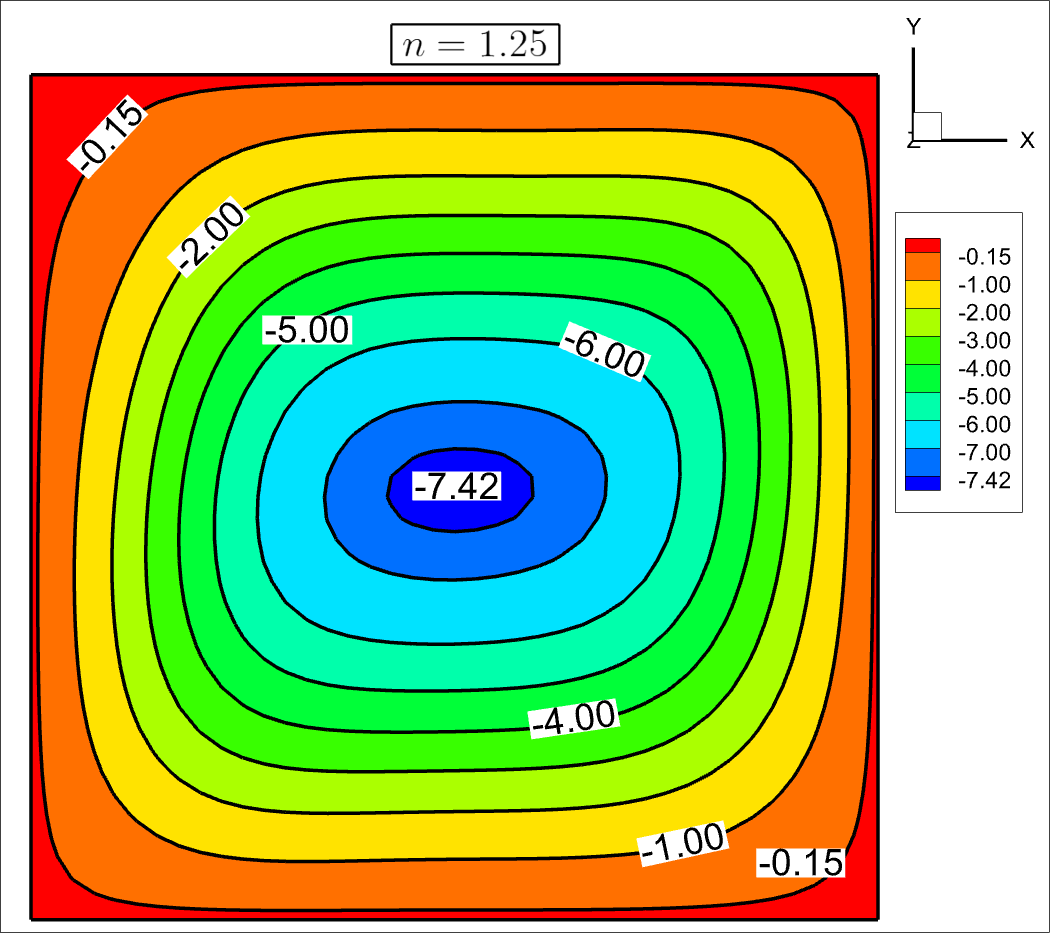}%
    \captionsetup{skip=2pt}%
    \caption{(l) $n=1.25, Ra = 10^5$}
    \label{fig:2D_streamlines_Ra_10^5_n_1_25.png}
  \end{subfigure}%
  \vspace*{1pt}%
  \hspace*{\fill}%
  \caption{Visualization of streamlines on the symmetric plane (z = 0.5) for different $Ra$ and $n$ values. Rows (a)–(c), (d)–(f), (g)–(i), and (j)–(l) show streamlines at \(Ra = 10^2\), \(10^3\), \(10^4\), and \(10^5\) for \(n = 0.75\), \(1.0\), and \(1.25\), respectively.
  }
  \label{fig:2D_streamlines}
\end{figure}

\subsection{Study of fluid flow and temperature field}
In Figure \ref{fig:2D_streamlines}, we present the streamline contours at the symmetric plane \( z=0.5 \) for shear-thinning (\( n=0.75 \)), shear-thickening (\( n=1.25 \)), and Newtonian fluids (\( n=1.0 \)) across various Rayleigh numbers (\( Ra \)) . The arrangement of the figure facilitates a row-wise comparison of the effect of the power law index ($n$) at fixed Rayleigh numbers ($Ra$) and a column-wise assessment of the effect of $Ra$ at fixed $n$ on the streamlines. At lower Rayleigh numbers (\( Ra=10^2 \) and \( Ra=10^3 \)), the streamline patterns appear smooth and circular for all values of \( n \). However, we observe that the maximum stream function values decrease with increasing \( n \) at both \( Ra=10^2 \) and \( Ra=10^3 \). The negative values of the stream function indicate a clockwise rotation of the fluid. As we increase the Rayleigh number to \( Ra=10^4 \), the influence of the power law index becomes more noticeable. For the shear-thinning case (\( n=0.75 \)), the circular streamline pattern transitions to an elliptical shape, signifying a higher convection rate. Similar changes are noted for \( n=1.0 \) and \( n=1.25 \); however, these alterations are less pronounced compared to the \( n=0.75 \) case at \( Ra=10^4 \), indicating that convection flow is notably higher for \( n=0.75 \). At \( Ra=10^5 \), we observe a significant transformation in the pattern for \( n=0.75 \). Instead of a single primary vortex, two distinct vortices emerge, further indicating a substantial increase in convection rates. For \( n=1.0 \), the streamline shape also changes from nearly round to more slender. Conversely, the effects are minimal for \( n=1.25 \), where the streamline pattern remains largely unchanged.

Overall, the stream function values consistently decrease with increasing \( n \) for any fixed \( Ra \). This occurs because, in shear-thinning fluids (\( n = 0.75 \)), the viscosity decreases with increasing shear rate, leading to stronger flow under thermal gradients and enhanced convective currents. Conversely, for shear-thickening fluids (\( n = 1.25 \)), the viscosity increases with shear rate, which results in reduced flow velocities and a weaker convective response to thermal gradients. Additionally, as \( Ra \) increases, the stream function values rise for any fixed \( n \), indicating stronger convection due to the increased buoyancy forces associated with higher $Ra$s.


\begin{figure}[htbp]
 \centering
 \vspace*{0pt}%
 \hspace*{\fill}%
\begin{subfigure}{0.33\textwidth}     
    \centering
    \includegraphics[width=\textwidth]{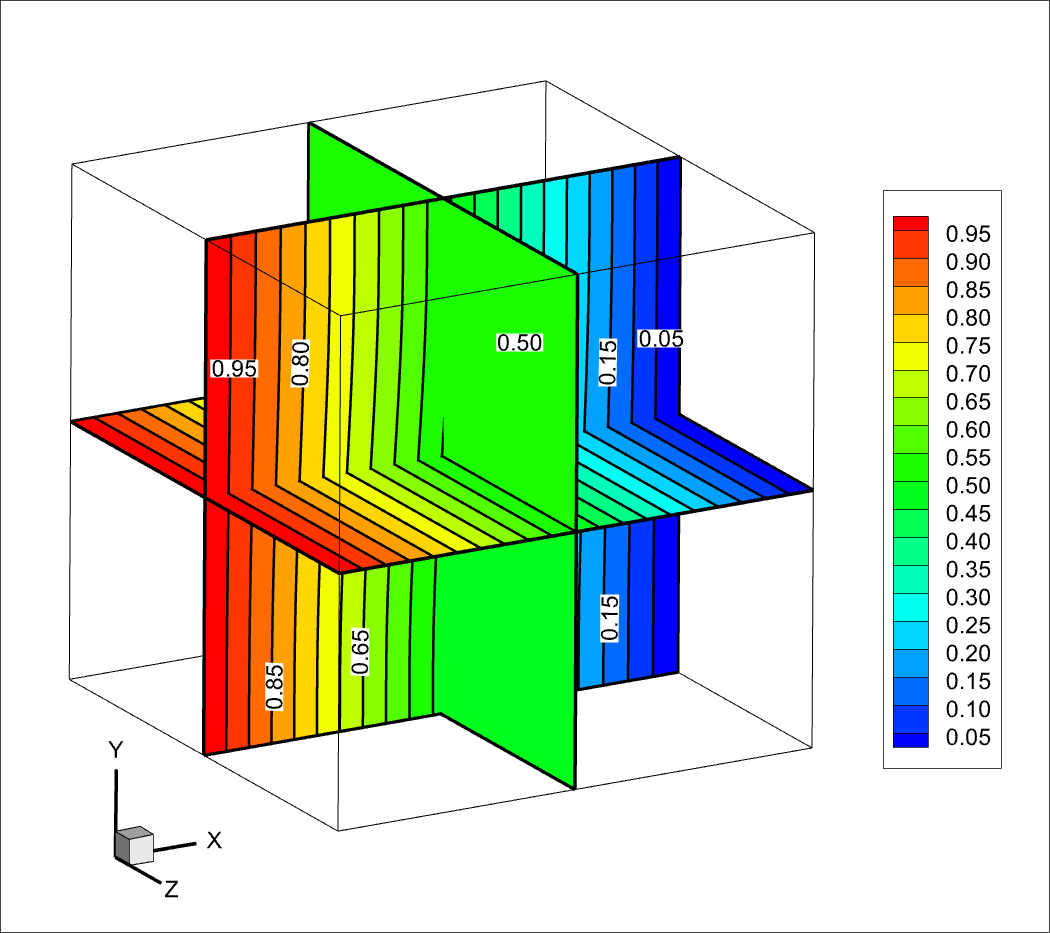}%
    \captionsetup{skip=2pt}%
    \caption{(a) $n=0.75, Ra = 10^2$ }
    \label{fig:3D_Isotherm_Ra_10^2_n_0_75.png}
  \end{subfigure}%
 \begin{subfigure}{0.33\textwidth}        
   \centering
    \includegraphics[width=\textwidth]{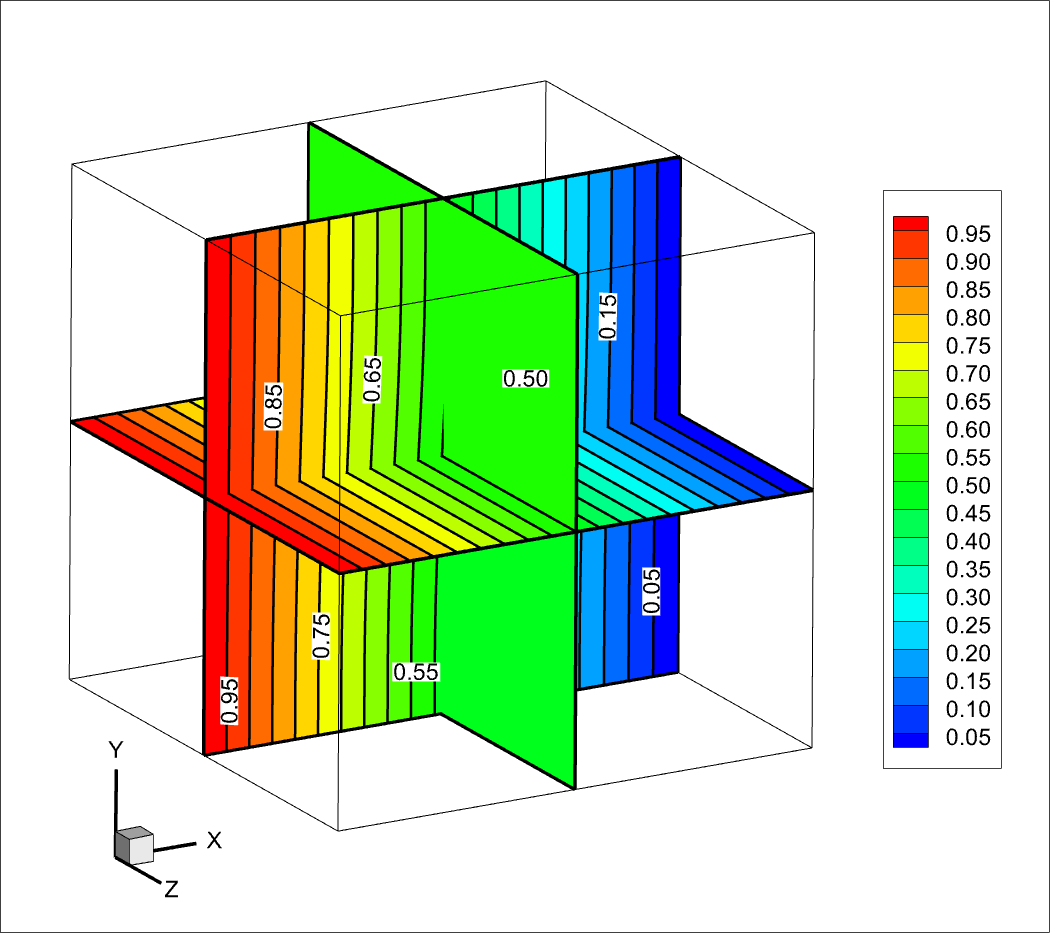}%
    \captionsetup{skip=2pt}%
    \caption{(b) $n=1.00, Ra = 10^2$}
    \label{fig:3D_Isotherm_Ra_10^2_n_1_00.png}
  \end{subfigure}
   \begin{subfigure}{0.33\textwidth}        
   \centering
    \includegraphics[width=\textwidth]{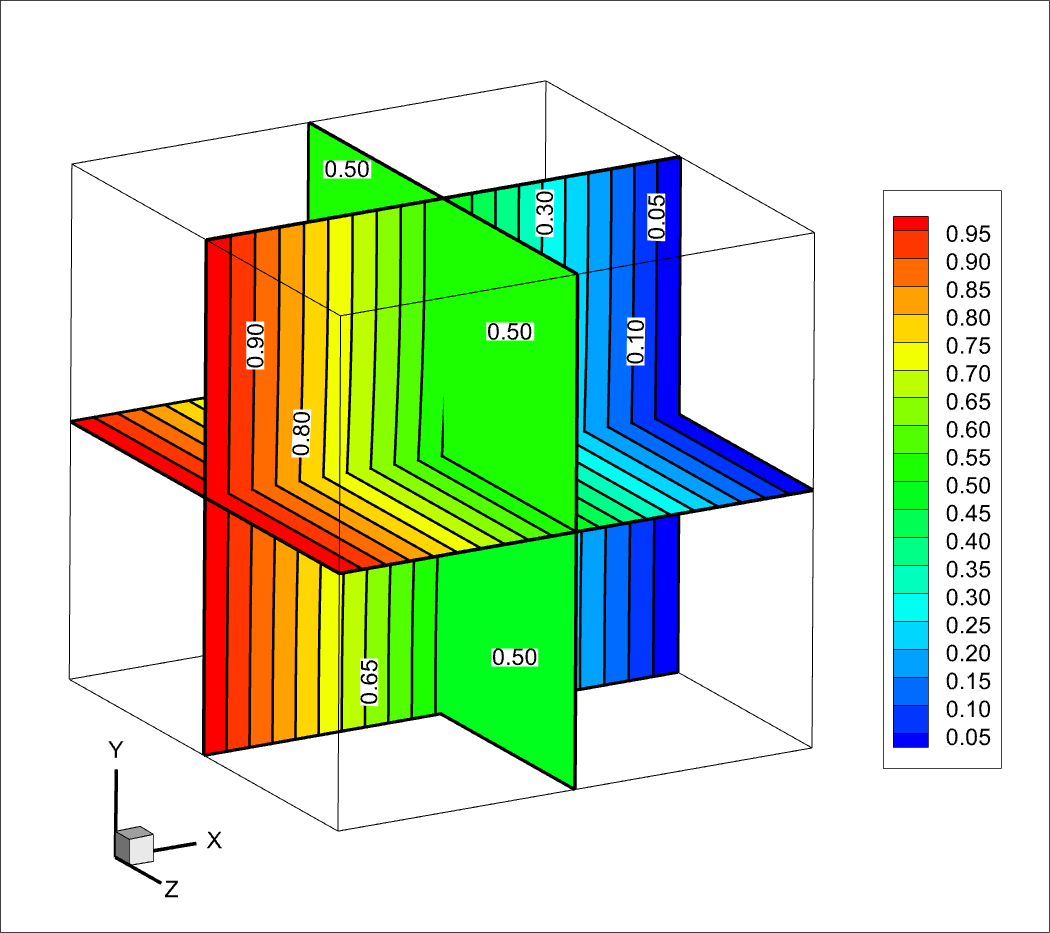}%
    \captionsetup{skip=2pt}%
    \caption{(c) $n=1.25, Ra = 10^2$}
    \label{fig:3D_Isotherm_Ra_10^2_n_1_25.png}
  \end{subfigure}%
  \hspace*{\fill}

  \vspace*{8pt}%
  \hspace*{\fill}%
  \begin{subfigure}{0.33\textwidth}     
    \centering
    \includegraphics[width=\textwidth]{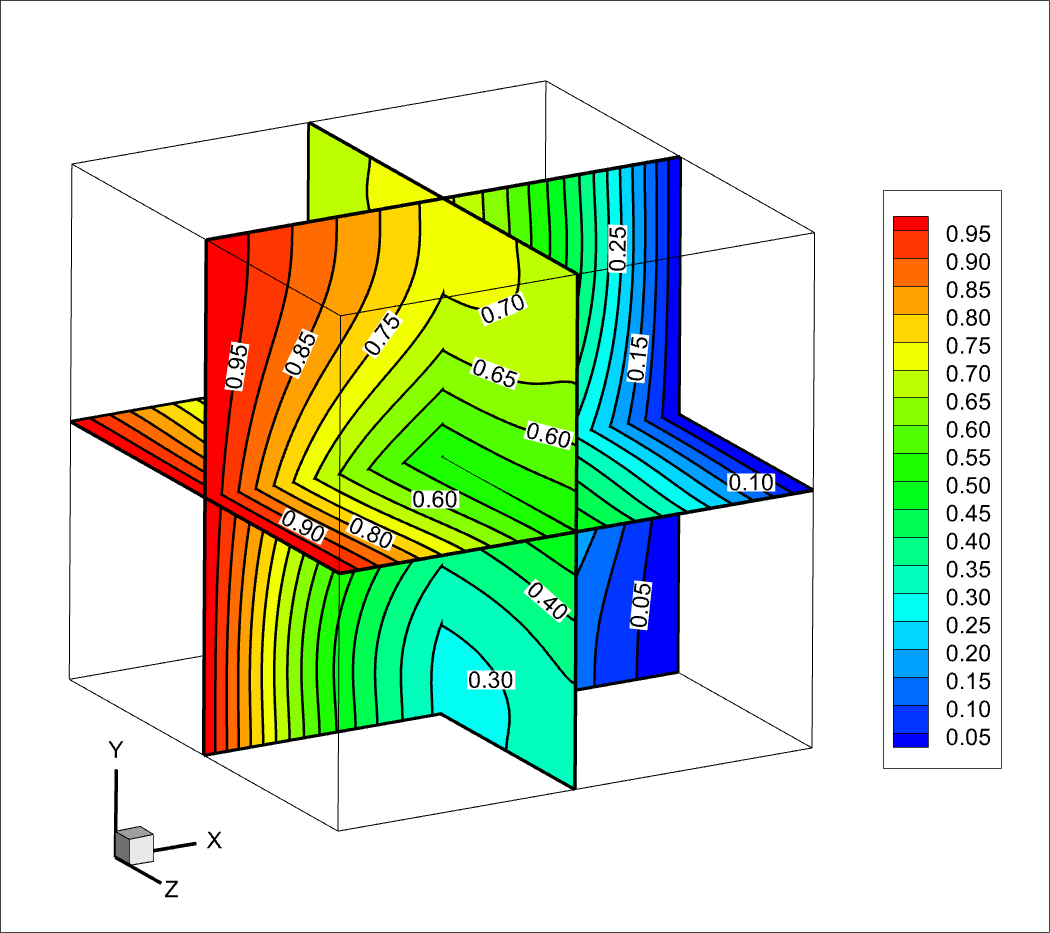}%
    \captionsetup{skip=2pt}%
    \caption{(d) $n=0.75, Ra = 10^3$}
    \label{fig:3D_Isotherm_Ra_10^3_n_0_75.png}
  \end{subfigure}%
 \begin{subfigure}{0.33\textwidth}        
   \centering
    \includegraphics[width=\textwidth]{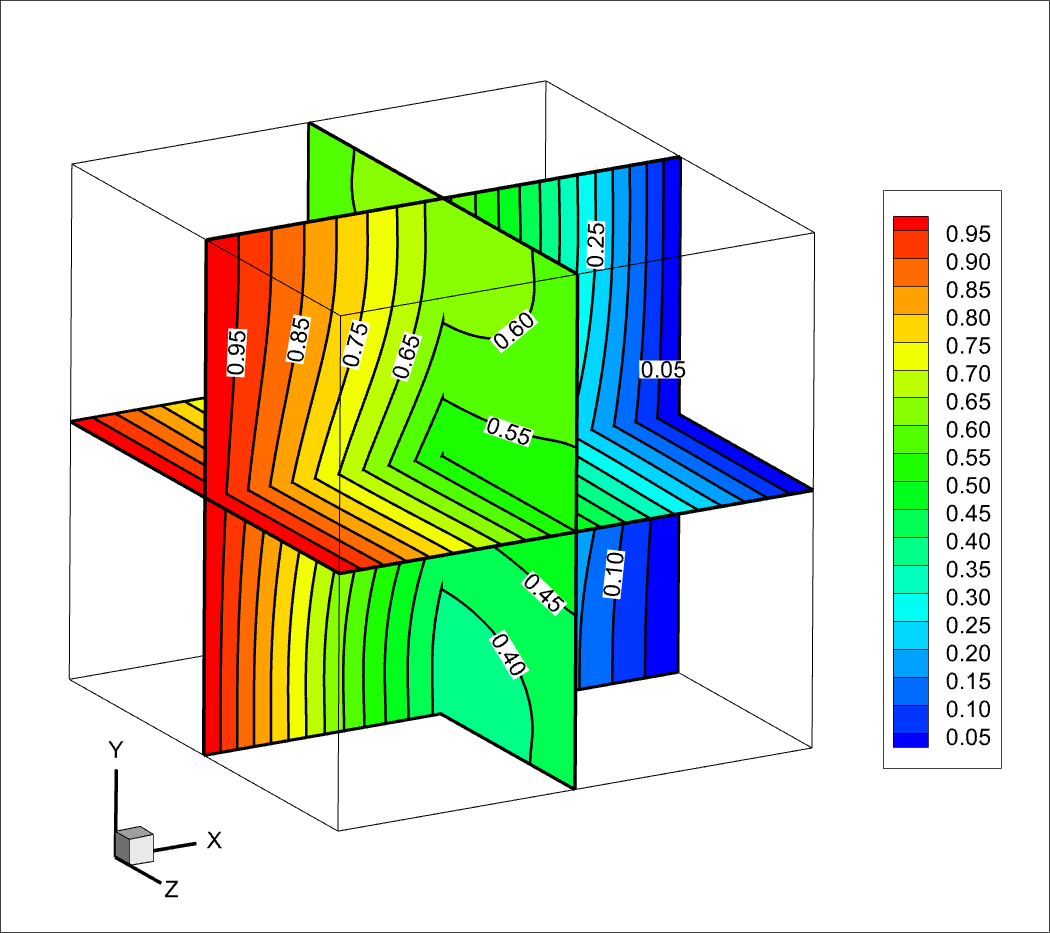}%
    \captionsetup{skip=2pt}%
    \caption{(e) $n=1.00, Ra = 10^3$}
    \label{fig:3D_Isotherm_Ra_10^3_n_1_00.png}
  \end{subfigure}
   \begin{subfigure}{0.33\textwidth}        
   \centering
    \includegraphics[width=\textwidth]{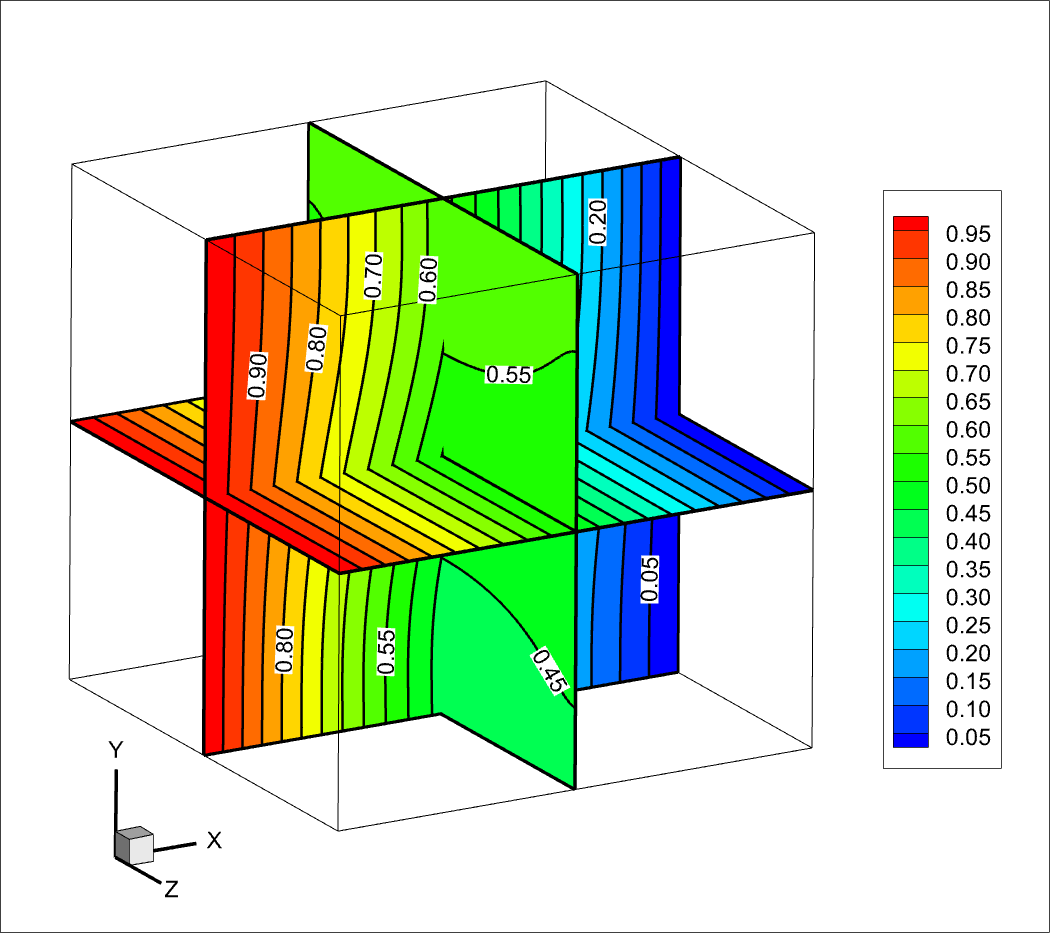}%
    \captionsetup{skip=2pt}%
    \caption{(f) $n=1.25, Ra = 10^3$}
    \label{fig:3D_Isotherm_Ra_10^3_n_1_25.png}
  \end{subfigure}%
  \hspace*{\fill}

  \vspace*{8pt}%
  \hspace*{\fill}%
  \begin{subfigure}{0.33\textwidth}     
    \centering
    \includegraphics[width=\textwidth]{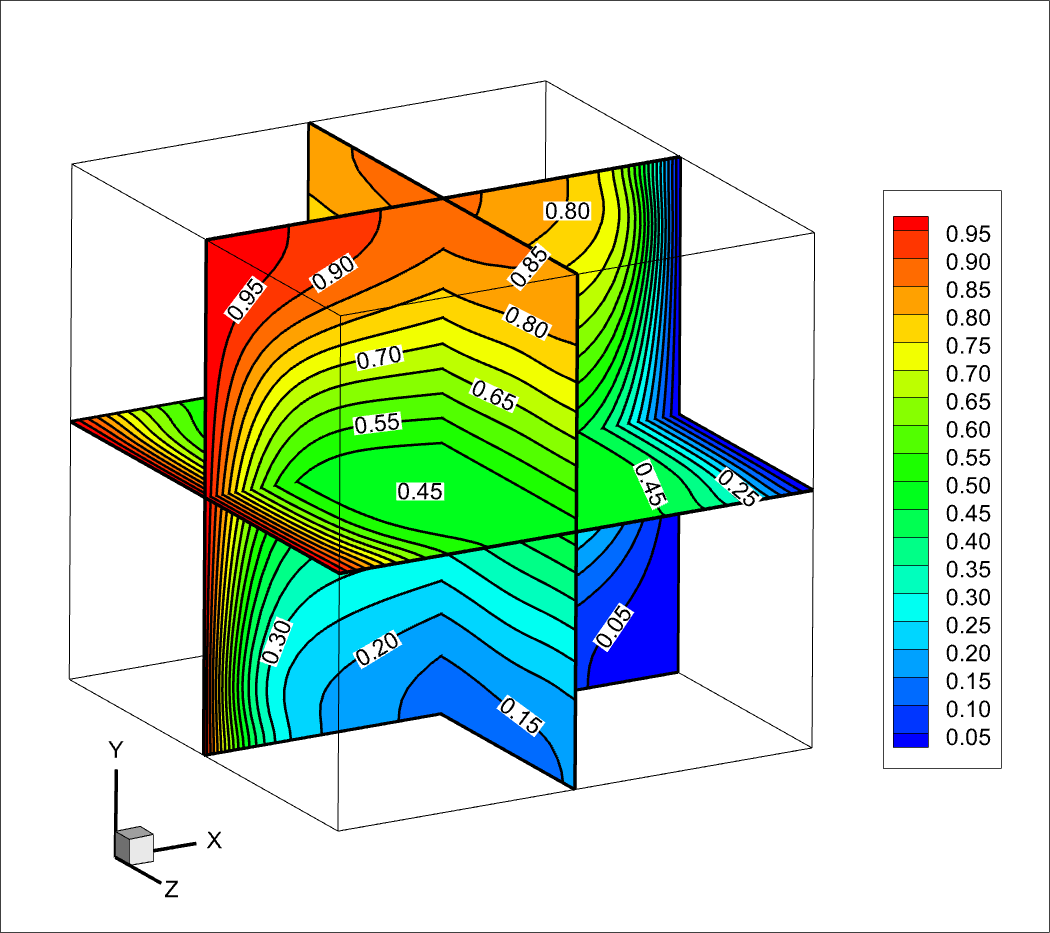}%
    \captionsetup{skip=2pt}%
    \caption{(g) $n=0.75, Ra = 10^4$}
    \label{fig:3D_Isotherm_Ra_10^4_n_0_75.png}
  \end{subfigure}%
 \begin{subfigure}{0.33\textwidth}        
   \centering
    \includegraphics[width=\textwidth]{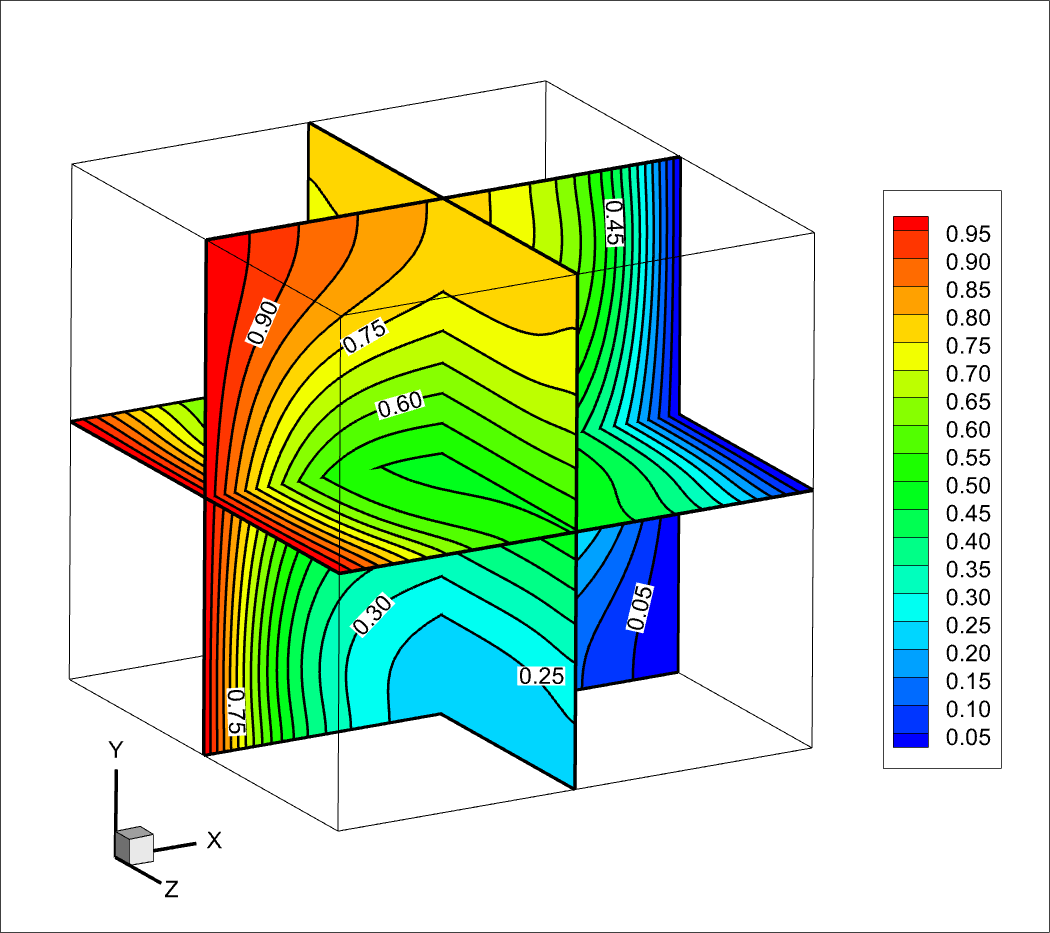}%
    \captionsetup{skip=2pt}%
    \caption{(h) $n=1.00, Ra = 10^4$}
    \label{fig:3D_Isotherm_Ra_10^4_n_1_00.png}
  \end{subfigure}
   \begin{subfigure}{0.33\textwidth}        
   \centering
    \includegraphics[width=\textwidth]{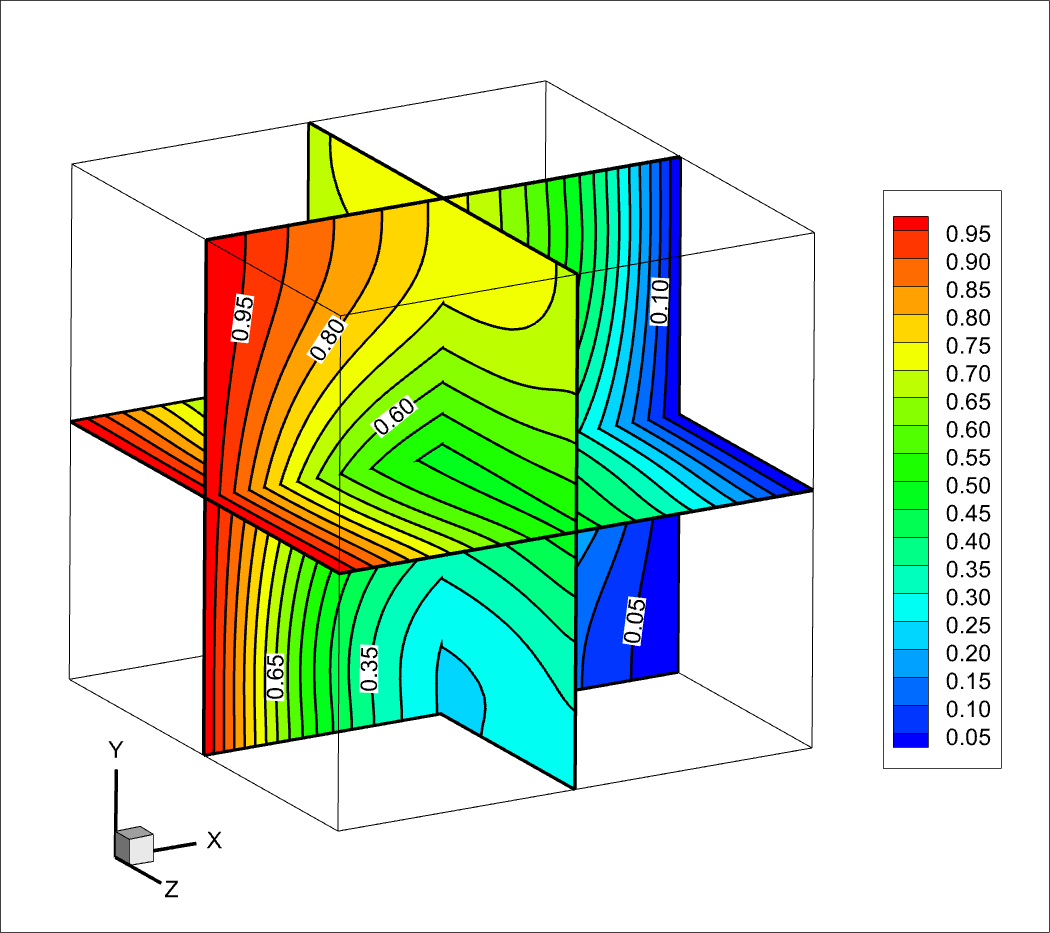}%
    \captionsetup{skip=2pt}%
    \caption{(i) $n=1.25, Ra = 10^4$}
    \label{fig:3D_Isotherm_Ra_10^4_n_1_25.png}
  \end{subfigure}%
  \hspace*{\fill}

  \vspace*{8pt}%
  \hspace*{\fill}%
  \begin{subfigure}{0.33\textwidth}     
    \centering
    \includegraphics[width=\textwidth]{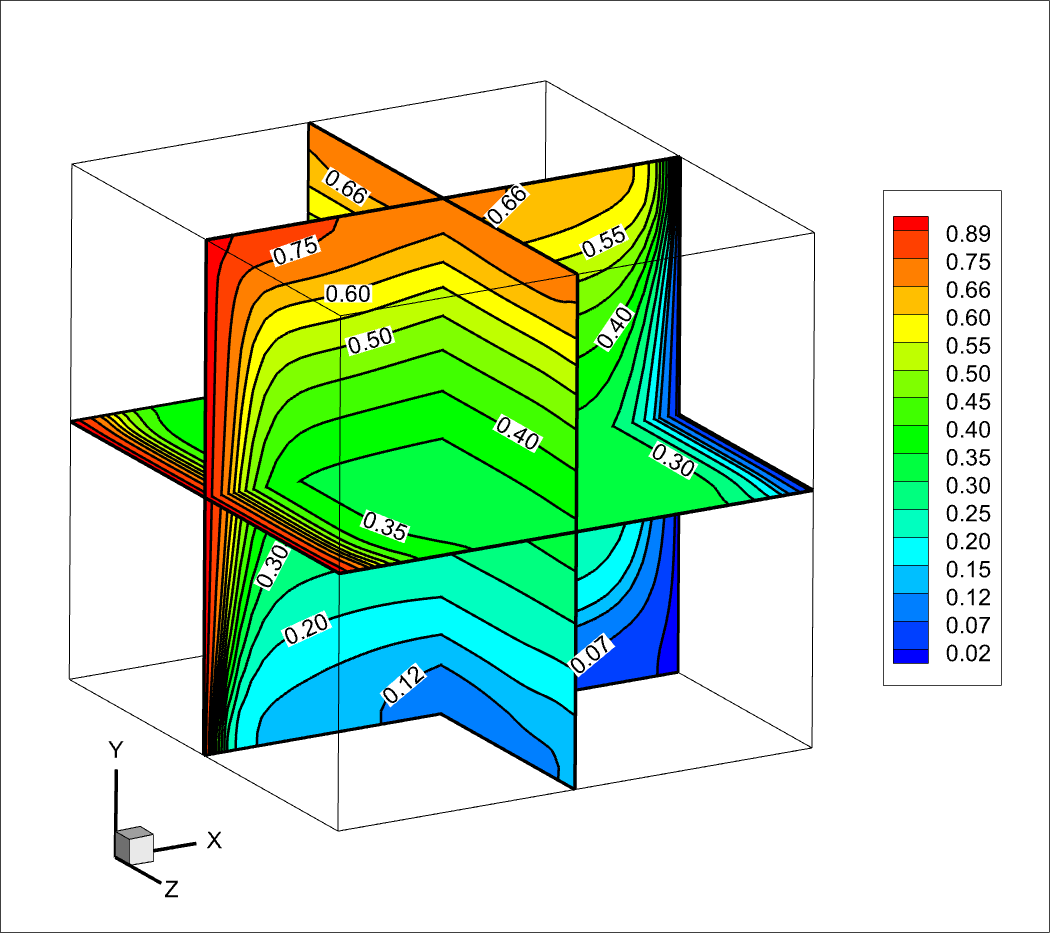}%
    \captionsetup{skip=2pt}%
    \caption{(j) $n=0.75, Ra = 10^5$}
    \label{fig:3D_Isotherm_Ra_10^5_n_0_75.png}
  \end{subfigure}%
 \begin{subfigure}{0.33\textwidth}        
   \centering
    \includegraphics[width=\textwidth]{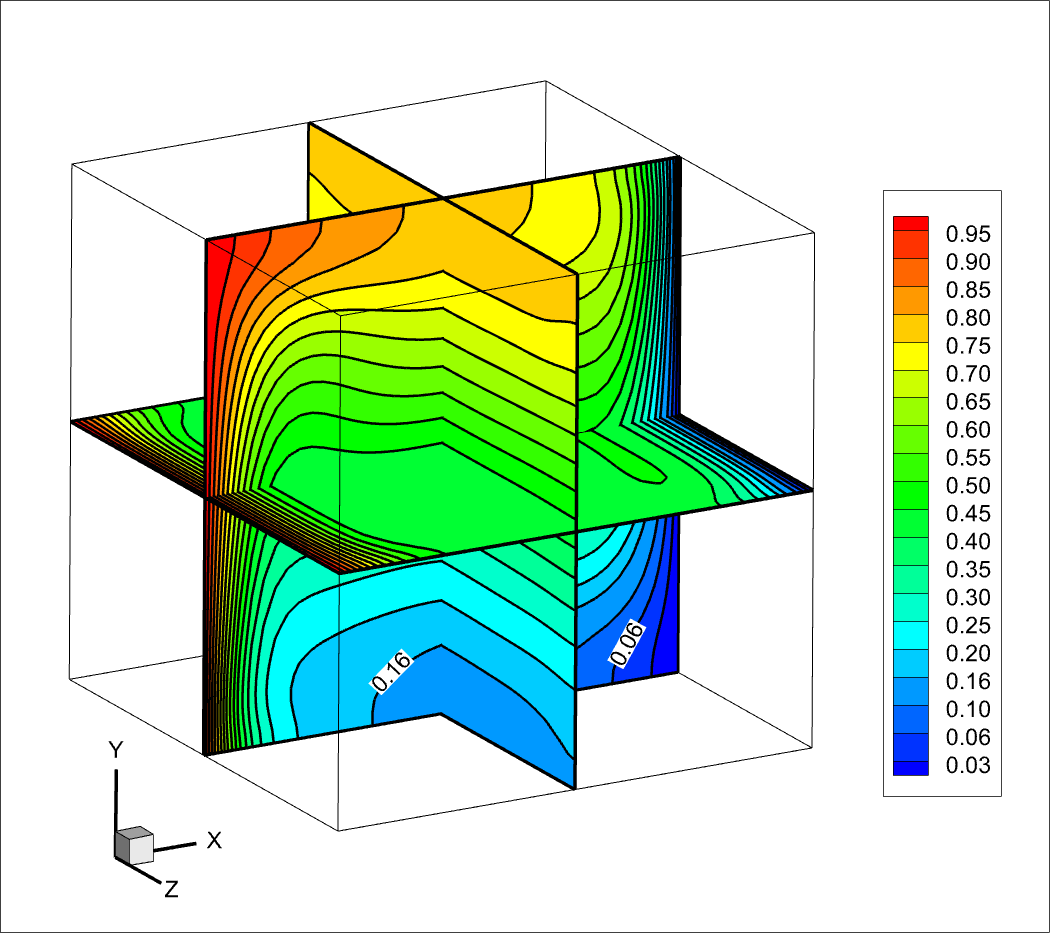}%
    \captionsetup{skip=2pt}%
    \caption{(k) $n=1.00, Ra = 10^5$}
    \label{fig:3D_Isotherm_Ra_10^5_n_1_00.png}
  \end{subfigure}
   \begin{subfigure}{0.33\textwidth}        
   \centering
    \includegraphics[width=\textwidth]{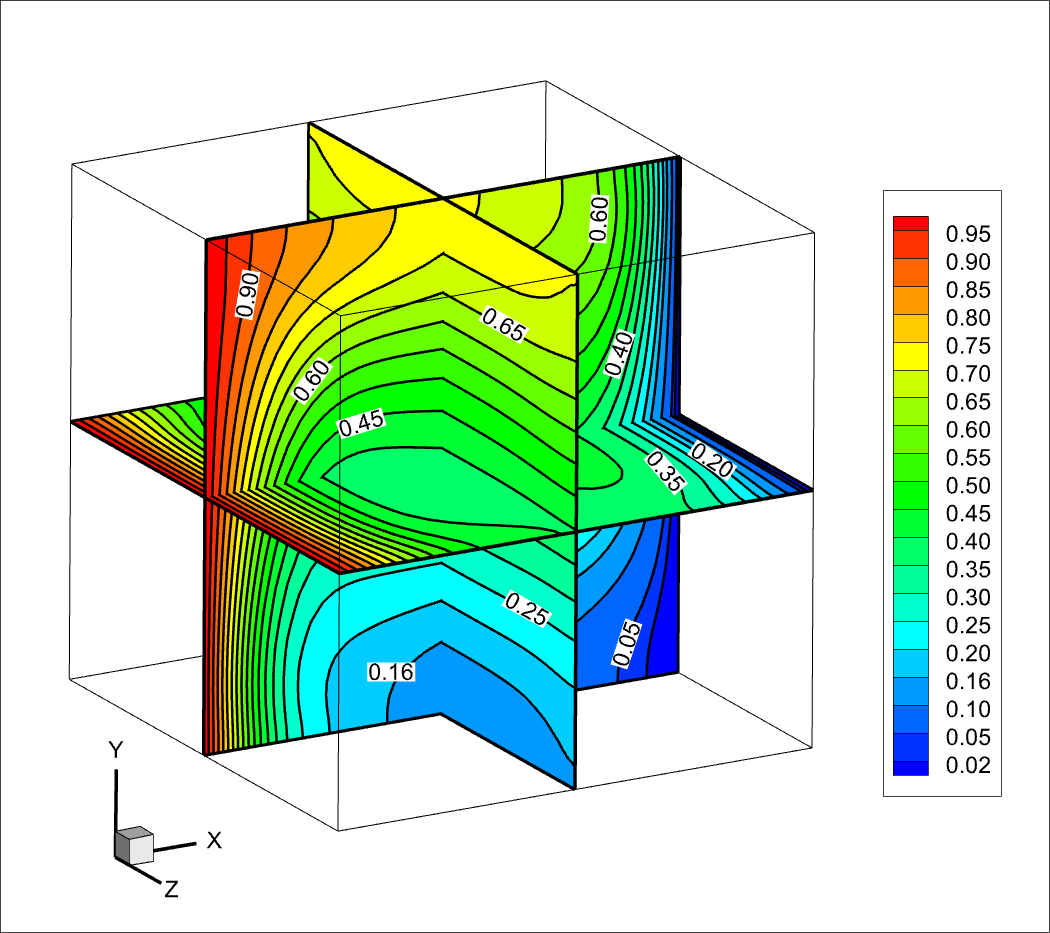}%
    \captionsetup{skip=2pt}%
    \caption{(l) $n=1.25, Ra = 10^5$}
    \label{fig:3D_Isotherm_Ra_10^5_n_1_25.png}
  \end{subfigure}%
  \vspace*{1pt}%
  \hspace*{\fill}%
  \caption{Isotherms on the three symmetric planes ($x=0.5, y=0.5, z=0.5)$ for different $Ra$ and $n$ values. Rows (a)–(c), (d)–(f), (g)–(i), and (j)–(l) show streamlines at \(Ra = 10^2\), \(10^3\), \(10^4\), and \(10^5\) for \(n = 0.75\), \(1.0\), and \(1.25\), respectively. }
  \label{fig:3D_streamlines}
\end{figure}


\begin{figure}[htbp]
 \centering
 \vspace*{0pt}%
 \hspace*{\fill}%
\begin{subfigure}{0.33\textwidth}     
    \centering
    \includegraphics[width=\textwidth]{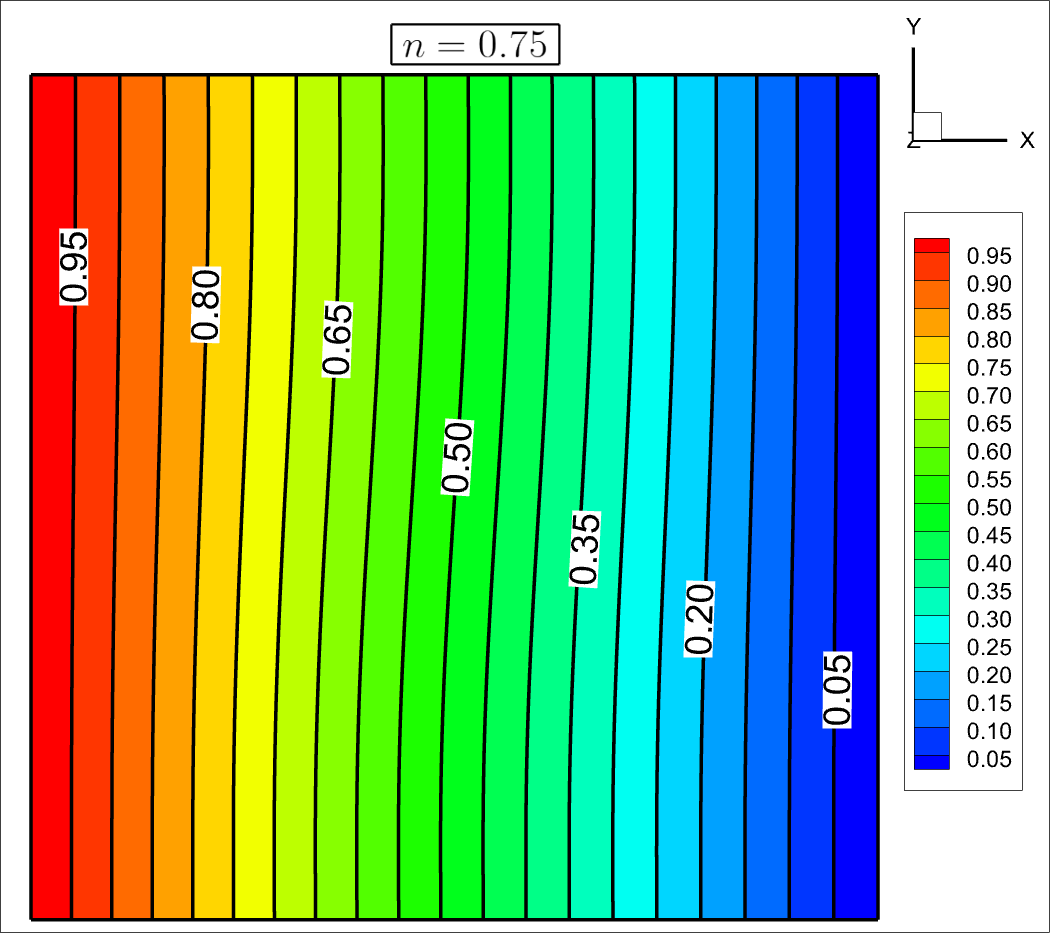}%
    \captionsetup{skip=2pt}%
    \caption{(a) $n=0.75, Ra = 10^2$ }
    \label{fig:2D_isotherm_Ra_10^2_n_0_75.png}
  \end{subfigure}%
 \begin{subfigure}{0.33\textwidth}        
   \centering
    \includegraphics[width=\textwidth]{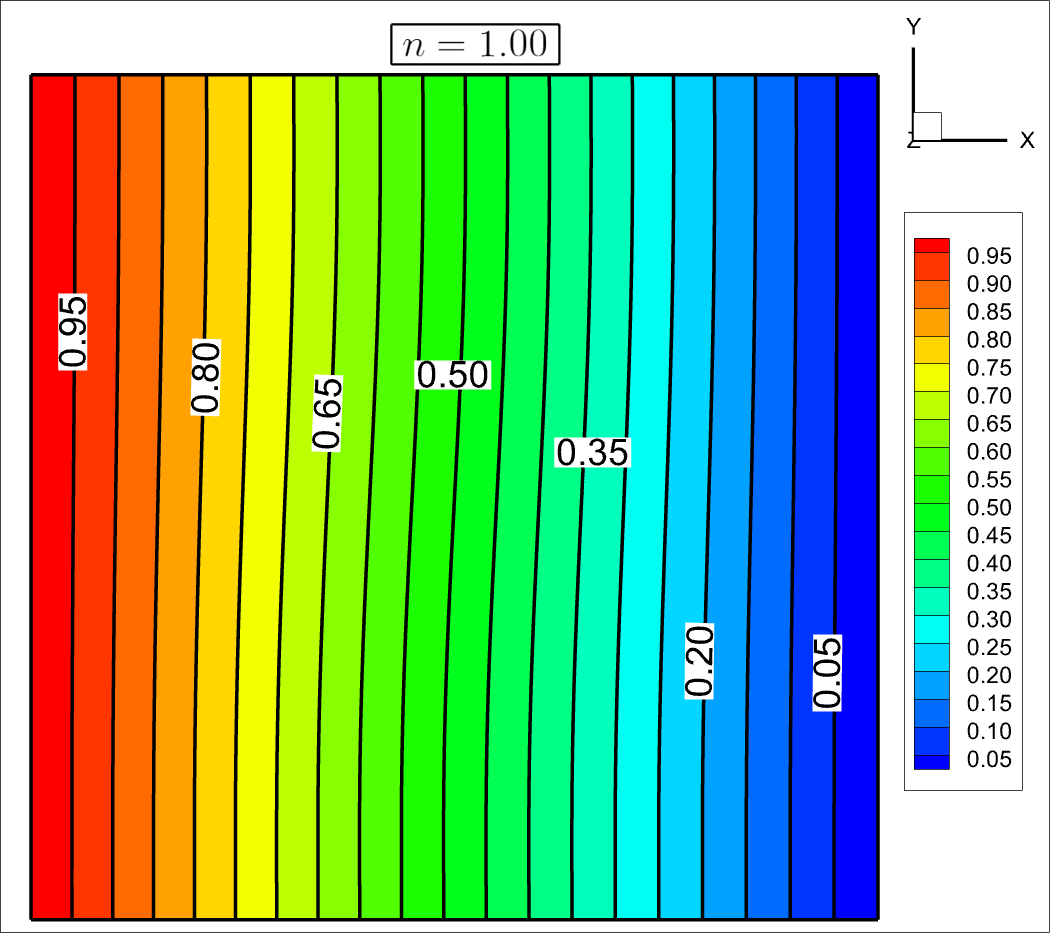}%
    \captionsetup{skip=2pt}%
    \caption{(b) $n=1.00, Ra = 10^2$}
    \label{fig:2D_isotherm_Ra_10^2_n_1.png}
  \end{subfigure}
   \begin{subfigure}{0.33\textwidth}        
   \centering
    \includegraphics[width=\textwidth]{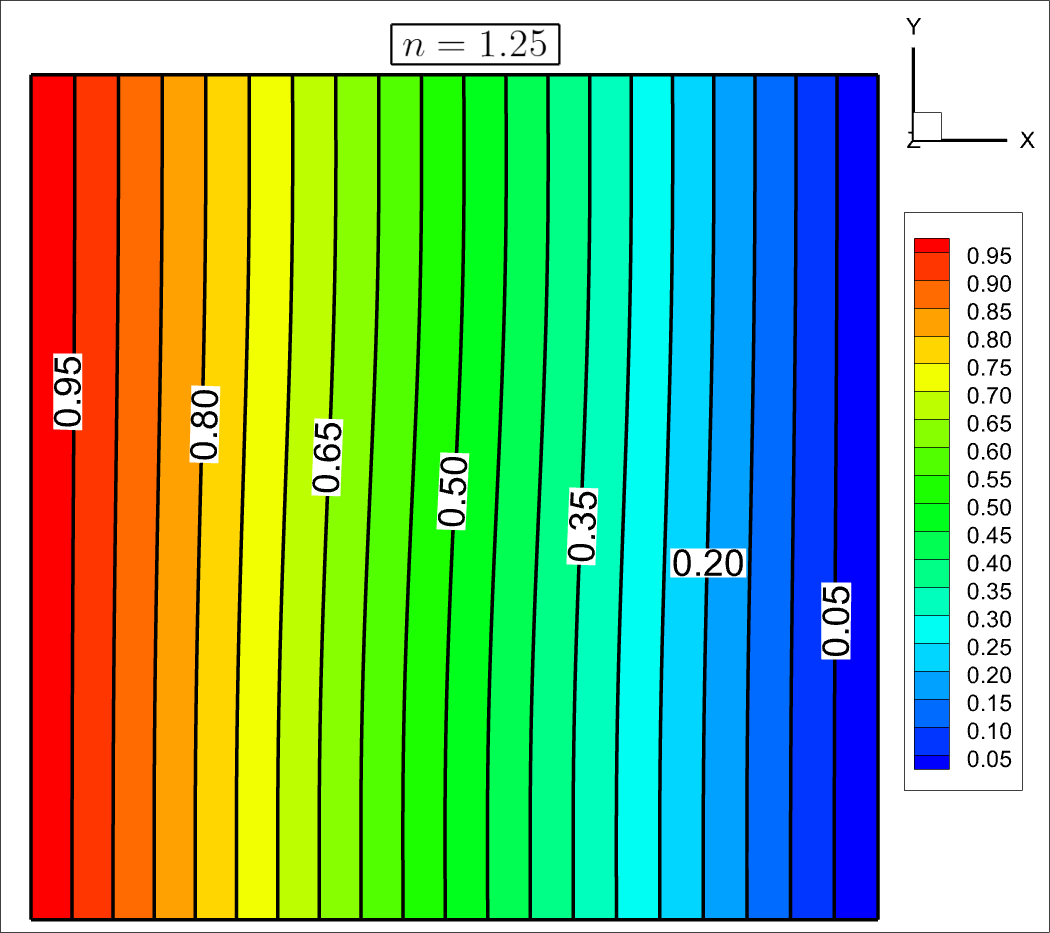}%
    \captionsetup{skip=2pt}%
    \caption{(c) $n=1.25, Ra = 10^2$}
    \label{fig:2D_isotherm_Ra_10^2_n_1_25.png}
  \end{subfigure}%
  \hspace*{\fill}

  \vspace*{8pt}%
  \hspace*{\fill}%
  \begin{subfigure}{0.33\textwidth}     
    \centering
    \includegraphics[width=\textwidth]{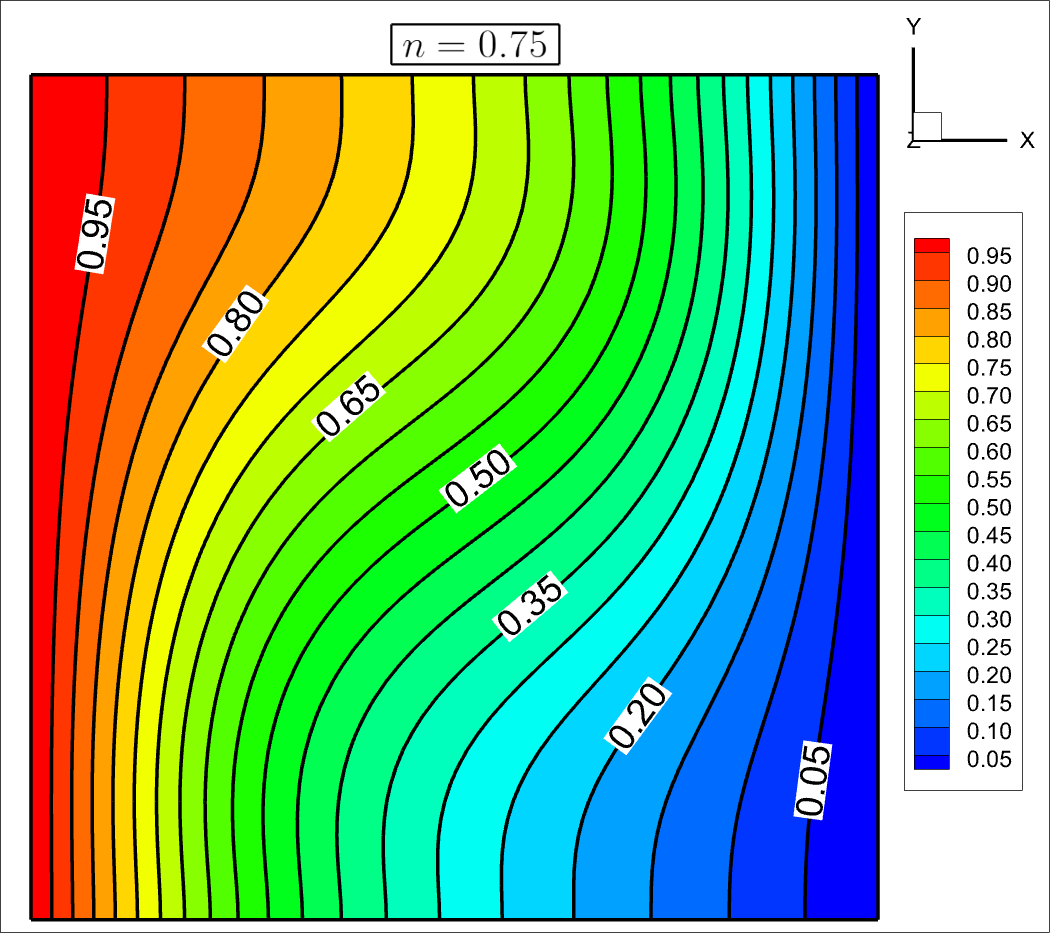}%
    \captionsetup{skip=2pt}%
    \caption{(d) $n=0.75, Ra = 10^3$}
    \label{fig:2D_isotherm_Ra_10^3_n_0_75.png}
  \end{subfigure}%
 \begin{subfigure}{0.33\textwidth}        
   \centering
    \includegraphics[width=\textwidth]{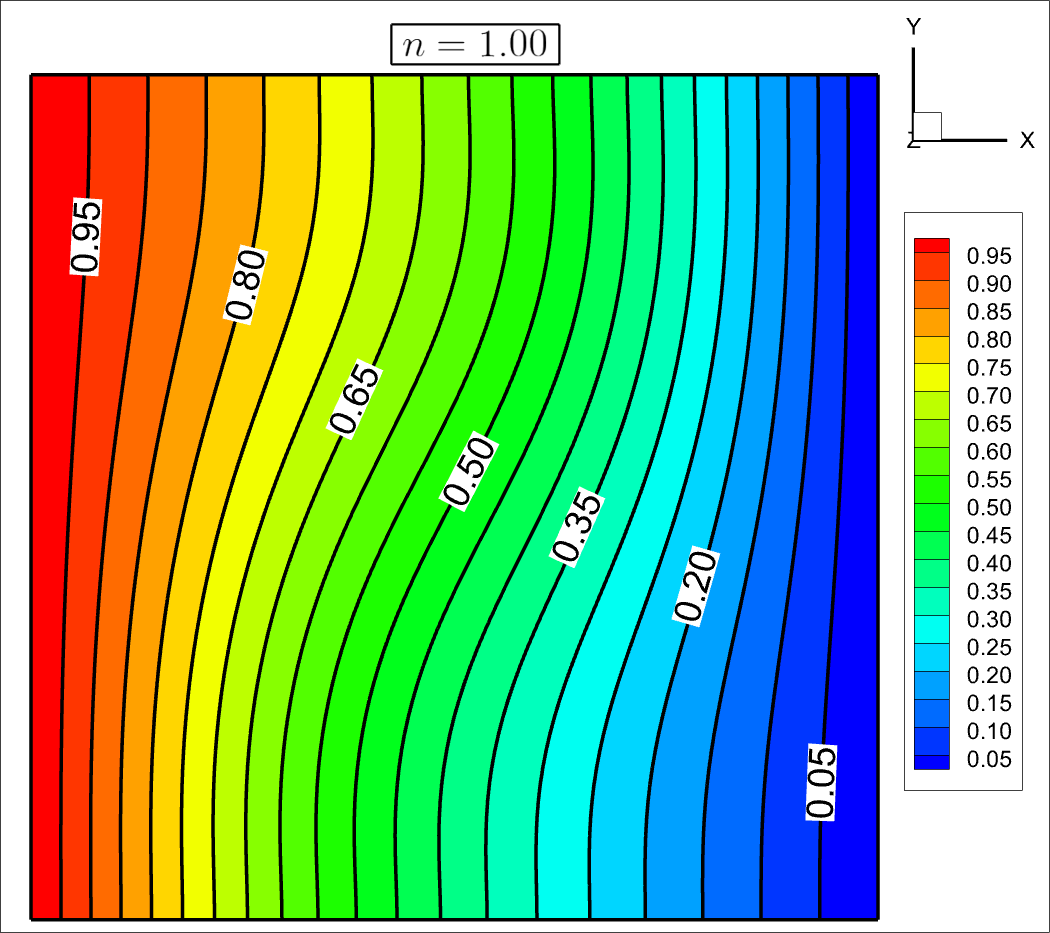}%
    \captionsetup{skip=2pt}%
    \caption{(e) $n=1.00, Ra = 10^3$}
    \label{fig:2D_isotherm_Ra_10^3_n_1_00.png}
  \end{subfigure}
   \begin{subfigure}{0.33\textwidth}        
   \centering
    \includegraphics[width=\textwidth]{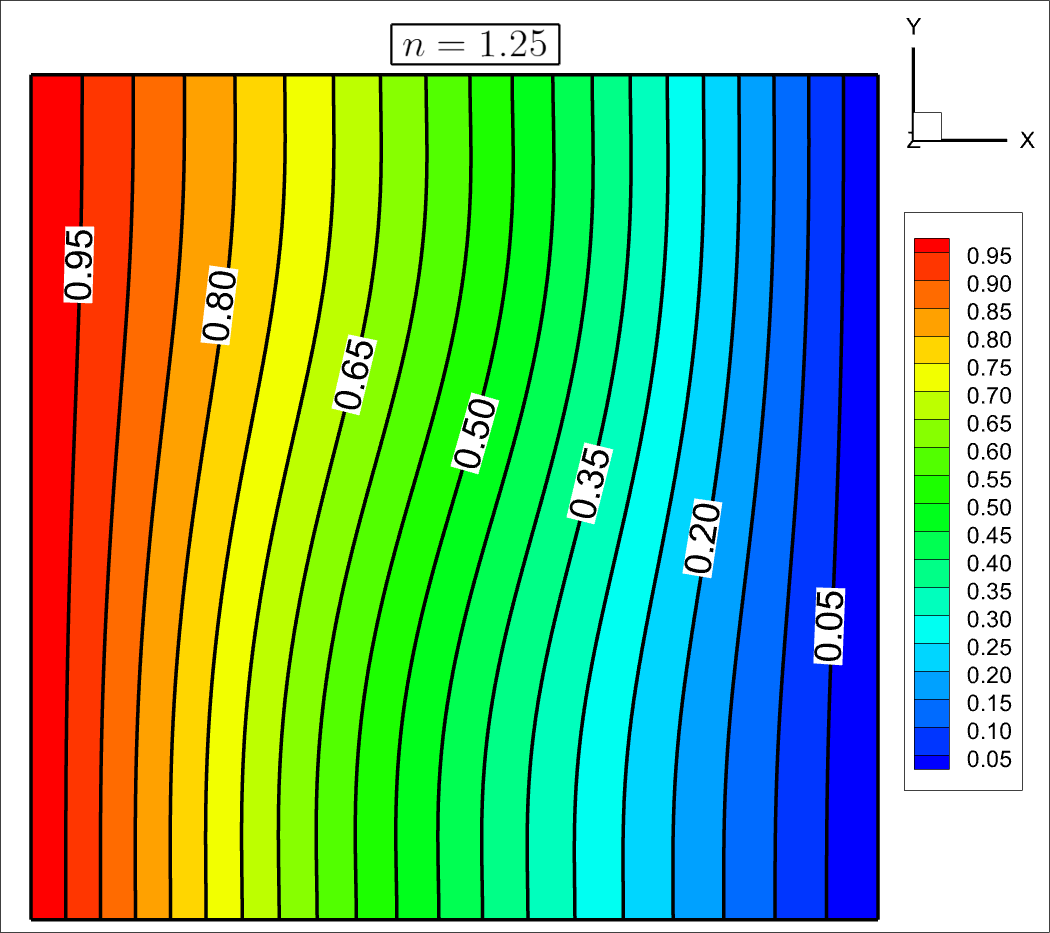}%
    \captionsetup{skip=2pt}%
    \caption{(f) $n=1.25, Ra = 10^3$}
    \label{fig:2D_isotherm_Ra_10^3_n_1_25.png}
  \end{subfigure}%
  \hspace*{\fill}

  \vspace*{8pt}%
  \hspace*{\fill}%
  \begin{subfigure}{0.33\textwidth}     
    \centering
    \includegraphics[width=\textwidth]{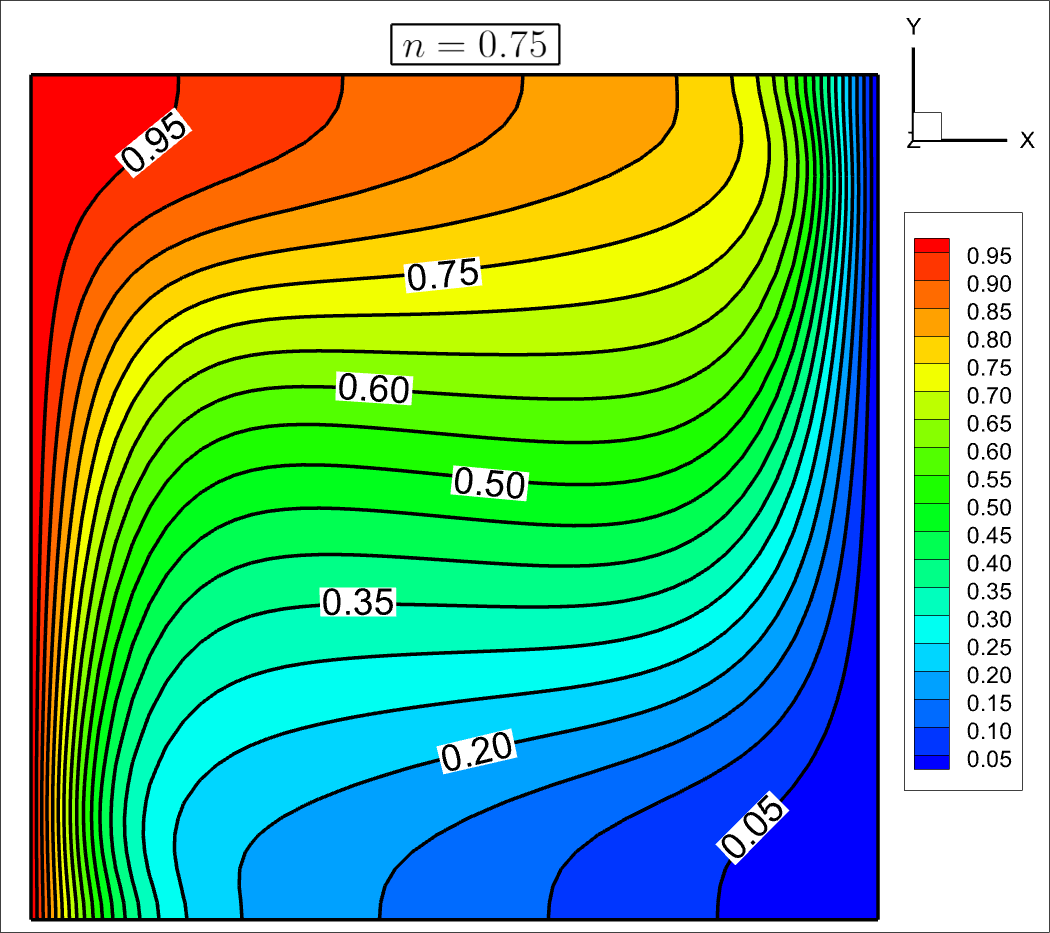}%
    \captionsetup{skip=2pt}%
    \caption{(g) $n=0.75, Ra = 10^4$}
    \label{fig:2D_isotherm_Ra_10^4_n_0_75.png}
  \end{subfigure}%
 \begin{subfigure}{0.33\textwidth}        
   \centering
    \includegraphics[width=\textwidth]{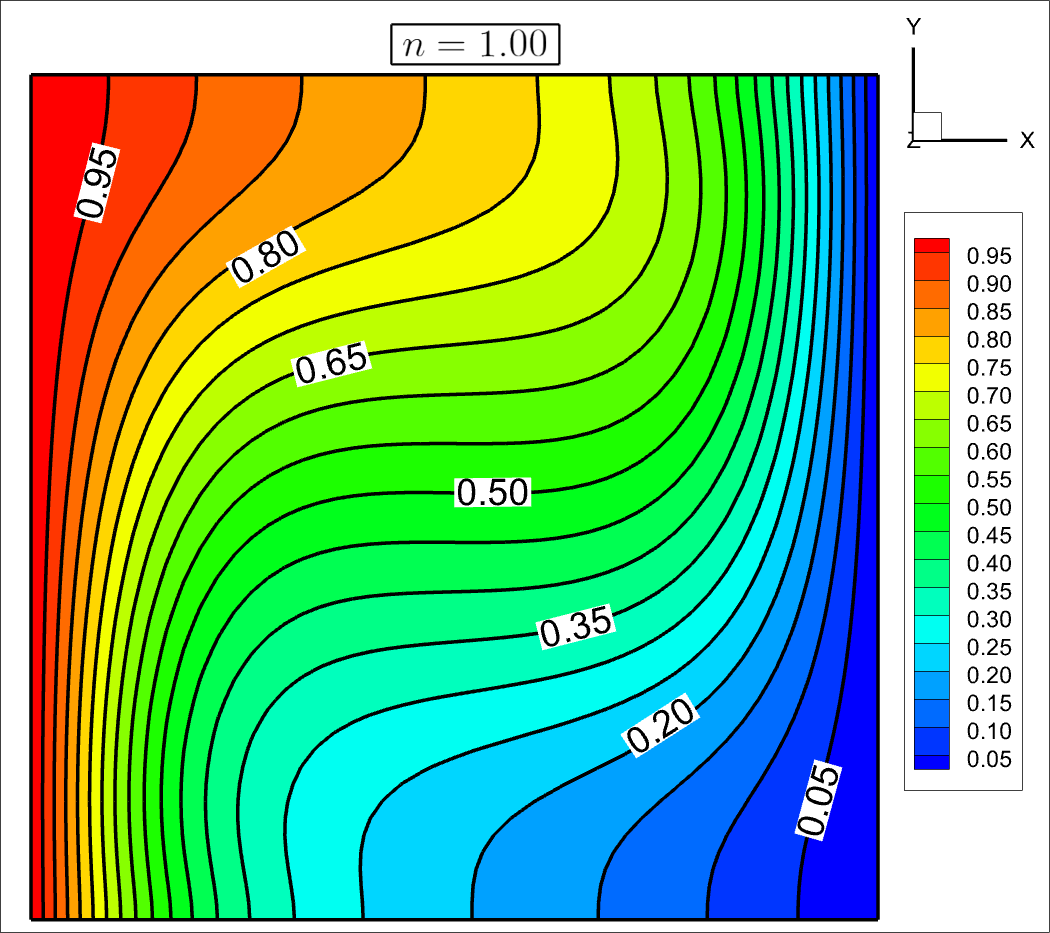}%
    \captionsetup{skip=2pt}%
    \caption{(h) $n=1.00, Ra = 10^4$}
    \label{fig:2D_isotherm_Ra_10^4_n_1_00.png}
  \end{subfigure}
   \begin{subfigure}{0.33\textwidth}        
   \centering
    \includegraphics[width=\textwidth]{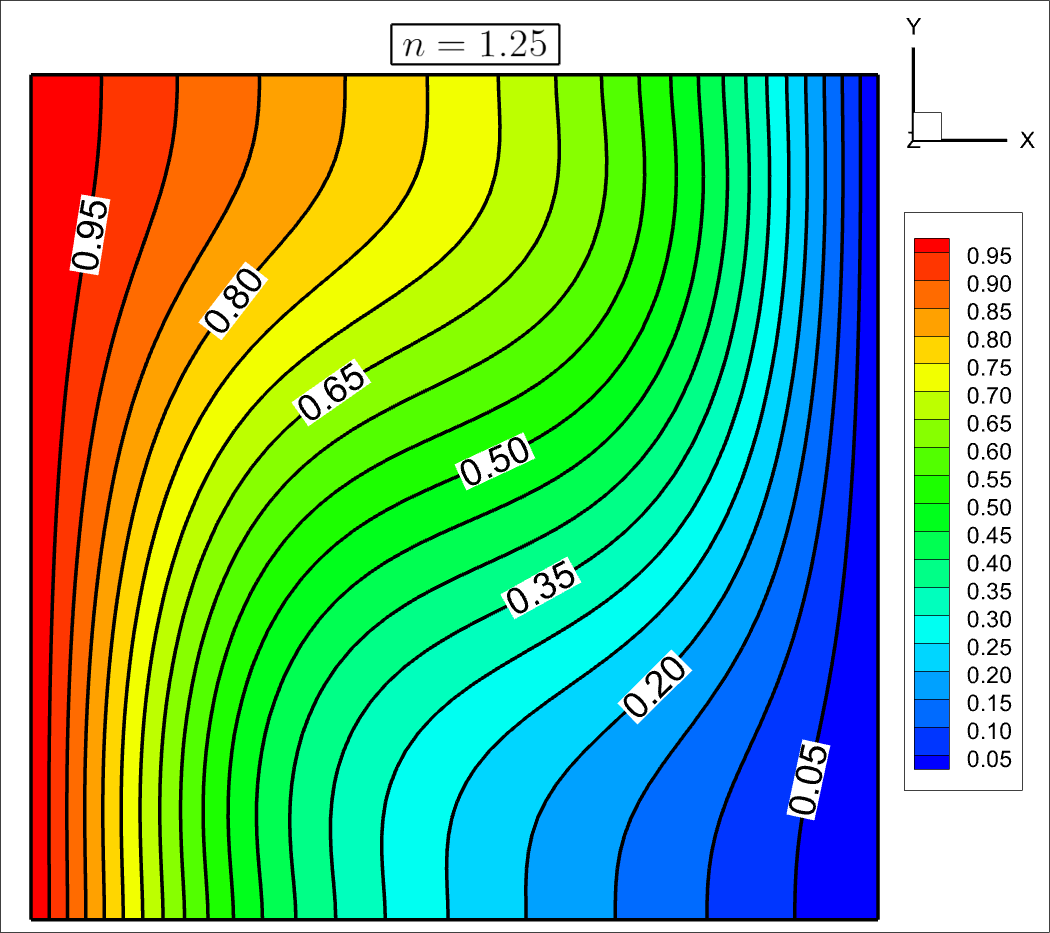}%
    \captionsetup{skip=2pt}%
    \caption{(i) $n=1.25, Ra = 10^4$}
    \label{fig:2D_isotherm_Ra_10^4_n_1_25.png}
  \end{subfigure}%
  \hspace*{\fill}

  \vspace*{8pt}%
  \hspace*{\fill}%
  \begin{subfigure}{0.33\textwidth}     
    \centering
    \includegraphics[width=\textwidth]{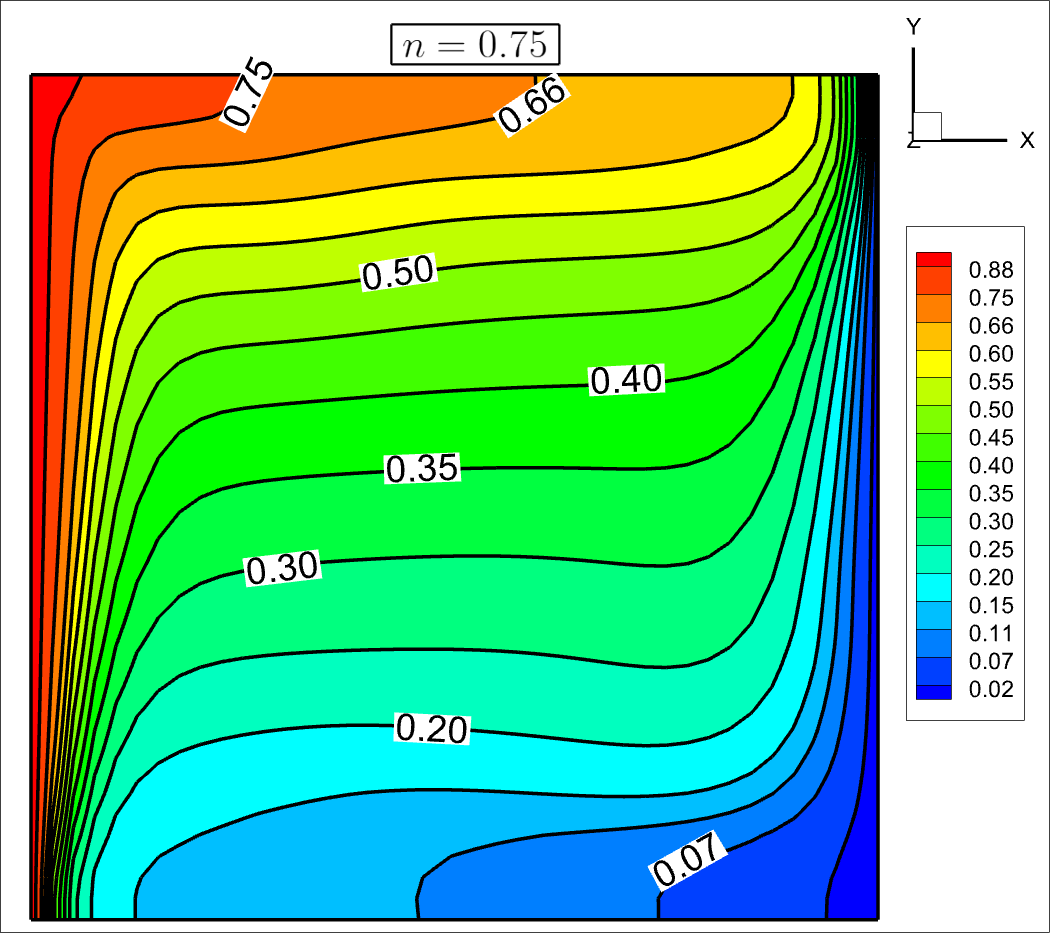}%
    \captionsetup{skip=2pt}%
    \caption{(j) $n=0.75, Ra = 10^5$}
    \label{fig:2D_isotherm_Ra_10^5_n_0_75.png}
  \end{subfigure}%
 \begin{subfigure}{0.33\textwidth}        
   \centering
    \includegraphics[width=\textwidth]{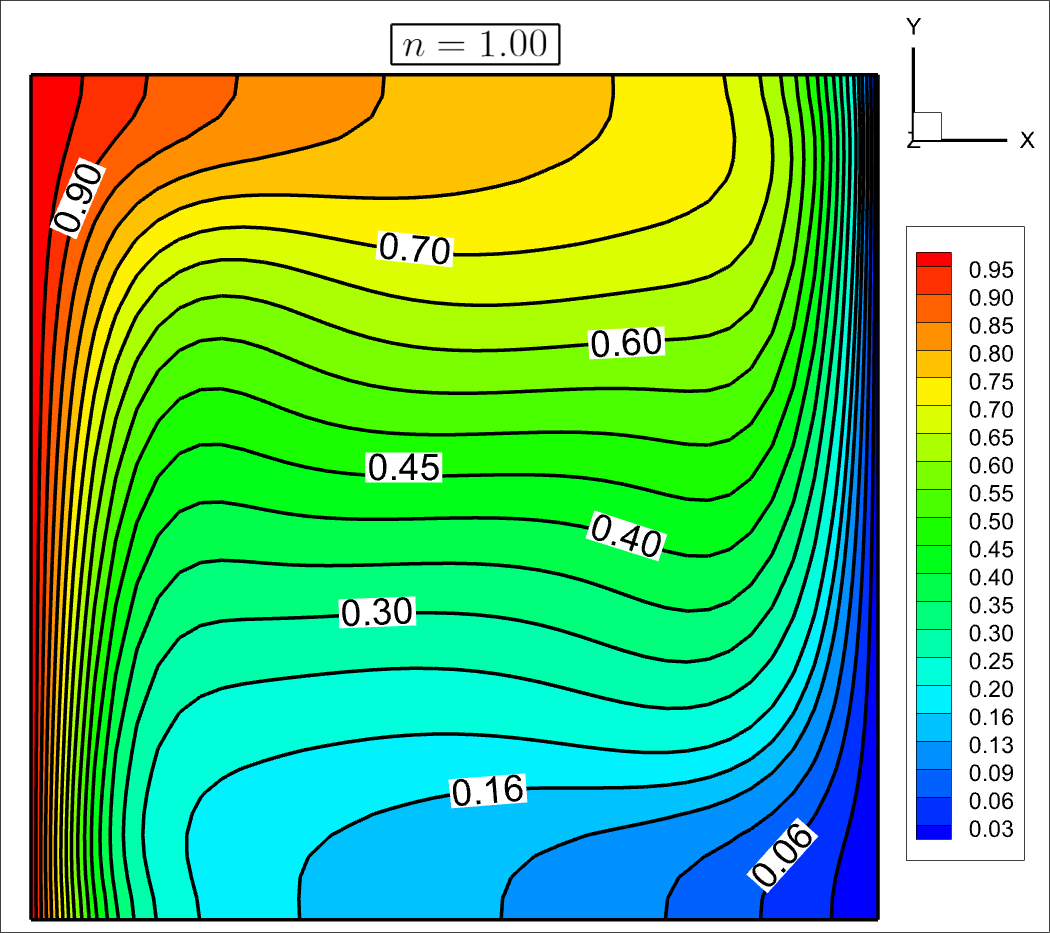}%
    \captionsetup{skip=2pt}%
    \caption{(k) $n=1.00, Ra = 10^5$}
    \label{fig:2D_isotherm_Ra_10^5_n_1.png}
  \end{subfigure}
   \begin{subfigure}{0.33\textwidth}        
   \centering
    \includegraphics[width=\textwidth]{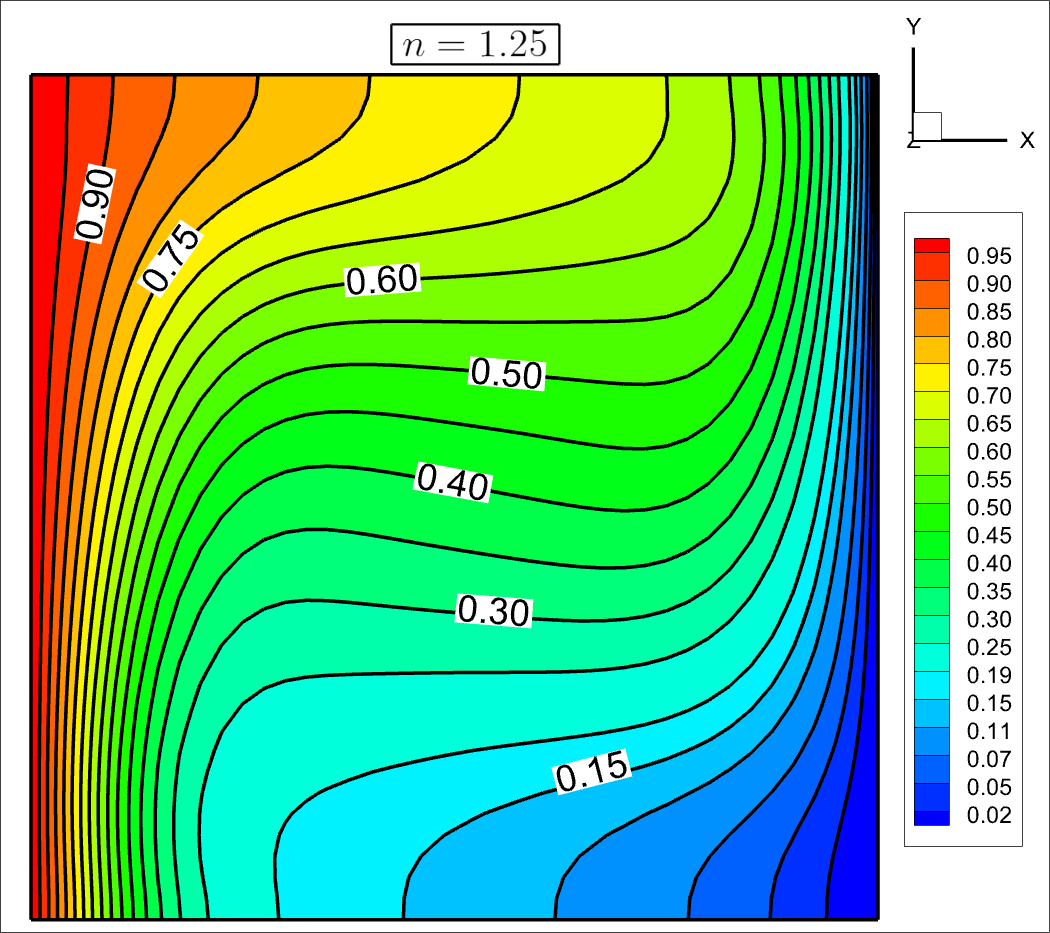}%
    \captionsetup{skip=2pt}%
    \caption{(l) $n=1.25, Ra = 10^5$}
    \label{fig:2D_isotherm_Ra_10^5_n_1_25.png}
  \end{subfigure}%
  \vspace*{1pt}%
  \hspace*{\fill}%
  \caption{Isotherms on the symmetric plane (z = 0.5) for different $Ra$ and $n$. Rows (a)–(c), (d)–(f), (g)–(i), and (j)–(l) show streamlines at \(Ra = 10^2\), \(10^3\), \(10^4\), and \(10^5\) for \(n = 0.75\), \(1.0\), and \(1.25\), respectively. }
  \label{fig:2D_isotherms}
\end{figure}

In Figure \ref{fig:3D_streamlines}, we present the isotherm contours at the symmetric planes \( x=0.5 \), \( y=0.5 \), and \( z=0.5 \). The most significant variations in the isotherms are observed on the \( z=0.5 \) plane, so a separate plot of the isotherms on this plane is provided in Figure \ref{fig:2D_isotherms}. At \( Ra=10^2 \), the isotherms appear (Figure \ref{fig:2D_isotherms}) nearly vertical and evenly spaced across all values of \( n \), indicating a conduction-dominant regime. As \( Ra \) increases to \( 10^3 \), the isotherms begin to curve, with the greatest curvature observed for \( n=0.75 \) (shear-thinning fluid), suggesting high convective activity compared to Newtonian (\( n=1.0 \)) and shear-thickening (\( n=1.25 \)) fluids. At \( Ra=10^4 \), the curvature of the isotherms intensifies for all values of \( n \), further signifying the transition from conduction to convection. At \( Ra=10^5 \), the buoyancy effects dominate, leading to the strongest natural convection. For \( n=0.75 \), the isotherms in the central region of the plane become nearly horizontal. Other values of \( n \) also show more twisted patterns, but the strongest convective effects are seen for \( n=0.75 \).
As \( Ra \) increases, the isotherms transition from almost vertical near the isothermal walls to more horizontal in the center of the plane \( z=0.5 \), confirming the shift from conduction-dominated to convection-dominated heat transfer. 
Regarding the power-law index \( n \), as \( n \) increases for any fixed \( Ra >10^2 \), the temperature boundary layers become thicker, indicating a weakening of convective heat transfer. The shear-thinning case \( n=0.75 \) consistently shows the highest variation in the isotherms, demonstrating that as \( n \) increases, the strength of convection diminishes. By examining the plane at \( x = 0.5 \)\ in Figure \ref{fig:3D_streamlines}, we observe a clear increase in the number of isotherm lines as the Rayleigh number (\( Ra \)) increases for any fixed power-law index (\( n \)), while a corresponding decrease in isotherm density is observed as \( n \) increases at any fixed \( Ra \). Additionally, the isotherms transition from curved to more linear patterns on this plane as \( Ra \) rises. Moreover, the temperature near the lateral adiabatic walls is still lower than that in the core area, as can be seen on the symmetry planes at \( x = 0.5 \) and \( y = 0.5 \), suggesting that the adiabatic walls restrict heat transmission. 
\subsection{Dynamics of Heat Transfer: Analyzing the Nusselt Number}
\begin{figure}[htbp]
 \centering
 \vspace*{5pt}%
 \hspace*{\fill}%
\begin{subfigure}{0.50\textwidth}     
    \centering
    \includegraphics[width=\textwidth]{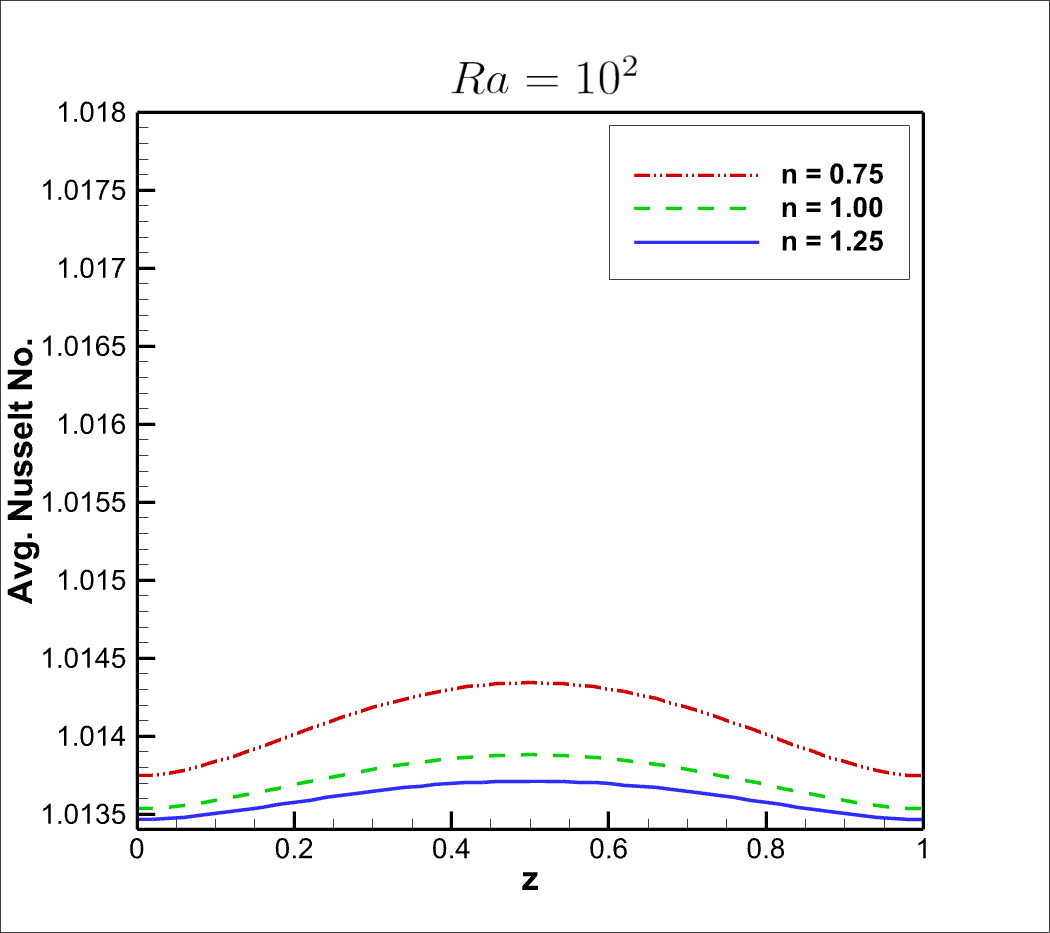}%
    \captionsetup{skip=5pt}%
    \caption{(a)}
    \label{fig:Ra_10^2_avg}
  \end{subfigure}%
 \begin{subfigure}{0.50\textwidth}        
   \centering
    \includegraphics[width=\textwidth]{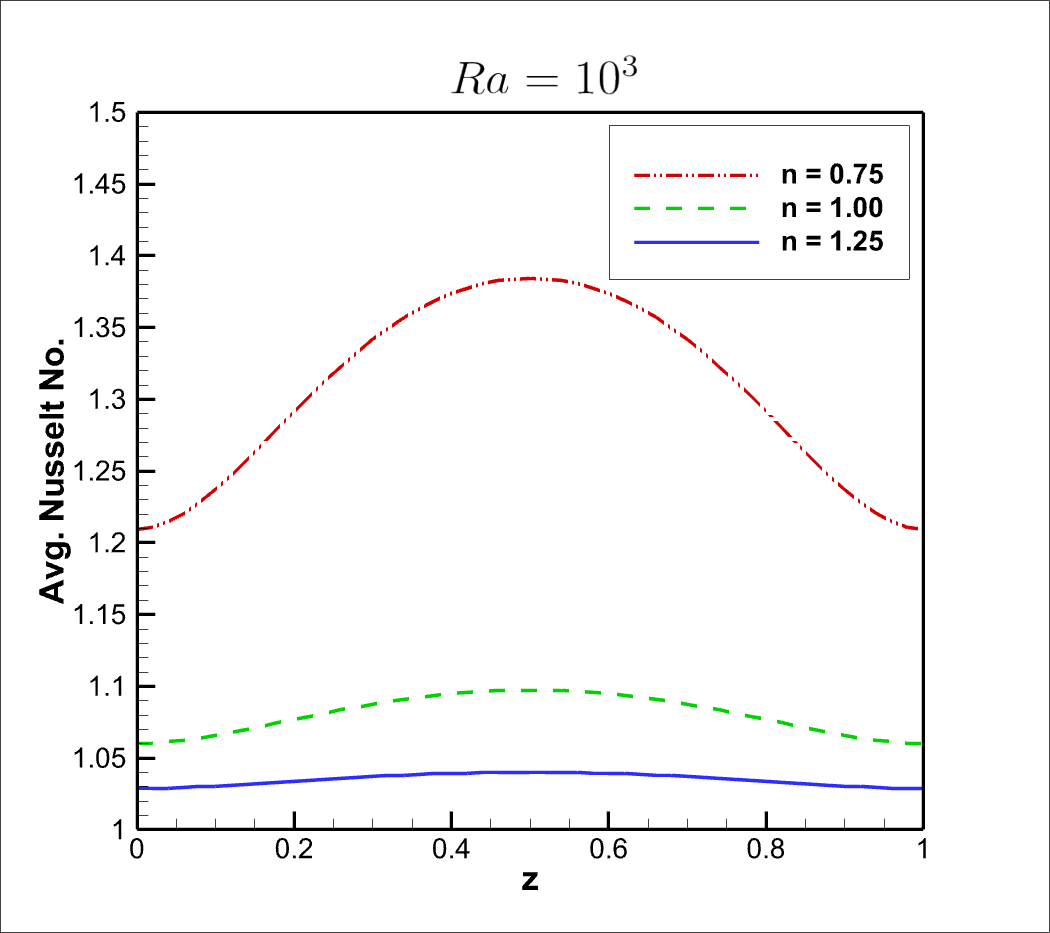}%
    \captionsetup{skip=5pt}%
    \caption{(b)}
    \label{fig:Ra_10^3_avg}
  \end{subfigure}
  \hspace*{\fill}

  \vspace*{8pt}%
  \hspace*{\fill}%
  \begin{subfigure}{0.50\textwidth}     
    \centering
    \includegraphics[width=\textwidth]{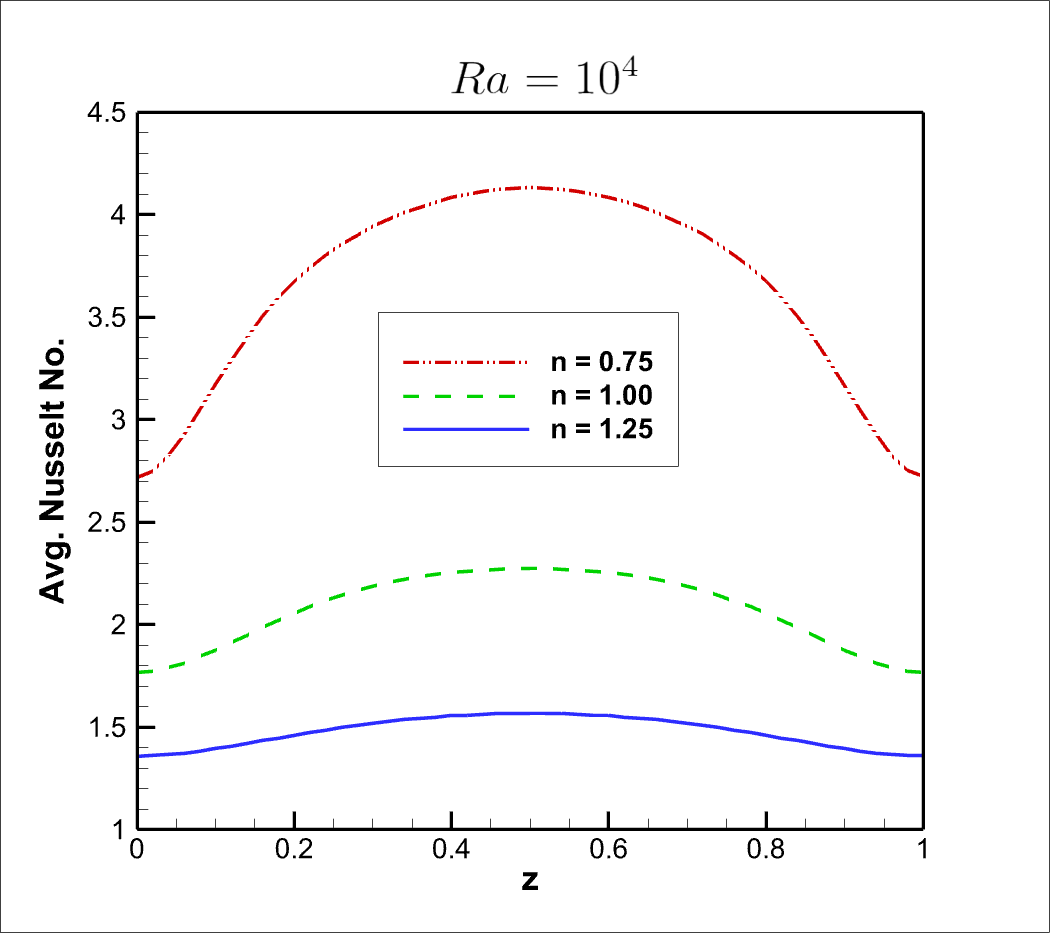}%
    \captionsetup{skip=5pt}%
    \caption{(c)}
    \label{fig:Ra_10^4_avg}
  \end{subfigure}%
  \begin{subfigure}{0.505\textwidth}     
    \centering
    \includegraphics[width=\textwidth]{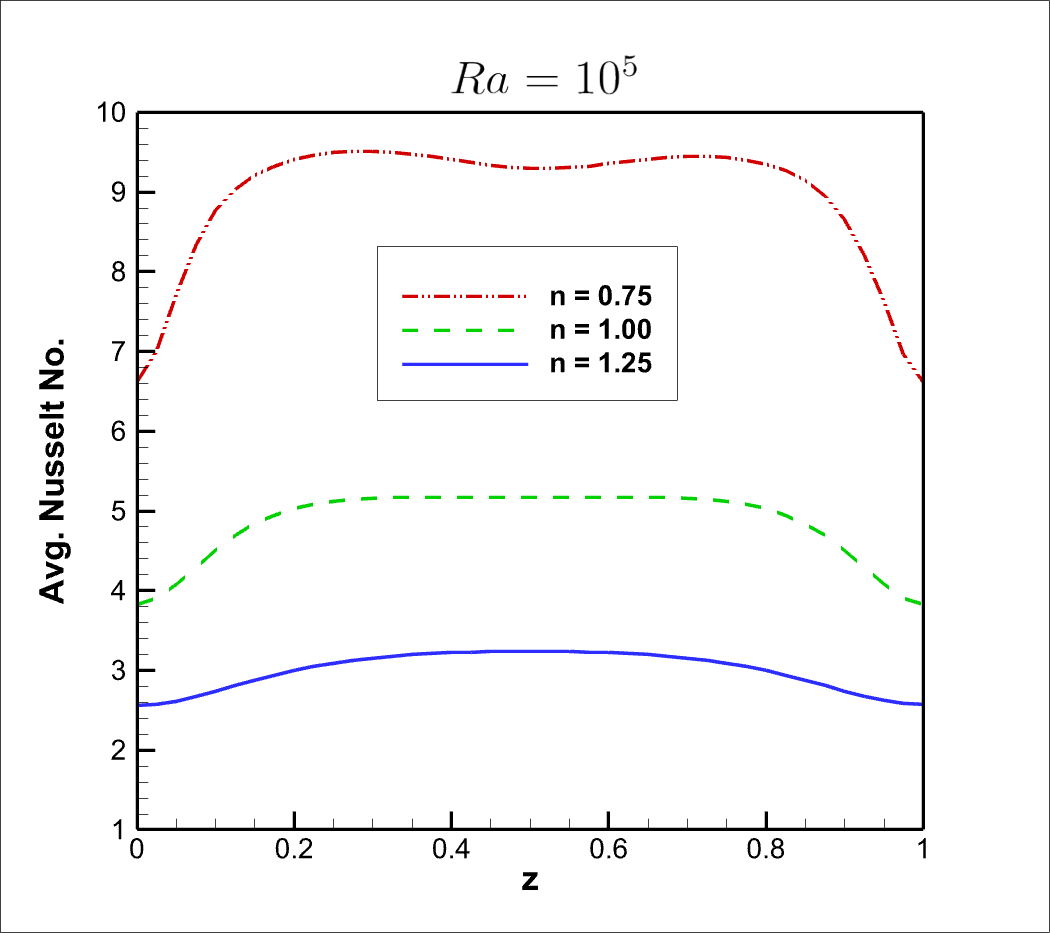}%
    \captionsetup{skip=5pt}%
    \caption{(d)}
    \label{fig:Ra_10^5_avg}
  \end{subfigure}%
   \caption{Variation of the average Nusselt number (\(Nu_{\text{Avg}}\)) with respect to different power-law indices (\(n\)) at various Rayleigh numbers: (a) \(Ra=10^2\), (b) \(Ra=10^3\), (c) \(Ra=10^4\), and (d) \(Ra=10^5\).}
  \label{fig:Average_Nusselt_Number_1}
\end{figure}

The efficiency of convective heat transmission between a fluid and a solid surface is represented by the Nusselt number ($Nu$), a dimensionless metric in fluid dynamics and heat transfer. Convection is the primary heat transmission mechanism when the Nusselt number is larger, whereas a lower value indicates that conduction plays a more significant role.
These quantities are defined as below:
$$Nu_{\text {local}}(y, z)= \frac{\partial \theta(y, z)}{\partial x} $$ at heated wall (x=0.0). Here, \( Nu_{\text{local}} \) denotes the local Nusselt number, while the average Nusselt number \( (Nu_{\text{avg}}) \) is expressed as:
\\ 
$$Nu_{\text {avg }}(z)=\int_0^1 N u_{\text {L}}(y, z) \mathrm{d} y$$ 
\begin{figure}
    \centering
    \includegraphics[width=\textwidth]{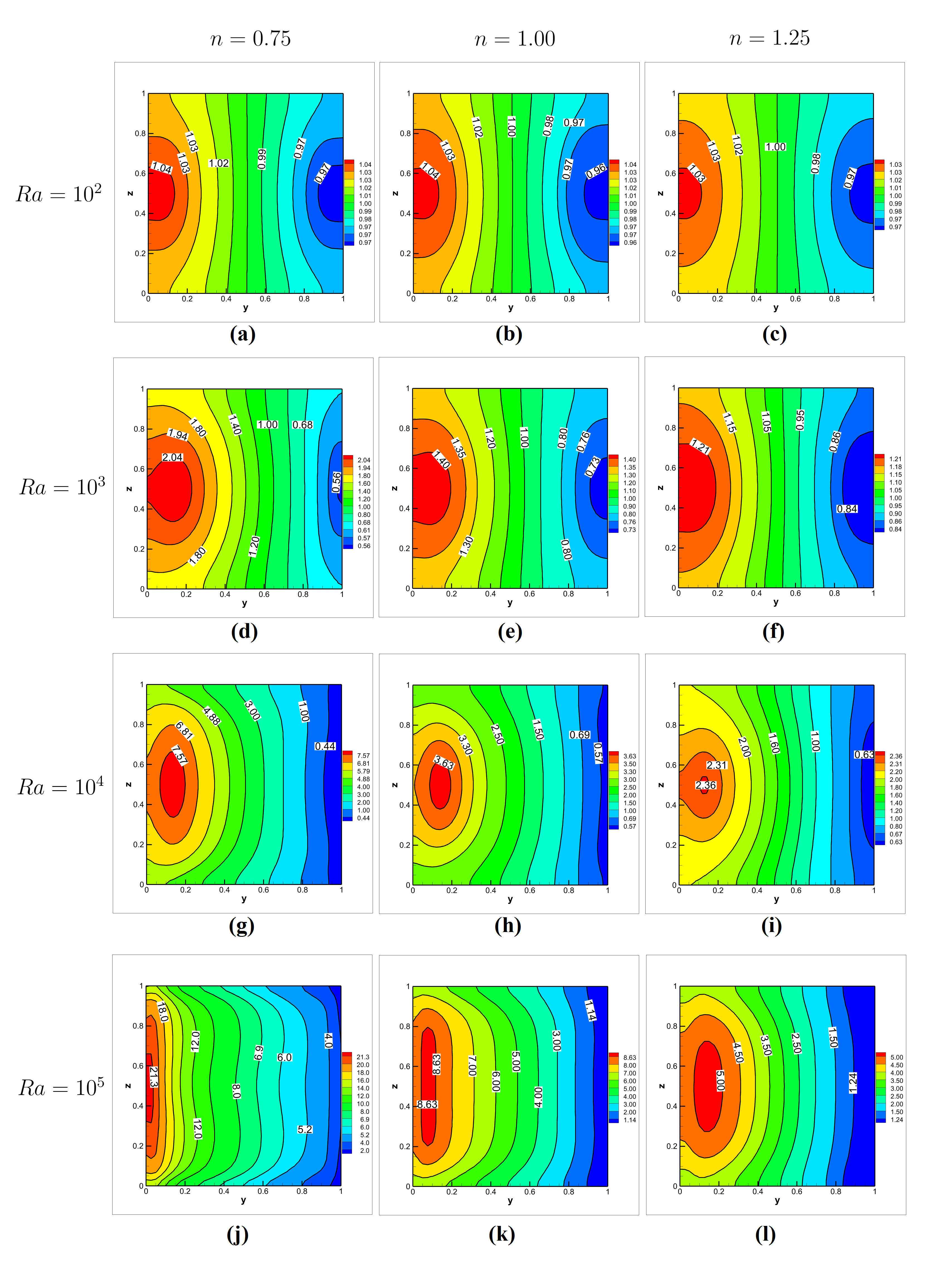}
    \caption{Local Nusselt number on the heated wall ($x=0$) for different $Ra$ and $n$. Rows (a)–(c), (d)–(f), (g)–(i), and (j)–(l) show streamlines at \(Ra = 10^2\), \(10^3\), \(10^4\), and \(10^5\) for \(n = 0.75\), \(1.0\), and \(1.25\), respectively.}
    \label{fig:Local_Nu}
\end{figure}
In Figure \ref{fig:Local_Nu}, we present the distribution of the local Nusselt number (\( Nu_{\text{local}} \)) along the heated wall (\(x = 0\)) of the cavity. At \(Ra = 10^2\), \( Nu_{\text{local}} \) exhibits a symmetric pattern about the line \(y = 0.5\) for all values of the power-law index \(n\), with the effect of \(n\) being negligible at this low $Ra$. However, as \(Ra\) increases, an asymmetry becomes evident with respect to \(y = 0.5\). As \( Ra \) increases to \( 10^4 \), the initially curved ‘C’-shaped contours near \( y = 1 \) begin to straighten slightly. With a further increase to \( Ra = 10^5 \), these contours gradually transform into a reverse ‘C’ shape, reflecting a significant shift in the heat transfer. The maximum value of \( Nu_{\text{local}} \) rises significantly with increasing \(Ra\) for any fixed \(n\), while it decreases as \(n\) increases for any fixed \(Ra\). As we progress from \(Ra = 10^2\) to \(Ra = 10^5\), the increase in the maximum \( Nu_{\text{local}} \) is approximately 1948\% for the shear-thinning fluid (\(n = 0.75\)), 729\% for the Newtonian fluid (\(n = 1\)), and 385\% for the shear-thickening fluid (\(n = 1.25\)). This indicates that the influence of Rayleigh number is most significant in shear-thinning fluids, followed by Newtonian and then shear-thickening fluids. These results highlight the critical interaction between thermal buoyancy forces and fluid rheology, with shear-thinning fluids exhibiting a much more enhanced convective heat transfer response as \(Ra\) increases.\\
Figure \ref{fig:Average_Nusselt_Number_1} illustrates the variation of the average Nusselt number ($Nu_{\text {avg }}$) for different values of \( n \) and \( Ra \). As the power-law index \( n \) increases, the $Nu_{\text {avg }}$ consistently decreases throughout the domain. However, when comparing the $Nu_{\text {avg }}$ with respect to \( Ra \) at any fixed \( n \), a noticeable increase is observed as \( Ra \) increases. For \( Ra = 10^2 \), \( 10^3 \), and \( 10^4 \), the $Nu_{\text {avg }}$ exhibits a consistent trend: it increases steadily along the domain, reaching a global maximum around \( z = 0.5 \), and then decreases beyond this point. This indicates that the most efficient heat transfer occurs near the middle of the cavity for these \( Ra \) values. 
However, at \( Ra = 10^5 \), a distinct shift in behavior is observed, as shown in Figure \ref{fig:Average_Nusselt_Number_1}(d). Instead of a single global maximum, two local maxima emerge in the case of shear-thinning fluids (\( n = 0.75 \)), highlighting the presence of highly intensified convective currents. This double peak pattern is less prominent in the cases of Newtonian (\( n = 1.0 \)) and shear-thickening fluids (\( n = 1.25 \)), where the effect of intense convection is not as strong at this high \( Ra \).
\subsection{Entropy Generation}
The analysis of entropy generation is crucial for understanding the effectiveness of heat transfer within fluids and quantifying energy losses caused by irreversible processes. Entropy generation results from thermal irreversibilities associated with temperature gradients and fluid friction irreversibilities linked to velocity gradients. In non-Newtonian fluids, the viscosity varies with shear rate, significantly influencing both fluid dynamics and heat transfer behavior. Examining entropy generation in these scenarios allows for the identification of irreversibility sources within the system, offering valuable insights into the complex relationship between heat transfer efficiency and viscous dissipation. The local entropy generation for the 3D non-Newtonian fluid in non-dimensional form can be expressed as \cite{Hasan_2024,Rahman_2022}:
\begin{equation}\label{eq_3_Entropy}
E_{\mathrm{local}}=E_{\mathrm{HT}}+E_{\mathrm{FF}}
\end{equation}
where, entropy generation due to heat transfer ($E_{\mathrm{HT}}$)  is 
\begin{equation}\label{eq_4_Entropy}
E_{\mathrm{HT}}=\left[\left(\frac{\partial \theta}{\partial z}\right)^2+\left(\frac{\partial \theta}{\partial Y}\right)^2 +\left(\frac{\partial \theta}{\partial x}\right)^2 \right]
\end{equation}
and entropy generation due to fluid friction ($E_{\mathrm{FF}}$) is 
\begin{equation}
\begin{aligned}
& E_{\mathrm{FF}} =\eta \chi^{*}\left[\left(\frac{\partial u}{\partial z}+\frac{\partial w}{\partial x}\right)^2 +\left(\frac{\partial u}{\partial y}+\frac{\partial v}{\partial x}\right)^2 +\left(\frac{\partial v}{\partial z}+\frac{\partial w}{\partial y}\right)^2\right] \\
&+ \eta\varphi^{*}\left[2\left(\frac{\partial w}{\partial z}\right)^2+2\left(\frac{\partial v}{\partial y}\right)^2+2\left(\frac{\partial u}{\partial x}\right)^2\right]
\end{aligned}\label{eq_5_Entropy}
\end{equation}
Here, the viscosity is indicated by \( \eta \) and the irreversibility distribution ratio, represented by \( \varphi^{*} \), is kept constant at \( 10^{-4} \) for every scenario \cite{Hasan_2024}. We can calculate the overall entropy generation \( E_{TE} \) by integrating the dimensionless \( E_{\mathrm{local}} \) across the system's whole volume.
$$
E_{total}=\int_{\vartheta} E_{\mathrm{local}} \mathrm{d} \vartheta
$$
The ratio of entropy produced by heat transfer to the total entropy arising from both fluid friction and heat transfer is known as the Bejan number (\( Be \)) in thermodynamics. It can be stated as follows:
$$
Be=\frac{E_{\mathrm{HT}}}{E_{\mathrm{local}}}
$$
A high Bejan number (\( Be > 0.5 \)) signifies that heat transfer accounts for a significant amount of entropy formation, demonstrating its dominance in the total entropy production. Conversely, a low Bejan number (\( Be < 0.5 \)) means that fluid friction generates more entropy than heat transmission.
\begin{figure}
    \centering
    \includegraphics[width=\textwidth]{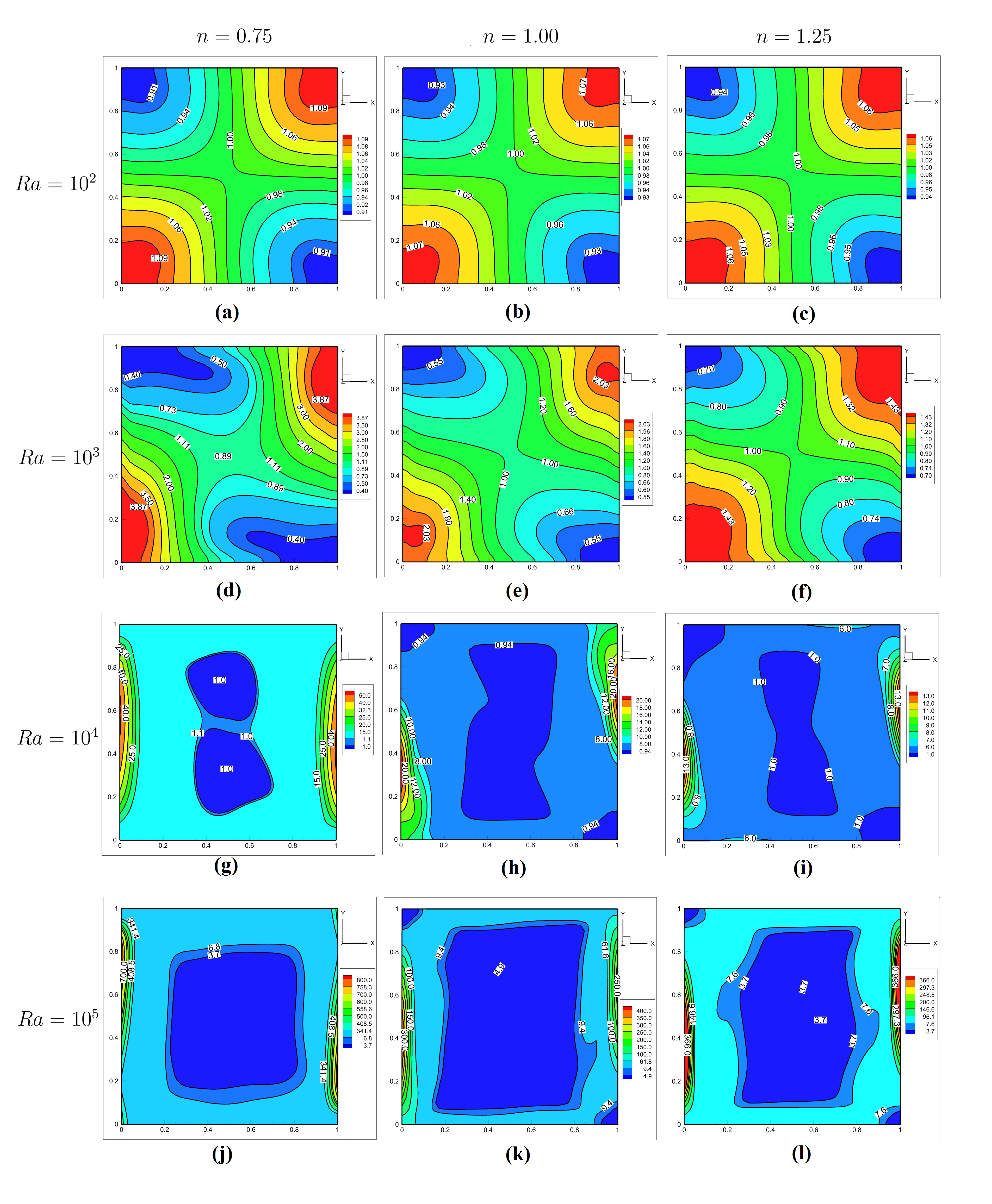}
    \caption{Local entropy generation for different $Ra$ and $n$. Rows (a)–(c), (d)–(f), (g)–(i), and (j)–(l) show streamlines at \(Ra = 10^2\), \(10^3\), \(10^4\), and \(10^5\) for \(n = 0.75\), \(1.0\), and \(1.25\), respectively.}
    \label{fig:Local_Entropy}
\end{figure}

Figure \ref{fig:Local_Entropy} presents the local entropy generation (\( E_{\mathrm{local}} \)) within the cavity for various values of \( Ra \) and \( n \). At lower Rayleigh numbers (\( Ra = 10^2 \) and \( 10^3 \)), the entropy generation is more uniformly distributed across the entire cavity for all values of \( n \). However, as \( Ra \) increases to \( 10^4 \) and \( 10^5 \), the entropy generation becomes increasingly localized near the thermally active walls. This transition can be attributed to the development of boundary layers at higher \( Ra \), where heat transfer and fluid friction are more intense, resulting in a concentrated entropy generation along the enclosure walls. Additionally, for all \( Ra \) and \( n \) values, the local entropy generation patterns exhibit near-diagonal symmetry across the cavity. 
The maximum value of the local entropy generation increases significantly with increasing \( Ra \), while it decreases as \( n \) increases. This indicates that shear-thinning fluids (\( n = 0.75 \)) tend to generate more entropy compared to Newtonian fluids (\( n = 1.0 \)) and shear-thickening fluids (\( n = 1.25 \)). This behavior highlights that higher convective flow, characteristic of shear-thinning fluids, leads to greater thermal and viscous dissipation, contributing to increased entropy generation.

\begin{figure}[htbp]
 \centering
 \vspace*{5pt}%
 \hspace*{\fill}%
\begin{subfigure}{0.5\textwidth}     
    \centering
    \includegraphics[width=\textwidth]{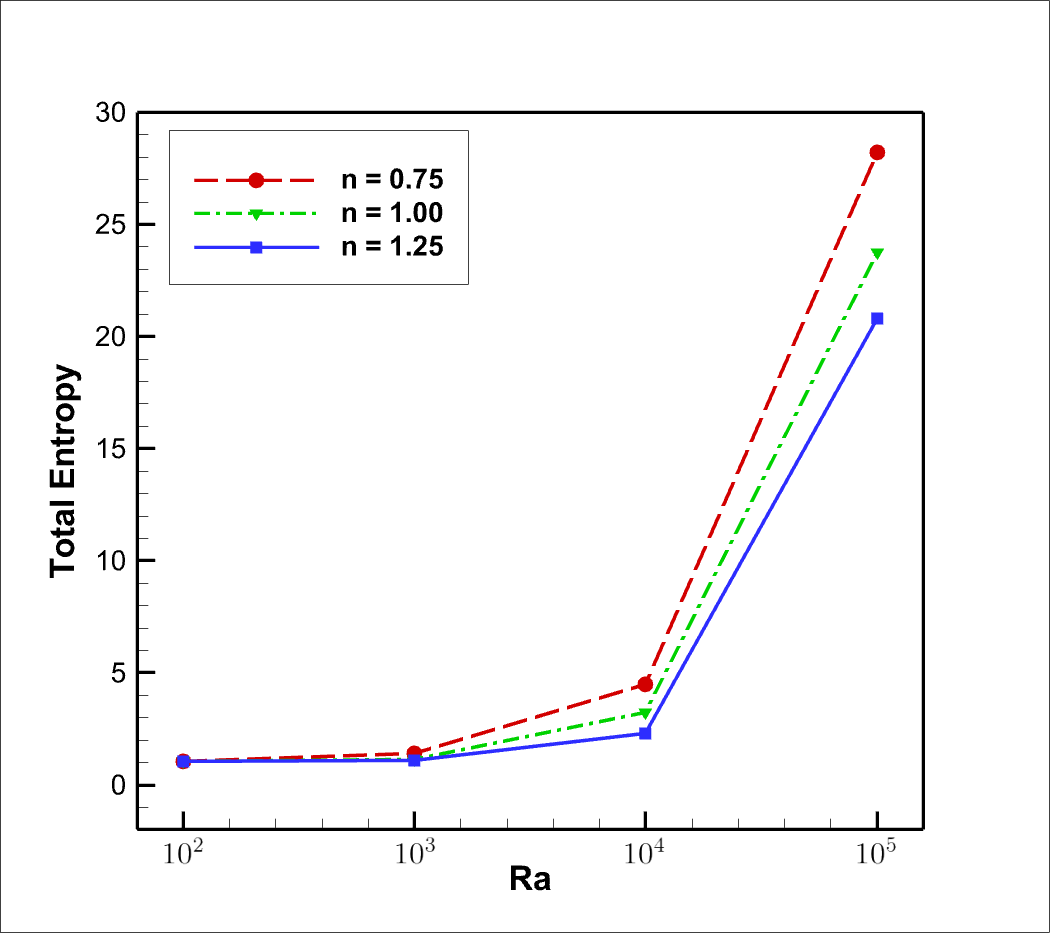}%
    \captionsetup{skip=5pt}%
    \caption{(a)}
    \label{fig:Total_ENTROPY}
  \end{subfigure}%
 \begin{subfigure}{0.5\textwidth}        
   \centering
    \includegraphics[width=\textwidth]{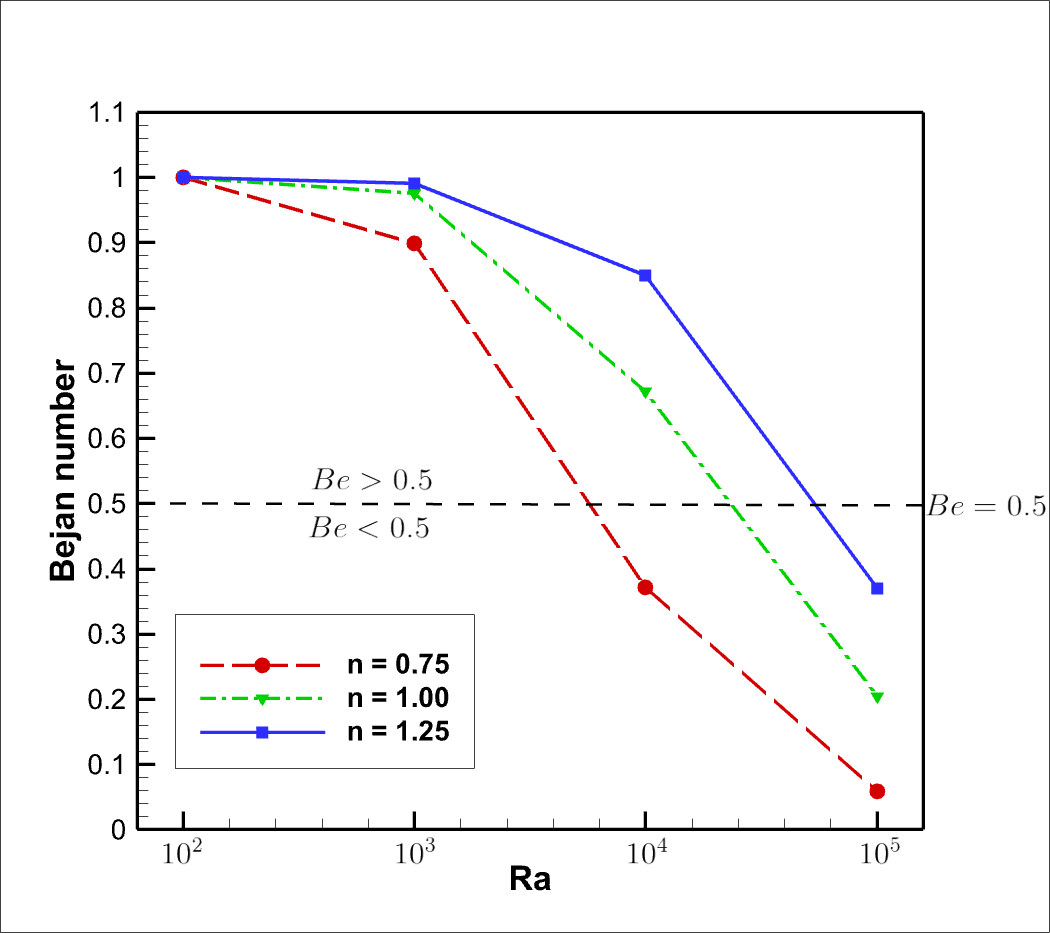}%
    \captionsetup{skip=5pt}%
    \caption{(b)}
    \label{fig:Bajan_Number}
  \end{subfigure}
  \hspace*{\fill}
  \vspace*{8pt}%
  \hspace*{\fill}%
  \caption{ Evolution of (a) Total entropy ($E_{TE}$) versus Rayleigh number ($Ra$)  (b)  Bejan number ($Be$) versus Rayleigh number ($Ra$)}
  \label{fig:Entropy_bajan_both}
\end{figure}

Figure \ref{fig:Entropy_bajan_both}(a) illustrates the influence of \( n \) and \( Ra \) on the total entropy generation ($E_{total}$). At \( Ra = 10^2 \), the $E_{total}$ remains nearly identical across all values of \( n \). However, as \( Ra \) increases, a clear trend emerges: the total entropy generation rises significantly with increasing \( Ra \), while it decreases as the power law index \( n \) increases. This inverse relationship between \( n \) and entropy generation highlights the stronger convective dissipation in shear-thinning fluids compared to Newtonian and shear-thickening fluids, particularly at higher Rayleigh numbers. The exact quantitative data for the total entropy is shown in Table \ref{Total_Entropy_with_Ra_and_n}.

The previous analysis focused on the total entropy generation, which encompasses both fluid friction and heat transfer irreversibilities. However, to understand the relative contributions of these factors, it is crucial to evaluate the Bejan number, \( Be \). Table \ref{Be_with_Ra_and_n} presents the $Be$ values for various $Ra$ and $n$ parameters. For visual comparison, the geometric representation is illustrated in Figure \ref{fig:Entropy_bajan_both}(b). As \( Ra \) increases, \( Be \) decreases, indicating that the contribution of fluid friction-induced entropy becomes more significant compared to that of heat transfer. A value of \( Be < 0.5 \) signifies that the system is predominantly influenced by fluid friction, meaning the losses due to viscosity are greater than the irreversibilities caused by temperature gradients. At \( Ra = 10^4 \), an intriguing result emerges: for \( n = 1 \) (Newtonian fluid) and \( n = 1.25 \) (shear-thickening fluid), \( Be > 0.5 \), suggesting that heat transfer irreversibilities dominate. However, for \( n = 0.75 \) (shear-thinning fluid), \( Be < 0.5 \), implying that fluid friction-induced entropy is more pronounced. This highlights the distinctive behavior of non-Newtonian fluids in influencing entropy generation mechanisms. At \( Ra = 10^5 \), \( Be < 0.5 \) for all values of \( n \), indicating that fluid friction irreversibilities dominate in all cases. However, the Bejan number increases with \( n \), showing that as the fluid becomes more shear-thickening, the relative contribution of heat transfer irreversibilities grows. This suggests that the influence of fluid friction becomes less dominant as the fluid's viscosity becomes more sensitive to shear rate.

{\small\begin{table}[htbp]
\caption{\small Total entropy generation for different $Ra$ and $n$ values. }\label{Total_Entropy_with_Ra_and_n}
\centering
 \begin{tabular}{cccccccccc}  \hline \hline
& Power law index&   $Ra=10^2$ & $Ra=10^3$  & $Ra=10^4$  &  $Ra=10^5$       \\ \hline 
& $n=0.75 $       &  1.042 & 1.414 & 4.503  & 28.205   \\
& $n=1.00$          &  1.041  & 1.143  & 3.237  &    23.758     \\
& $n=1.25$     &  1.041 & 1.082 & 2.284 & 20.796   \\

\hline\hline
 \end{tabular}
\end{table}
}

{\small\begin{table}[htbp]
\caption{\small Bejan number ($Be$) for different $Ra$ and $n$ values. }\label{Be_with_Ra_and_n}
\centering
 \begin{tabular}{cccccccccc}  \hline \hline
& Power law index&   $Ra=10^2$ & $Ra=10^3$  & $Ra=10^4$  &  $Ra=10^5$       \\ \hline 
& $n=0.75 $       &  0.999527  & 0.899659  & 0.371309   & 0.058813    \\
& $n=1.00$          &  0.999708  & 0.975242   & 0.672082   &    0.204057      \\
& $n=1.25$     &  0.999779  & 0.991450  & 0.849350  & 0.370746    \\

\hline\hline
 \end{tabular}
\end{table}
}

\newpage
\section{Conclusion}
In this study, we introduced a novel Higher-Order Super-Compact finite difference scheme to investigate 3D natural convection of non-Newtonian power law fluids within a cubic cavity. Present results provide significant insights into the thermal and fluid flow behavior of shear-thinning, Newtonian, and shear-thickening fluids under varying Rayleigh numbers ($Ra$) and power-law indices ($n$). To the best of our knowledge, this is the first higher-order finite difference scheme to study the 3D non-Newtonian natural convection, and it has demonstrated remarkable accuracy and efficiency in capturing complex flow structures and heat transfer characteristics.

Our results demonstrate that the power-law index ($n$) significantly affects both the flow and thermal fields. For shear-thinning fluids (\(n = 0.75\)), the convective flow is enhanced due to the reduction in viscosity with increasing shear rate, leading to stronger natural convection currents. In contrast, shear-thickening fluids (\(n = 1.25\)) exhibit weaker flow responses due to an increase in viscosity with shear rate, resulting in a more restrained convective behavior. The stream function values consistently decrease with increasing \(n\), reflecting the diminished convective effects in shear-thickening fluids. Isotherm analysis further confirms these trends, with the transition from conduction-dominant to convection-dominant heat transfer being highly sensitive to both \(n\) and \(Ra\). The local Nusselt number distribution reveals that the maximum heat transfer occurs near the thermally active walls and is significantly influenced by both \(Ra\) and \(n\), with shear-thinning fluids exhibiting the largest enhancement in heat transfer. Entropy generation analysis also indicates the complex interplay between heat transfer irreversibilities and fluid friction. As \(Ra\) increases, entropy generation due to fluid friction becomes more prominent, particularly for shear-thinning fluids, where the Bejan number \(Be < 0.5\), indicating the dominance of viscous effects over thermal gradients. For Newtonian and shear-thickening fluids, the entropy generation remains more influenced by heat transfer irreversibilities.

Overall, the HOSC scheme demonstrates its effectiveness and reliability in analyzing 3D non-Newtonian natural convection. This scheme can be further extended to investigate 3D heat transfer in non-Newtonian fluids by incorporating magnetohydrodynamics (MHD), nanofluids or hybrid nanofluids, porous media, and thermal radiation effects. These extensions hold significant potential for a wide range of practical applications. 
\vspace{11pt}\\
{\textbf{Author Declaration}}\\
The authors have no conflicts to disclose.
\vspace{11pt}\\
{\textbf{Data Availability}}\\
The data that support the findings of this study are available from the corresponding author upon reasonable request.
\vspace{11pt}\\
{\textbf{Funding}}\\
This work is partially supported by the \textbf{Scheme for Promotion of Academic and Research Collaboration (SPARC)} under the Ref No. \textbf{SPARC/2024-2025/CAMS/P2742 (May 2024)}.

\end{document}